\documentclass[aps,prd,reprint,11pt,onecolumn,amsmath,amssymb,nofootinbib,groupedaddress,superscriptaddress,floatfix,tightenlines,eqsecnum]{revtex4-2}

\usepackage{macros}

\begin{document}

\title{Massive neutrinos and cosmic composition}

\author{Marilena Loverde}\email[]{mloverde@uw.edu}
\affiliation{Department of Physics, University of Washington, Seattle, Washington, 98195, USA}
\author{Zachary J. Weiner}\email[]{zweiner@perimeterinstitute.ca}
\affiliation{Department of Physics, University of Washington, Seattle, Washington, 98195, USA}
\affiliation{Perimeter Institute for Theoretical Physics, Waterloo, Ontario N2L 2Y5, Canada}

\date{\today}
\begin{abstract}
Cosmological data probe massive neutrinos via their effects on the geometry of the Universe and
the growth of structure, both of which are degenerate with the late-time expansion history.
We clarify the nature of these degeneracies and the individual roles of both probes in neutrino mass
inference.
Geometry is strongly sensitive to neutrino masses: within $\Lambda$CDM, the primary cosmic
microwave background anisotropies alone impose that the matter fraction $\Omega_m$ must increase
fivefold with increasing neutrino mass.
Moreover, large-scale structure observables, like weak lensing of the CMB, are dimensionless and
thus depend not on the matter density (as often quoted) but in fact the matter fraction.
We explore the consequential impact of this distinction on the interplay between probes of
structure, low-redshift distances, and CMB anisotropies.
We derive constraints on the neutrino's masses independently from their suppression of structure and
impact on geometry, showing that the latter is at least as important as the former.
While the Dark Energy Spectroscopic Instrument's recent baryon acoustic oscillation data place
stringent bounds largely deriving from their geometric incompatibility with massive neutrinos, all
recent type Ia supernova datasets drive marginal preferences for nonzero neutrino masses because
they prefer substantially larger matter fractions.
Recent CMB lensing data, however, neither exclude neutrinos' suppression of structure nor constrain
it strongly enough to discriminate between mass hierarchies.
Current data thus evince not a need for modified dynamics of neutrino perturbations or structure
growth but rather an inconsistent compatibility with massive neutrinos' impact on the expansion
history.
We identify two of DESI's measurements that strongly influence its constraints, and we also discuss
neutrino mass measurements in models that alter the sound horizon.
\end{abstract}

\maketitle
\makeatletter
\def\l@subsubsection#1#2{}
\makeatother
\tableofcontents

\section{Introduction}
\label{sec:introduction}

On both theoretical and observational grounds, cosmology is no longer well described by the
six-parameter $\Lambda$ cold dark matter (\LCDM{}) model.
While current data from any single probe---cosmic microwave background (CMB) anisotropies, baryon
acoustic oscillations (BAO), low-redshift distances from type Ia supernovae (SNe), and large-scale
structure (LSS)---are \emph{individually} well described by \LCDM{}, many of these observables
prefer mutually inconsistent parameter space.
Moreover, laboratory experiments definitively establish that neutrinos are not massless, and
cosmological datasets are already sensitive to the gravitational effects of their nonrelativistic
dynamics.
Neglecting the neutrinos' masses or fixing them to some particular value imposes a strong prior that
is neither theoretically nor experimentally justified.
And while the critical issue remains corroborating whether discrepancies between datasets are
cosmological or systematic in nature, their consistency must be assessed in the full parameter space
that describes known physics---including neutrino masses.

On the one hand, the CMB~\cite{Planck:2018vyg, Planck:2019nip, Planck:2018lbu} infers (within
\LCDM{}) an amplitude of matter clustering larger than that measured by some probes of
LSS~\cite{Ivanov:2019pdj, KiDS:2020suj, DES:2022qpf, Abdalla:2022yfr, Chen:2024vuf, DESI:2024rsk}.
But massive neutrinos constitute hot dark matter---their thermal velocities permit them to stream
in and out of CDM overdensities rather than cluster in tandem.
The suppression of net structure growth by neutrinos, which accumulates over a large interval of
expansion, is typically considered the most promising observable with which to measure their
masses~\cite{Hu:1997mj}, motivating future surveys of CMB lensing~\cite{Lesgourgues:2005yv,
CMB-S4:2016ple, SimonsObservatory:2018koc}, redshift-space distortions~\cite{Font-Ribera:2013rwa,
Dore:2019pld}, weak lensing of galaxies~\cite{Archidiacono:2016lnv, DES:2022ccp,
LSSTDarkEnergyScience:2012kar, Euclid:2024imf}, cluster abundances~\cite{Mantz:2014paa}, and other
probes of the late time matter field~\cite{Dvorkin:2019jgs, Ferraro:2022cmj, Gerbino:2022nvz}.
The possibility that massive neutrinos could reconcile current determinations of clustering is
complicated in part by a preference in \Planck{}'s 2018 CMB data~\cite{Planck:2018vyg,
Planck:2019nip, Planck:2018lbu} for even greater smearing of the acoustic peaks induced by
gravitational lensing than it predicts without massive neutrinos.
Adding to the intrigue, a phenomenological analysis of peculiar velocity measurements and
redshift-space distortions finds evidence that structure grows more slowly than \LCDM{}
predicts~\cite{Nguyen:2023fip}.

On the other hand, calibrating the local distance ladder with Cepheids~\cite{Riess:2016jrr,
Riess:2019cxk, Riess:2021jrx, Freedman:2024eph} measures a present-day expansion rate $H_0$ severely
larger than that inferred (again, within \LCDM{}) by the distance to the last scattering surface via
CMB data~\cite{Planck:2018vyg}.
Alternative calibrations using the tip of the red giant branch~\cite{Freedman:2019jwv,
Freedman:2020dne, Freedman:2021ahq, Freedman:2023jcz, Freedman:2024eph} yield results that are less
discrepant but still shifted toward larger $H_0$.
Massive neutrinos, however, increase the density (and so decrease the horizon) of the Universe in
the matter-dominated era, requiring a compensatory decrease in density during the dark-energy epoch
to preserve the distance to last scattering.
Without additional physics, cosmologies with massive neutrinos thus only permit smaller Hubble
constants than those without.
Moreover, because dark energy domination began rather recently, fixing the distance to last
scattering requires so disproportionate a decrease in the dark energy density that the matter
fraction $\Omega_m$ must increase \emph{fivefold} compared to the neutrinos' relative contribution
to the matter density.\footnote{
    Extending late-time cosmology beyond the flat \LCDM{} model we focus on introduces additional
    freedom that can break this relationship; such extensions may be tested directly by low-redshift
    distance measurements.
}
Direct measurements of the late-time expansion history can break this degeneracy and stand to
provide powerful, independent constraints on the neutrino masses.
While both structure growth and geometric probes are sensitive to possible dynamics of dark energy,
low-redshift distances can directly constrain the possibility.
Furthermore, geometric probes suffer no degeneracy with the optical depth to reionization, a primary
obstacle to uniquely inferring neutrino masses from the amplitude of
structure~\cite{Kaplinghat:2003bh, CMB-S4:2016ple}.

\begin{figure}[t!]
\begin{centering}
    \includegraphics[width=\textwidth]{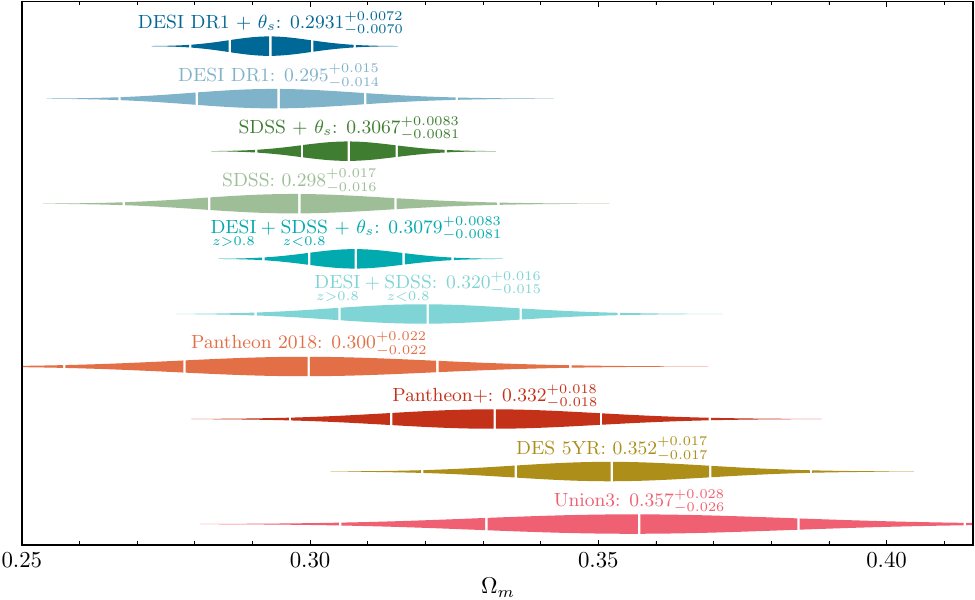}
    \caption{
        Recent measurements of the matter fraction $\Omega_m$ inferred within flat \LCDM{} cosmology
        from various probes of late-time geometry (described in the text).
        White lines indicate the median and $\pm 1$ and $2 \sigma$ values, and each posterior is
        labeled by its median and $\pm 1 \sigma$ deviations.
    }
    \label{fig:matter-fraction-measurements}
\end{centering}
\end{figure}

More recently, a number of datasets that directly measure the late-time expansion history with
standardizable cosmological distances yield discrepant inference of the Universe's composition.
A great success of \LCDM{} in the previous decade is its concordant inference of both $H_0$ and the
present matter fraction $\Omega_m$ from CMB anisotropies and via the BAO feature measured in galaxy
surveys at low redshift.
But the most recent SNe datasets---Pantheon+~\cite{Brout:2022vxf, Scolnic:2021amr}, the five-year
sample from the Dark Energy Survey (DES)~\cite{DES:2024tys}, and Union3~\cite{Rubin:2023ovl}---all
prefer a Universe comprising a greater fraction in matter than any BAO dataset (see
\cref{fig:matter-fraction-measurements}).
The sensitive increase of $\Omega_m$ with increasing neutrino mass (at fixed distance to last
scattering) in principle provides a means to reconcile CMB and SNe data.
But cosmological distances measured by SNe and BAO---even marginalizing over the calibration of their
brightness and the sound horizon, respectively---directly trace the expansion history over the same
interval of redshift, so it is unclear what cosmological (rather than systematic) effect could
account for this apparent violation of the distance-duality relation~\cite{etherington}.

The first BAO measurements released by the Dark Energy Spectroscopic Instrument
(DESI)~\cite{DESI:2024mwx, DESI:2024lzq, DESI:2024uvr} yield neutrino mass bounds so stringent as to
suggest incompatibility with the minimal expectation from neutrino oscillation
experiments~\cite{Esteban:2020cvm, deSalas:2020pgw}.
These tighter bounds derive not from improved precision---DESI's data has not yet exceed that of the
Sloan Digital Sky Survey (SDSS) and its Extended Baryon Oscillation Spectroscopic Survey (eBOSS)
measurements~\cite{eBOSS:2020yzd, eBOSS:2020lta, eBOSS:2020hur, Ross:2014qpa}---but rather from a
preference for even smaller matter fractions than SDSS~\cite{DESI:2024mwx} that further preclude
sizable neutrino masses when combined with CMB data.
Data from luminous red galaxies that are discrepant with SDSS's measurements at similar redshifts
play an important role in this trend; \cref{fig:matter-fraction-measurements} shows that exchanging
DESI's measurements for SDSS's at low redshift yields inference of larger matter fractions more
compatible with those from SNe (and also with heavier neutrinos).
Moreover, when also including the analogous information from \Planck{} (i.e., the angular extent of
the photon sound horizon, $\theta_s$), this combination agrees extremely well with SDSS's, but both
results are offset from DESI's.
If merely due to statistical fluctuations, subsequent data releases could counterintuitively produce
weaker constraints on neutrino masses in spite of improved precision.

The advent of precision cosmology has thus yielded precision discrepancies with crucial implications
for inferences of the neutrino mass scale.
Identifying genuine signatures of massive neutrinos---or the absence of ones that are
expected---requires a rigorous theoretical understanding of what cosmological observables measure,
what degeneracies with \LCDM{} parameters arise, and how those degeneracies are broken by combining
datasets.
In this work we clarify the manner in which the various effects of massive neutrinos are degenerate
with the composition of the Universe.
Emphasizing that (nearly) all cosmological observables measure dimensionless quantities, we
demonstrate that the signatures of massive neutrinos---quantified by their nonrelativistic
contribution to the matter density relative to that from baryons and cold dark matter---are
degenerate with matter's fractional contribution to the Universe's energy density, not its absolute
density.
\Cref{sec:signatures} shows how this degeneracy arises in neutrinos' effects on both the geometry
and the large-scale structure of the Universe.
CMB data are crucial in breaking these degeneracies; \cref{sec:mass-independent} reviews what
information (and what dimensionless combinations of \LCDM{} parameters) are robustly measured by CMB
anisotropies independent of the neutrino masses.

In \cref{sec:results} we apply the theoretical description of \cref{sec:signatures} to interpret
constraints on neutrino masses from current cosmological data.
We identify combinations of datasets that yield constraints on both neutrinos' geometric effects and
their suppression of structure independently of the other; we specialize to CMB lensing as a probe
of structure for concreteness and because of current observations' important role in neutrino mass
inference.
\Cref{sec:geometric-tensions} studies the former---measurements of geometry from uncalibrated SNe
data and from BAO data.
We show how recent SNe datasets combine with geometric information from \Planck{} to detect a
nonzero neutrino mass sum with significance as high as $3 \sigma$, in striking contrast to the
strong upper limits from DESI data.
In \cref{sec:desi} we investigate the origin of DESI's constraints, exploring what subsets of DESI's
data and combinations with SDSS yield weaker constraints on the neutrino masses as well as
inferences of the matter fraction consistent with SNe.
\Cref{sec:lensing-constraints} then derives bounds purely on neutrinos' suppression of structure
using CMB lensing data from \Planck{} and the Atacama Cosmology Telescope (ACT)~\cite{ACT:2023kun,
ACT:2023dou}.
Prior bounds derived from recent CMB lensing data always include BAO data as well; the results we
present that exclude this data better represent the independent constraining power provided by CMB
lensing as a probe of structure.
We then present constraints from various combinations of probes of geometry and structure, showing
that a modest preference for neutrinos with mass sums $\approx 0.14~\mathrm{eV}$ persists with DES's
SNe data even when combined with all \Planck{} data and the aforementioned CMB lensing data.

Finally, in \cref{sec:discussion} we discuss the implications of our results in interpreting current
datasets and on possible new physics.
\Cref{sec:tensions} surveys the implications of our results for the numerous tensions between
cosmological datasets.
We discuss cosmologies that alter the sound horizon in \cref{sec:modified-cosmology}: neutrino-mass
bounds are largely robust in models that increase the density at recombination with new degrees of
freedom, but early-recombination scenarios effectively remove any geometric information about
neutrino masses.
In \cref{sec:modifications-to-neutrinos} we comment on the interpretation of current data as
preferring neutrinos with ``negative mass'' and on new physics in the neutrino sector that might
explain discrepant signatures in the Universe's geometry and structure.
We conclude in \cref{sec:conclusions}.
A set of appendices describe the derivation of a geometric likelihood from \Planck{}
(\cref{app:bao-from-cmb}), present a number of supplementary figures
(\cref{app:supplemental-results}), and demonstrate the importance of nonlinear structure growth on
inference of neutrino masses from current CMB lensing data (\cref{app:nonlinear}).

\section{Cosmology of massive neutrinos}
\label{sec:signatures}

We begin by reviewing the separate cosmological impact of massive neutrinos on the geometry
(\cref{sec:geometry}) and large-scale structure (\cref{sec:structure}) of the
Universe.
In each case we establish formalism and illustrate how relevant observables depend only on
dimensionless quantities that characterize the composition of the Universe (namely, the matter
fraction $\Omega_m$ in flat \LCDM{} cosmologies).
Within even the most conservative current limits, neutrinos become nonrelativistic only after
recombination---otherwise, their transition from radiationlike to matterlike behavior hastens the
onset of matter domination and alters the early integrated Sachs-Wolfe (ISW) effect to a
unacceptable degree~\cite{Planck:2018vyg, Hou:2012xq}.
In the regime in which their impact on the ISW effect is negligible, massive neutrinos affect
cosmological observables only by altering distances (including that to last scattering) and
suppressing the late-time growth of structure (which distorts the visible CMB via gravitational
lensing).
Before proceeding, we establish notation and (in \cref{sec:mass-independent}) discuss what
information the CMB robustly measures independently of the neutrino masses.

A single neutrino species with mass $m_{\nu_i}$ and temperature $T_\nu$ becomes semirelativistic
(i.e., on average over phase space) at a redshift
\begin{align}
    z_{\nu_i} + 1
    = \frac{m_{\nu_i}}{3.15 T_\nu(a_0)}
    = 113 \,
        \frac{m_{\nu_i}}{0.06~\mathrm{eV}}
    \label{eqn:massive-neutrino-z-NR}
\end{align}
and has a present-day density\footnote{
    The contributions of various species to the present-day energy density are parametrized by
    $\omega_X = \bar{\rho}_{X, 0} / 3 H_{100}^2 \Mpl^2$, i.e., $X = \gamma$, $b$, $c$, $\nu$,
    and $\Lambda$ denoting photons, baryons, cold dark matter (CDM), neutrinos, and cosmological
    constant, respectively.
    We parametrize the Hubble constant by
    $H_0 = h \cdot 100~\mathrm{Mpc}^{-1} \mathrm{km} / \mathrm{s} \equiv h H_{100}$.
}
of
\begin{align}
    \omega_{\nu_i}
    = \frac{3 \zeta(3)}{2 \pi^2} \frac{m_{\nu_i} T_\nu(a_0)^3}{3 H_{100}^2 \Mpl^2}
    &= \frac{m_{\nu_i}}{93.1~\mathrm{eV}}.
    \label{eqn:massive-neutrino-density}
\end{align}
\Cref{eqn:massive-neutrino-z-NR} takes $3.15 T_\nu$ as the average momentum for a
Fermi-Dirac--distributed species, and the right-hand sides of
\cref{eqn:massive-neutrino-z-NR,eqn:massive-neutrino-density}
account for phase space distortions generated when neutrinos decouple from the
plasma~\cite{Mangano:2005cc}.
Massive neutrinos therefore contribute to the matter density below a redshift and with an abundance
both proportional to their mass.
Neutrino oscillation experiments limit the sum of the neutrino masses to at least
$0.0588~\mathrm{eV}$ or $0.099~\mathrm{eV}$ in the normal and inverted hierarchies,
respectively~\cite{deSalas:2020pgw, Esteban:2020cvm}.
Since the visibility function peaks around a photon temperature of $0.26~\mathrm{eV}$ (or redshift
$z_\star \approx 1089$), the aforementioned requirement that neutrinos become nonrelativistic after
recombination limits each species to be lighter than about $0.58~\mathrm{eV}$ (barring any
nonminimal neutrino dynamics).
Unless otherwise stated, we approximate the mass hierarchy with three equal-mass neutrinos per
standard practice~\cite{Archidiacono:2020dvx}, as their mass splittings only have a minor impact on
their signatures and only for mass sums close to the minimum.

The early-time energy density of neutrinos is typically quantified in terms of the effective number
of degrees of freedom $N_\mathrm{eff}$ as
\begin{align}
    \bar{\rho}_\nu(z \gg z_{\nu_i})
    &= \frac{7}{8} \left( \frac{4}{11} \right)^{4/3} N_\mathrm{eff} \cdot \bar{\rho}_\gamma.
    \label{eqn:omega-nu-omega-gamma}
\end{align}
Following convention, the total radiation density parameter denotes the sum of the relativistic
neutrino and photon contributions, $\omega_r \equiv \omega_\gamma + \omega_\nu$, as if the neutrinos
remained relativistic today.
We denote the total matter density (i.e., at late times) as
$\omega_m(z \ll z_{\nu_i}) = \omega_b + \omega_c + \omega_\nu$; $\omega_\nu$ here denotes the
density in nonrelativistic neutrinos, which comprise a fraction
\begin{align}
    f_\nu
    &\equiv \frac{\omega_\nu}{\omega_m}
    = \frac{\omega_\nu}{\omega_b + \omega_c + \omega_\nu}
    \approx \frac{\summnu}{13.2~\mathrm{eV}}
        \left( \frac{\omega_b + \omega_c}{0.142} \right)^{-1}
    \label{eqn:neutrino-fraction}
\end{align}
of the total matter density.
Here $\summnu = \sum_i m_{\nu_i}$ is the neutrino mass sum.
When modeling $\mathcal{N}_\nu$ massive neutrino states, we fix the total early-time radiation
density to correspond to $N_\mathrm{eff} = 3.044$
[\cref{eqn:omega-nu-omega-gamma}]~\cite{Akita:2020szl, Froustey:2020mcq, Bennett:2020zkv} by
including an ultrarelativistic fluid with
$N_\mathrm{eff} - \mathcal{N}_\nu T_\nu^4 / (4/11)^{4/3}$ effective degrees of freedom.

\subsection{Cosmological information independent of neutrino masses}
\label{sec:mass-independent}

The primary CMB anisotropies are generated by the dynamics of the photon-baryon plasma leading up to
hydrogen recombination.
Recombination happens to occur shortly after the radiation density drops below the total matter
density, granting the primary anisotropies sensitivity to dark matter perturbations via their
gravitational effects on the plasma.
As established above, before recombination Standard Model neutrinos are still relativistic but are
also collisionless; their free-streaming behavior while relativistic also has characteristic effects
on, e.g., the location of the CMB acoustic peaks~\cite{Bashinsky:2003tk, Baumann:2015rya}, but these
signatures are independent of their mass.
To describe the physical effects constrained by the CMB, we follow the notation and arguments of
Ref.~\cite{Baryakhtar:2024rky}, which took as time coordinate the scale factor relative to that at
recombination, $x = a / a_\star$, and assumed the energy content of the early Universe is well
described by photons, baryons, relativistic neutrinos, and cold dark matter.

The most robust observable in the CMB power spectrum is the structure and location of its peaks,
which reflect the oscillations of acoustic waves traveling in the plasma at the time of
recombination.
The relative peak heights reflect the shift in the zero point of the density perturbations'
oscillations due to the weight of the baryons relative to the photons' radiation pressure;
with their precise measurements of the first several peaks, current datasets strongly constrain the
baryon-to-photon ratio at recombination,
\begin{align}
    R_\star
    &\equiv \frac{3 \rho_{b, \star}}{4 \rho_{\gamma, \star}}
    \propto \frac{3 \omega_b a_\star}{4 \omega_\gamma}.
    \label{eqn:R-star}
\end{align}
Acoustic waves propagate a distance (the so-called sound horizon) determined by the sound speed of
the tightly coupled photon-baryon plasma and the comoving horizon of the Universe.
Because recombination occurs in the beginning of the matter era, the evolution of the sound horizon
depends on when matter-radiation equality occurs relative to recombination,
\begin{align}
    x_\mathrm{eq}
    &\equiv \frac{a_\mathrm{eq}}{a_\star}
    = \frac{\rho_{r, \star}}{\rho_{m, \star}}
    = \frac{\omega_r}{\left( \omega_b + \omega_c \right) a_\star}.
    \label{eqn:x-eq}
\end{align}
Cold dark matter makes gravitational potentials deeper during recombination to an extent that is
also fully parametrized by $x_\mathrm{eq}$, determining both of the so-called radiation driving and
early ISW effects~\cite{Hu:1996mn, Hu:1995en}.
When $R_\star$ and $x_\mathrm{eq}$ are held fixed, the ratio of the sound horizon to the comoving
horizon at equality $k_\mathrm{eq} \equiv a_\mathrm{eq} H_\mathrm{eq}$ is invariant.
The shape of the acoustic peaks therefore measures the abundances of both baryons and cold dark
matter, but only at recombination and relative to the photon and radiation abundances,
respectively.\footnote{
    To be clear, the density ratios evaluated at recombination are fully sufficient to
    \emph{parametrize} the impact of baryons and CDM on the CMB, assuming (as we are) that
    their evolution with redshift is unchanged.
    The primary CMB does depend on the conditions of the plasma over a finite duration in time about
    recombination (determined by the width of the visibility function) and on the evolution of
    metric potentials at all times (via the ISW effect).
}

Translating these dimensionless quantities to dimensionful, present-day baryon and CDM abundances
requires knowledge of the photon and radiation densities and the scale factor at recombination.
Since the baryon-to-photon ratio encodes the effect of their interactions, inferring the absolute
density $\omega_b$ in this manner should be robust to any impact of massive neutrinos.
Sufficiently heavy neutrinos could confound the inference of $\omega_c$ as
$\omega_r / a_\star x_\mathrm{eq} - \omega_b$ via their own transition to matterlike behavior, but
only for masses that are safely excluded by current data.
We emphasize that \cref{eqn:x-eq} depends only on whatever contributors to the matter density are
present at recombination.

One of the few measured, dimensionful parameters relevant to the CMB is its present-day temperature,
$T_0$.
The scale factor at recombination relative to today, of course, is given by the inverse ratio of the
corresponding photon temperatures, $a_\star / a_0 = T_0 / T_\star$.
The temperature dependence of recombination largely arises relative to the ionization energy
of hydrogen $E_I$, which of course is also measured.
Assuming atomic parameters are unchanged at early times (i.e., no time variation of the fundamental
constants) thus anchors the redshift of recombination.\footnote{
    Recombination is marginally sensitive to the expansion history, but $T_\star$ varies at below
    the permille level within the limits of current CMB data if recombination is described by the SM.
}
Finally, the (relativistic) neutrino abundance is predicted precisely in the Standard
Model~\cite{Mangano:2005cc, Akita:2020szl, Froustey:2020mcq, Bennett:2020zkv, Cielo:2023bqp};
assuming no new physics again, measurements of $T_0$ and $E_I$ calibrate a prediction for the
\emph{dimensionful} photon and radiation densities at recombination.
The shape of the acoustic peaks then measures the dimensionful abundances of baryons and CDM.

Notably, calibrating $\omega_b$ and $\omega_c$ in \LCDM{} does not hinge even on the angular scale
of the acoustic peaks.
The actual extent of these features on the sky instead measures the ratio of the sound horizon and
the distance to last scattering.
The former may be computed analytically in matter-radiation Universes~\cite{Eisenstein:1997ik} and
is a function of the matter densities only via $R_\star$ and $x_\mathrm{eq}$:
\begin{align}
    r_{s}(a)
    &= \frac{a}{\sqrt{\omega_r}}
        \frac{2 \sqrt{ x_\mathrm{eq} / 3 R_\star } }{H_{100} / c}
        \ln \left(
            \frac{
                \sqrt{R_\star} \sqrt{x + x_\mathrm{eq}}
                + \sqrt{1 + R_\star x}
            }{
                1 + \sqrt{R_\star x_\mathrm{eq}}
            }
        \right)
    \label{eqn:rs-in-matter-radiation-ito-wr-R_star}
\end{align}
where again $x = a / a_\star$.
The distance to last scattering, on the other hand, depends on the expansion history in the matter
and dark-energy eras; its dependence on the matter density is therefore not specified by the
combination $x_\mathrm{eq}$.
Moreover, massive neutrinos become nonrelativistic during this period and alter the late-time
expansion history in a mass-dependent manner, a geometric signature we discuss in
\cref{sec:geometry}.

Finally, we note several additional physical processes to which (primary) CMB polarization and
small-scale anisotropies are uniquely sensitive.
Small-scale power is suppressed by photon diffusion~\cite{Hu:1995em, Zaldarriaga:1995gi} and is
sensitive to the ratio of the photon and total radiation densities, for example.
That recombination is not instantaneous---namely, that CMB photons visible today last scattered over
a nonzero interval in time---both suppresses small-scale power~\cite{Zaldarriaga:1995gi,
Weinberg:2008zzc} and determines the amplitude of the polarization
spectrum~\cite{Zaldarriaga:1995gi, Hu:1997hv}.
Reference~\cite{Baryakhtar:2024rky} discusses the parametrization of these effects and parameter
degeneracies in extended models in detail.
That current measurements of polarization and small-scale anisotropies are entirely consistent with
\LCDM{} (and the SM) without the need for additional parameter freedom is a definitive success of
the model.

\subsection{Geometric signatures}
\label{sec:geometry}

The classical cosmological probes of geometry use astrophysical measurements of a quantity that is
dimensionful but supposed to be uniform (or at least predictively determined) across redshift.
Any observed variability of these otherwise standard quantities is then associated to a cosmological
distance, and associating these observations with measured redshifts then traces out the expansion
history~\cite{SupernovaCosmologyProject:1996grv}.
Historically, the luminosities of supernovae (``standard candles'') provided the first probe of
cosmic distances, but the spatial correlation of the matter and photon distributions due to acoustic
oscillations also provides a standard distance scale (or ``ruler'')~\cite{Hu:2003ti,
Seo:2003pu,Blake:2003rh, 2dFGRS:2005yhx, Eisenstein:2006nk}.
Distance measurements thus constrain the geometry of the Universe---both spatial curvature and the
evolution of the scale factor.

The fundamental distance measure is the transverse comoving distance~\cite{Hogg:1999ad},
\begin{align}
    D_M(a)
    &\equiv \chi(a) \sinc \left[ \sqrt{-\Omega_k} \frac{\chi(a)}{1 / H_0} \right],
    \label{eqn:transverse-distance}
\end{align}
where $\Omega_k$ is the curvature parameter and the line-of-sight comoving distance is
\begin{align}
    \chi(a)
    &= \int_{a}^{a_0} \frac{\ud \tilde{a}}{\tilde{a}} \,
        \frac{1}{\tilde{a} H(\tilde{a})}.
    \label{eqn:comoving-distance}
\end{align}
In flat Universes (with $\Omega_k = 0$), $D_M(a) = \chi(a)$.
Specializing to \LCDM{}, whose expansion history at late times is fully characterized by the scale
factor at matter-$\Lambda$ equality
\begin{align}
    \amL
    &= \sqrt[3]{\frac{\omega_m}{\omega_\Lambda}}
    = \sqrt[3]{\frac{\Omega_m}{1 - \Omega_m}}
    \label{eqn:aml-ito-Wm}
\end{align}
and the Hubble constant $h = \sqrt{\omega_m + \omega_\Lambda}$, the comoving distance
\cref{eqn:comoving-distance} integrates to~\cite{Baes:2017rfj}
\begin{align}
    \chi(a)
    &= \frac{2}{\sqrt{\omega_m} H_{100} / c}
        \left[ F_M(1; \amL) - F_M(a; \amL) \right],
    \label{eqn:comoving-distance-matter-Lambda}
\end{align}
where
\begin{align}
    F_M(a; \amL)
    &\equiv \sqrt{a} \cdot {}_2{F}_1(1/6, 1/2; 7/6, - [a / \amL]^3)
    \label{eqn:matter-lambda-distance-function}
\end{align}
and ${}_2{F}_1$ is a hypergeometric function.
The hypergeometric function asymptotes to unity for $a \ll \amL$, i.e., it encodes the correction
from $\Lambda$ to the result for a pure matter Universe.

\subsubsection{Baryon acoustic oscillations}\label{sec:bao}

The acoustic oscillations of the baryonic plasma also leave an oscillatory feature in the matter
correlation function; the local maximum thereof (the ``BAO scale'') effectively measures the
comoving sound horizon when baryons decoupled, $r_\mathrm{d}$~\cite{Thepsuriya:2014zda,
Schoneberg:2019wmt}.
This so-called drag scale is slightly larger than $r_{s, \star}$ because baryons decouple slightly
after recombination ($a_\mathrm{d} > a_\star$)~\cite{Eisenstein:1997ik}.
Galaxy surveys extract the BAO scale in various forms: the transverse scale (as an angular extent on
the sky),
\begin{subequations}\label{eqn:theta-bao}
\begin{align}
    \theta_{\mathrm{BAO}, \perp}(a)
    &\equiv \frac{r_\mathrm{d}}{D_M(a)}
    \label{eqn:theta-bao-perp}
    ,
\end{align}
the line-of-sight scale,\footnote{
    Note that $\theta_{\mathrm{BAO}, \parallel}$ is written as a ratio of comoving and physical
    length scales following convention; the stray factor of $a$ is irrelevant for BAO data
    because their redshifts are measured.
}
\begin{align}
    \theta_{\mathrm{BAO}, \parallel}(a)
    &\equiv \frac{r_\mathrm{d}}{c H(a)^{-1}}
    \label{eqn:theta-bao-parallel}
    ,
\end{align}
and a volume average of the two,
\begin{align}
    \theta_\mathrm{BAO}(a)
    &\equiv \sqrt[3]{
            \frac{
                \theta_{\mathrm{BAO}, \perp}(a)^2
                \theta_{\mathrm{BAO}, \parallel}(a)
            }{
                \left( 1 / a - 1 \right)
            }
        }
    \label{eqn:theta-bao-DV}.
\end{align}
\end{subequations}
Substituting \cref{eqn:comoving-distance-matter-Lambda} for the comoving distance in
\cref{eqn:theta-bao-perp}
and
\begin{align}
    H(a)
    &= \sqrt{\omega_m} H_{100} \sqrt{a_{m-\Lambda}^{-3} + a^{-3}}
    \label{eqn:matter-lambda-hubble-rate}
\end{align}
into \cref{eqn:theta-bao-parallel} shows that BAO distances depend on $H_0$ only through an overall
scaling in the combination $r_\mathrm{d} \sqrt{\omega_m} = r_\mathrm{d} \sqrt{\Omega_m} H_0$:
\begin{subequations}\label{eqn:theta-bao-flat-lcdm}
\begin{align}
    \theta_{\mathrm{BAO}, \perp}(a)
    &= \frac{r_\mathrm{d} \sqrt{\omega_m}}{2 c / H_{100}}
        \left[ F_M(1; \amL) - F_M(a; \amL) \right]^{-1}
    \label{eqn:theta-bao-perp-flat-lcdm}
    \\
    \theta_{\mathrm{BAO}, \parallel}(a)
    &= \frac{r_\mathrm{d} \sqrt{\omega_m}}{c / H_{100}}
        \sqrt{a_{m-\Lambda}^{-3} + a^{-3}}
    \label{eqn:theta-bao-parallel-flat-lcdm}
    .
\end{align}
\end{subequations}
Parametrizing the ``amplitude'' as $r_\mathrm{d} \sqrt{\omega_m}$ rather than $h r_\mathrm{d}$ both
yields simpler expressions and better facilitates interpreting the implications of BAO measurements
for neutrino masses later on.

\Cref{eqn:theta-bao-perp-flat-lcdm,eqn:theta-bao-parallel-flat-lcdm} both depend on scale factor
through a (nonlinear) function depending on cosmological parameters only via $\amL$ (or equivalently
$\Omega_m$).
At very early times, the scaling of the BAO angles with the matter fraction asymptotes to
$\theta_{\mathrm{BAO}, \perp}(a) \propto r_\mathrm{d} \sqrt{\omega_m} \Omega_m^{-0.1}$
and $\theta_{\mathrm{BAO}, \parallel}(a) \propto r_\mathrm{d} \sqrt{\omega_m} a^{-3/2}$.
At very low redshift, both scale nearly with $r_\mathrm{d} h$ (since they only depend on the very
recent expansion history); combining measurements at moderate or high redshift with ones at very low
redshift can therefore measure both $r_\mathrm{d} \sqrt{\omega_m}$ and $\Omega_m$ individually.
Measuring both $\theta_{\mathrm{BAO}, \perp}(a)$ and $\theta_{\mathrm{BAO}, \parallel}(a)$ at a
single redshift also breaks the degeneracy, but only weakly because they scale with powers of
$\Omega_m$ differing by no more than $0.15$ to $0.2$ in magnitude at any redshift (unless $\Omega_m$
is extremely small).

The BAO feature is measured by tracers from small redshift to redshifts of order a couple, well
after the heaviest neutrinos become nonrelativistic.\footnote{
    The BAO scale is extracted by marginalizing over smooth features in the two-point
    function~\cite[see, e.g.,][]{DESI:2024uvr, DESI:2024lzq}, and should therefore be estimated in
    a manner insensitive to the scale-dependent suppression of structure by massive neutrinos.
}
Taking $\omega_m$ to include all of the mass density in neutrinos at late times is thus an excellent
approximation.
BAO data on their own, however, cannot identify how the matter abundance is partitioned between
baryons, CDM, and neutrinos; nor can they even infer the overall density without calibrating the
drag horizon $r_\mathrm{d}$ (i.e., measuring a dimensionful scale).
Using the information in the shape of the CMB power spectra (\cref{sec:mass-independent}) to
calibrate the absolute baryon and CDM densities (i.e., with assumptions about early-Universe
physics), within flat \LCDM{} cosmologies one can attribute to massive neutrinos any remaining
matter density measured by BAO data.

\subsubsection{Cosmic microwave background}\label{sec:cmb-geometry}

The angular location of the acoustic peaks in the CMB is one of its most robust features, measuring
the ratio of the sound horizon at recombination to the comoving distance to last scattering,
$\theta_s \equiv r_s(a_\star) / D_M(a_\star)$.
This geometric information is precisely analogous to the transverse BAO scale (and is largely robust
to changes in the shape of the CMB power spectrum); in fact, though the CMB is most sensitive to the
sound horizon at photon decoupling (being a measurement of the photon distribution), within \LCDM{}
and common extensions the evolution of the sound horizon between photon and baryon decoupling varies
minimally with parameters~\cite{Lin:2021sfs}.
The CMB therefore constrains the transverse BAO scale \cref{eqn:theta-bao-perp} nearly as precisely
as it does $\theta_s$; we describe this inference in more detail in \cref{app:bao-from-cmb} and,
since they contain the same physical information, discuss the impact of massive neutrinos on
$\theta_s$ next.

Again excluding neutrinos so heavy as to become nonrelativistic before recombination, the abundance
of all matter relevant at late times is necessarily larger than that relevant at early times (i.e.,
$\omega_b + \omega_c$) by an amount determined by the neutrino's masses.
In both the normal and inverted hierarchies, the neutrino species that comprise the majority of the
neutrino mass density necessarily become nonrelativistic long before matter-$\Lambda$ equality.
The integral that defines the distance to last scattering $\chi_\star \equiv \chi(a_\star)$
[\cref{eqn:comoving-distance}] has most of its support around matter-$\Lambda$ equality, since $1/a H$ increases in
the matter era and decreases during dark-energy domination; for illustrative purposes, $\chi_\star$
may be approximated by simply taking $\omega_m$ to always include the full mass density of the
neutrinos.
For any mass hierarchy, the resulting error is comparable to the standard deviation of \Planck{}'s
measurement of $\theta_s$ [$\sim 4 \times 10^{-4}$; see \cref{eqn:planck-theta-perp-measurement}]
for the minimal mass sum and quadruples by $\summnu = 1~\mathrm{eV}$.
\Planck{} measures the angular drag scale so precisely, however, that this error (or even many
multiples of it) is negligible compared to the precision of any other probe.

To understand the parametric dependence of the angle $\theta_s$, note that
$F_M(1; \amL) \approx 1.014 \, \Omega_m^{0.094}$ is accurate at the subpercent level for
$\Omega_m > 0.1$ and the subpermille for $0.23 < \Omega_m < 0.39$.\footnote{
    Beware that these errors are larger than that with which \Planck{} measures $\theta_s$ and,
    moreover, are smaller than that from neglecting the contribution of radiation to
    $\chi(a_\star)$.
    Avoiding these errors, which would clearly be overly pedantic for our present, illustrative
    purposes, is imperative for quantitative analyses (see \cref{app:bao-from-cmb}).
}
Then
\begin{align}
    \theta_s
    &\approx \frac{r_s \sqrt{\omega_m} / (2 c / H_{100})}{1.014 \Omega_m^{0.094} - \sqrt{a_\star}}
\end{align}
which, neglecting $F_M(a_\star, \amL) \approx \sqrt{a_\star} \sim 10^{-2}$ in the denominator
and taking $r_s \propto r_\mathrm{d}$ (see \cref{app:bao-from-cmb}), means the CMB's geometric
degeneracy in flat \LCDM{} cosmologies lies along
\begin{align}
    \left. \Omega_m \right\vert_{\theta_s}
    \propto \left( \omega_m r_\mathrm{d}^2 \right)^{5.32},
    \label{eqn:cmb-geometric-degeneracy-general}
\end{align}
independent of early-time physics.
To infer the implications of \cref{eqn:cmb-geometric-degeneracy-general} for cosmological parameters
(including neutrino masses), write the ratio of the late- and early-time matter densities in terms
of the neutrino fraction \cref{eqn:neutrino-fraction}:
\begin{align}
    \frac{\omega_b + \omega_c + \omega_\nu}{\omega_b + \omega_c}
    &= \frac{1}{1 - f_\nu}
    .
    \label{eqn:fnu-ito-bcnu}
\end{align}
Replacing $\omega_m$ in \cref{eqn:comoving-distance-matter-Lambda} with
$\omega_r / (1 - f_\nu) a_\star x_\mathrm{eq}$ and substituting the sound horizon
\cref{eqn:rs-in-matter-radiation-ito-wr-R_star} yields
\begin{align}
    \theta_s
    &\approx
        \frac{1}{1.014 \Omega_m^{0.094} - \sqrt{a_\star}}
        \sqrt{\frac{a_\star}{1 - f_\nu}}
        \sqrt{ \frac{1}{3 R_\star} }
        \ln \left(
            \frac{
                \sqrt{R_\star} \sqrt{1 + x_\mathrm{eq}}
                + \sqrt{1 + R_\star}
            }{
                1 + \sqrt{R_\star x_\mathrm{eq}}
            }
        \right)
    ,
    \label{eqn:theta-s-degeneracy}
\end{align}
The terms in \cref{eqn:theta-s-degeneracy} that depend on $R_\star$ and $x_\mathrm{eq}$ alone
are measured no less precisely than $\sim 0.300 \pm 0.001$ in \LCDM{} or any of the extended
cosmologies discussed in \cref{app:bao-from-cmb}.
The CMB thus tightly constrains the combination $\Omega_m^{1/5.32} (1 - f_\nu) / a_\star$.
That is, fixing $\theta_s$ within flat \LCDM{} cosmologies requires the late-time matter fraction to
scale as
\begin{align}
    \left. \Omega_m \right\vert_{\theta_s, \, x_\mathrm{eq}, \, R_\star}
    &\propto \left( a_\star \left[ 1 + f_\nu \right] \right)^{5.32}.
    \label{eqn:matter-fraction-fnu-degeneracy-cmb}
\end{align}
The resulting variation of the Hubble constant may be inferred by writing
$\Omega_m = (\omega_b + \omega_c) / (1 - f_\nu) h^2 \approx (1 + f_\nu) \omega_r / h^2 a_\star x_\mathrm{eq}$,
such that~\cite{Baryakhtar:2024rky}
\begin{align}
    \left. h \right\vert_{\theta_s, \, x_\mathrm{eq}, \, R_\star}
    &\propto \frac{
            \sqrt{\omega_r}
        }{
            a_\star^{3.16}
            \left( 1 + f_\nu \right)^{2.16}
        }
    .
    \label{eqn:hubble-fnu-degeneracy-cmb}
\end{align}
Holding the angular extent of the sound horizon fixed thus requires not just a smaller Hubble
constant but also a substantial distortion to the shape of the late-time expansion history when
varying the neutrino mass sum (or the redshift of recombination).\footnote{
    Quantitatively, the scaling with $1 + f_\nu$ is slightly shallower than that reported in
    \cref{eqn:matter-fraction-fnu-degeneracy-cmb,eqn:hubble-fnu-degeneracy-cmb} (about $5$ rather
    than $5.32$ for $\Omega_m$), simply because these results assume the neutrinos are matterlike
    since last scattering; the interim period before they actually become nonrelativistic makes a
    small contribution to the distance to last scattering.
}
The scaling of $\Omega_m$ in \cref{eqn:matter-fraction-fnu-degeneracy-cmb} is especially steep
compared to the trivial linear scaling from holding $h$ fixed,
$\left. \Omega_m \right\vert_{h, \, \omega_b, \, \omega_c} \propto 1 + f_\nu$.
But no quantification of the size of the Universe comes close to the CMB's subpermille measurement
of $\theta_s$ in precision or robustness; without invoking new physics to modify recombination
(i.e., $a_\star$) or the dynamics of dark energy [such that
\cref{eqn:matter-fraction-fnu-degeneracy-cmb} is invalid], the matter fraction necessarily increases
steeply with neutrino mass.
Geometric probes of late-time cosmology thus offer a substantial (and perhaps underappreciated)
opportunity to measure the neutrino masses.

\subsubsection{Type Ia supernovae}\label{sec:supernova}

Type Ia supernovae (SNe Ia) are thought to occur at a characteristic mass that yields a
standardizable luminosity suitable for inferring distances.
The apparent magnitude of a SN Ia is parametrized as~\cite{SupernovaCosmologyProject:1996grv,
Pan-STARRS1:2017jku, Brout:2022vxf, Scolnic:2021amr}
\begin{align}
    m
    &= 5 \log_{10} \left[
            \frac{a}{a_{\mathrm{hel}}}
            \frac{D_L(a)}{10~\mathrm{pc}}
        \right]
        + M_B
    ,
    \label{eqn:apparent-magnitude}
\end{align}
where $M_B$ is the fiducial magnitude of an SN Ia and $D_L$ the luminosity distance,
$D_L(a) = D_M(a) / a$; the extra factor of $a$ simply accounts for the diminishing luminosity of a
distant source relative to its (more fundamental) comoving distance $D_M(a)$.
In addition, the ratio of the CMB-frame ($a$) and heliocentric ($a_{\mathrm{hel}}$) redshifts
inside the logarithm corrects for peculiar velocities~\cite{Davis:2010jq, Davis:2019wet,
Carr:2021lcj}; they are cosmology independent and may as well be absorbed into the apparent
magnitude via the definition $\hat{m} \equiv m - 5 \log_{10} (a / a_{\mathrm{hel}})$.
Rearranging \cref{eqn:apparent-magnitude} into the apparent brightness yields a ``luminosity angle''
in analogy to the BAO angles \cref{eqn:theta-bao}:
\begin{align}
    \theta_L(a)
    \equiv \frac{10~\mathrm{pc} \cdot 10^{- M_B / 5}}{D_L(a)}
    &= 10^{- \hat{m} / 5}.
    \label{eqn:luminosity-angle}
\end{align}
The numerator of the middle expression is a standardized luminosity distance.

Like BAO distances, calibration is required to directly measure $H_0$ with SNe distances, i.e.,
calibration of the fiducial magnitude $M_B$ using, e.g., Cepheids~\cite{Riess:2016jrr,
Riess:2019cxk, Riess:2021jrx, Freedman:2024eph} or the tip of the red giant
branch~\cite{Freedman:2019jwv, Freedman:2020dne, Freedman:2021ahq, Freedman:2023jcz,
Freedman:2024eph}.
The ``luminosity angle'' [\cref{eqn:luminosity-angle}] factorizes in the same manner as
\cref{eqn:theta-bao-perp-flat-lcdm}:
\begin{align}
    \theta_L(a)
    &= \frac{10^{\log_{10} \sqrt{\omega_m} - M_B / 5}}{c~\mathrm{s} / \mathrm{m}}
        \frac{
            a
        }{
            2 \left[ F_M(1; \amL) - F_M(a; \amL) \right]
        }.
    \label{eqn:luminosity-angle-lcdm}
\end{align}
The Hubble constant is only constrained via the combination $\log_{10} h - M_B / 5$
[which is linearly related to the normalization $\log_{10} \sqrt{\omega_m} - M_B / 5$ in
\cref{eqn:luminosity-angle-lcdm}].
To the extent that their luminosity is indeed standard, SNe distances can still robustly probe the
shape of the expansion history independent of (i.e., marginalized over) its calibration.
In particular, by measuring the matter fraction, SNe distances can break the CMB's geometric
degeneracy via \cref{eqn:matter-fraction-fnu-degeneracy-cmb}.

\subsection{Signatures in structure}
\label{sec:structure}

While geometric observables only probe background cosmology, large-scale structure is sensitive to
the dynamics of neutrino perturbations because neutrinos do not cluster efficiently on scales
shorter than their free-streaming length~\cite{Bond:1980ha, Hu:1997mj}.
When studying how external probes break the degeneracies of neutrino free streaming with other
parameters, care must be taken to identify what geometric quantities are constrained; per
\cref{sec:geometry}, \LCDM{} has the parameter freedom to hold only one of $\Omega_m$, $h$, and
$\theta_s$ fixed when varying the neutrino masses (and taking the baryon and CDM densities to be
constrained by, e.g., the shape of the CMB anisotropies as described in
\cref{sec:mass-independent}).
The interplay of lensing and geometric constraints can be especially subtle in extended models that
modify the sound horizon.
We next review the well-established formalism describing probes of weak gravitational lensing in
order to elucidate these relationships and identify what parameter combinations are best constrained
by probes of structure.

Weak lensing observables may be defined in terms of lensing potentials $\phi_X$ given by a weighted
integral of the Weyl potential $\Psi$ over the line of sight distance $\chi$~\cite{Lewis:2006fu}:
\begin{align}
    \phi_X(\hat{n})
    &\equiv - 2 \int_0^\infty \ud \chi \,
        W_X(\chi) \Psi(\eta(\chi), \chi \hat{n})
    \label{eqn:def-lensing-potential}
\end{align}
where the dimensionless lensing kernel is
\begin{align}
    W_X(\chi)
    &\equiv
        \int_\chi^\infty \ud \chi_1 \,
        n_X(\chi_1)
        \left( 1 - \frac{\chi}{\chi_1} \right)
    \label{eqn:def-weight-function}
\end{align}
and $n_X(\chi_1)$ the (normalized) distribution of sources.
The angular power spectrum for two such fields $X$ and $Y$ is likewise given by a weighted,
line-of-sight integral over the dimensionless power spectrum of the Weyl potential,
\begin{align}
    \Delta_{\Psi}^2(\eta, k_1)
    (2 \pi)^3 \delta^{3}(\mathbf{k}_1 + \mathbf{k}_2)
    &\equiv \frac{k_1^3}{2 \pi^2}
        \left\langle \Psi(\eta, \mathbf{k}_1) \Psi(\eta, \mathbf{k}_2) \right\rangle,
\end{align}
as
\begin{align}
    C_\ell^{X Y}
    &= \frac{2 \pi^2}{L(\ell)^3}
        \int_0^\infty \frac{\ud \chi}{\chi} \, W_X(\chi) W_Y(\chi)
        \Delta_{\Psi}^2(\eta(\chi), L(\ell) / \chi)
    \label{eqn:C-ell-XY-limber}
    .
\end{align}
\Cref{eqn:C-ell-XY-limber} takes the Limber approximation~\cite{Limber:1954zz, LoVerde:2008re},
which sets $L(\ell) = \sqrt{\ell (\ell + 1)}$ and is generally accurate at the permille level for
multipoles $\ell \gtrsim 10$.\footnote{
    CLASS and CAMB both use $L(\ell) \approx \ell + 1/2$, which approximates the CMB lensing
    spectrum with precision better than a part in $10^{-3}$ for $\ell \gtrsim 22$; using the full
    expression $L(\ell) = \sqrt{\ell (\ell + 1)}$ instead achieves the same precision for
    $\ell \gtrsim 12$.
}
Weak lensing probes are typically characterized by derivatives of $\phi_X$ with respect to angular
displacements (i.e., orthogonal to the line of sight), like the lensing convergence
$\kappa_X \equiv - \nabla^2 \phi_X / 2$.

The statistical properties of weak lensing observables [like \cref{eqn:C-ell-XY-limber}] are
fundamentally quantified by the dimensionless angular correlations (and possibly the redshift) of
dimensionless fields, such as angular deflections or density contrasts.
As such, they can only constrain ratios of dimensionful quantities.
In the remainder of this section we summarize the growth of structure in \LCDM{}
(\cref{sec:growth-with-neutrinos}) and its application to CMB lensing (\cref{sec:cmb-lensing}),
taking care to phrase results in terms of the most pertinent dimensionless ratios of densities and
scales.
We briefly comment on the secondary effects of massive neutrinos on the CMB temperature and
polarization power spectra in \cref{sec:peak-smearing-isw}.

\subsubsection{Growth of structure with massive neutrinos}\label{sec:growth-with-neutrinos}

The evolution of metric potentials is conventionally factored into a transfer function $T_\Psi(k)$
encoding dynamics of modes with wave number $k$ from horizon crossing until the matter era and a
growth factor $D_+(a)$ accounting for scale-independent growth during and after the matter era.
We promote the growth factor to depend on $k$ in order to schematically describe the additional
scale dependence from freely streaming neutrinos on structure growth in that epoch.
The power spectrum of the Weyl potential is thus related to that of the initial comoving curvature
perturbation $\mathcal{R}$ by
\begin{align}
    \Delta_{\Psi}^2(\eta, k)
    &\equiv \left[ \frac{6}{5} \frac{D_+(a(\eta), k)}{a} T_\Psi(k) \right]^2
        \Delta_{\mathcal{R}}^2(k).
\end{align}
The early-time dynamics encoded by the transfer function $T_\Psi(k)$ largely depend on wave number
only via $k / k_\mathrm{eq}$~\cite{Bardeen:1985tr, Eisenstein:1997ik}.
The effects of acoustic oscillations and anisotropic stress introduce additional scale dependence
that, though crucial to the shape of CMB anisotropies, are relatively unimportant to weak lensing
observables, which are projected over a range of scales.
The transfer function thus deviates from unity at $k \gtrsim k_\mathrm{eq}$, where it falls off
$\sim (k / k_\mathrm{eq})^{-2} \ln (k / k_\mathrm{eq})$.

The late-time dynamics of observable modes (i.e., those that enter the horizon during or before the
matter era) are stratified by the neutrino free-streaming scale~\cite{Lesgourgues:2006nd},
\begin{align}
    k_\mathrm{fs}
    \equiv \sqrt{\frac{3}{2}} \frac{a H}{c_{\nu_i}}
    &= \frac{1.75 \times 10^{-2}~\mathrm{Mpc}^{-1}}{\sqrt{1 + z}}
        \frac{m_{\nu_i}}{0.06~\mathrm{eV}}
        \sqrt{\frac{\omega_m}{0.143}}
        \sqrt{1 + ([1 + z] \amL)^{-3}}
    \label{eqn:neutrino-free-streaming-scale}
\end{align}
where $c_{\nu_i} = T_\nu(a) / m_{\nu i}$ is a measure of the velocity of the $i$th neutrino species.
In flat \LCDM{} cosmologies, the linear growth factor for the fraction $1 - f_\nu$ of matter that
does cluster is
\begin{align}
    D_\pm(a, k)
    &\propto C_\pm
        a^{\gamma_\pm}
        \cdot {}_2{F}_1(\gamma_\pm / 3, \gamma_\pm / 3 + 2/3, 2 \gamma_\pm / 3 + 7/6, - [a / \amL]^3),
    \label{eqn:growth-factor-lcdm-neutrinos}
\end{align}
where $\gamma_\pm$ are the growth rates in a matter-dominated Universe:
\begin{align}
    \gamma_\pm
    = - \frac{1}{4} \pm \frac{5}{4} \sqrt{1 - \frac{24}{25} f_\nu}
    = \begin{dcases}
            1 - \frac{3}{5} f_\nu & +
            \\
            - \frac{3}{2} + \frac{3}{5} f_\nu & -
        \end{dcases}
        + \mathcal{O}(f_\nu^2)
    .
    \label{eqn:growth-rate-mixed}
\end{align}
The $k$ dependence of \cref{eqn:growth-factor-lcdm-neutrinos} is implicit in the interpretation of
$f_\nu$ as a function of scale (and time).
In application to massive neutrinos, \cref{eqn:growth-factor-lcdm-neutrinos} holds at $z < z_{\nu_i}$
[\cref{eqn:massive-neutrino-z-NR}] and on scales $k_\mathrm{fs} \ll k \ll k_\mathrm{nl}(a)$ [where
$k_\mathrm{nl}(a)$ is the scale at which nonlinear evolution becomes important];
at earlier times and/or on larger scales, the same result holds but with $f_\nu$ set to
zero.\footnote{
    None of these results account for the small, residual impact of radiation, whether from
    neutrinos while they remain relativistic or from photons.
}

In summary, at times well after matter-radiation equality and on linear scales that cross the
horizon before the dark-energy era, the Weyl potential depends on cosmological parameters via
\begin{align}
    \Delta_{\Psi}^2(\eta, k)
    &\approx \left[
            \frac{6}{5}
            \frac{D_+(a(\eta) / \amL, k / k_\mathrm{fs}; f_\nu)}{a}
            T_\Psi(k / k_\mathrm{eq}) \right]^2
        A_s \left( \frac{k}{k_\mathrm{p}} \right)^{n_s - 1},
    \label{eqn:weyl-power-functional-dependence}
\end{align}
implicitly changing the arguments to $D_+$ and $T_\Psi$ to reflect the dimensionless numbers (ratios
of wave numbers and scale factors and the neutrino fraction) that by and large fully parametrize
them.
Recall that $\amL$ is specified by $\Omega_m$ [\cref{eqn:aml-ito-Wm}].
\Cref{eqn:weyl-power-functional-dependence} also inserted the canonical parametrization of the
primordial power spectrum in terms of a tilt $n_s$ and the amplitude $A_s$ at an arbitrary pivot
scale $k_\mathrm{p}$, which is typically taken to be $0.05~\mathrm{Mpc}^{-1}$ but may be freely
redefined to, e.g., $k_\mathrm{eq}$ by redefining $A_s$.
For sources with some fixed (i.e., externally measured) distribution $n_X$,
\cref{eqn:weyl-power-functional-dependence} encodes nearly all of the cosmological dependence of
weak lensing observables [\cref{{eqn:C-ell-XY-limber}}].\footnote{
    Typically the distribution of sources is measured over redshift; the conversion to line-of-sight
    distances $\chi$ in \cref{eqn:def-weight-function} does then depend on cosmology.
    For CMB lensing, which we specialize to in \cref{sec:cmb-lensing}, the line-of-sight distance
    to last scattering is essentially fixed by the precise constraint on $\theta_s$ in cosmologies
    that don't alter the sound horizon.
}
Before applying these results to a concrete example, we discuss pertinent numerical estimates for
the dependence of the growth factor [and therefore \cref{eqn:weyl-power-functional-dependence}] on
the parameters of late-time cosmology.

The hypergeometric function in the growth factor [\cref{eqn:growth-factor-lcdm-neutrinos}]
approaches unity for $a \ll \amL$, i.e., it encodes the deviation from the matter-era result
$\propto a^{\gamma_\pm}$ due to the onset of dark-energy domination.
For $f_\nu = 0$, this growth deviation factor is $0.8725$ at matter--dark-energy equality, while its
present-day value is approximated reasonably well by
$0.78 \left( \Omega_m / 0.3 \right)^{0.23}$ for $0.2 \lesssim \Omega_m \lesssim 0.6$.
At fixed $\theta_s$, increasing the neutrino mass sum increases the total matter fraction and thus
delays the dark energy era; from \cref{eqn:matter-fraction-fnu-degeneracy-cmb}, the present-day
suppression factor in the growth factor scales as $\sim (1 + f_\nu)^{1.15}$.
The delayed onset of dark-energy domination thus offsets part of the suppression of structure
accumulated while neutrinos do not cluster (as encoded by the factor $a^{\gamma_+}$ in $D_+$),
which itself may be approximated by
\begin{align}
    D_+(a) / a
    &\approx 1 + (\gamma_+ - 1) \ln[(1 + z_{\nu_i}) / (1 + z)]
    \approx 1 - \frac{3}{5} f_\nu \ln[(1 + z_{\nu_i}) / (1 + z)]
    \label{eqn:growth-neutrinos}
\end{align}
in the matter era (with $z_{\nu_i}$ equal for all mass eigenstates for the degenerate hierarchy).
For the minimum mass sum, the growth factor deviates from the $f_\nu = 0$ result by $- 1.1 f_\nu$ to
$- 2.2 f_\nu$ between redshifts $5$ and $0$; the coefficients of $f_\nu$ increase to
$-2.4$ and $-3.4$, respectively, for $\summnu = 0.5~\mathrm{eV}$.

While the individual components of \cref{eqn:weyl-power-functional-dependence}---the initial
conditions, transfer function, and growth factor---are fairly straightforward to study
(semi)analytically, the translation to the angular correlations of an observable
[\cref{eqn:C-ell-XY-limber} in the Limber approximation] involves a weighted integral over
$\Delta_\Psi^2$ evaluated on scales and at times both specified by the line-of-sight distance.
To concretely illustrate the true dependence of LSS observables on dimensionless cosmological
parameters, we specialize to CMB lensing next.

\subsubsection{CMB lensing}\label{sec:cmb-lensing}

For CMB lensing the source distribution is the CMB visibility function, which peaks narrowly at
recombination and may be approximated by a delta function at $\chi_\star$.
The weight function is simply $W(\chi) = 1 - \chi / \chi_\star$,\footnote{
    Technically, the lensing kernel averages the inverse comoving distance over the visibility
    function, which deviates from $1/\chi_\star$ (i.e., evaluated at peak visibility) by
    about $0.1\%$.
    The use of $\chi_\star$ in the kernel in both CLASS and CAMB incurs an error of the same order,
    growing to $0.3\%$ at $\ell \geq 1000$.
} so
\begin{align}
    C_\ell^{\kappa \kappa}
    &= \frac{2 \pi^2 \ell^2 (\ell + 1)^2}{L(\ell)^3}
        \int_0^{\chi_\star} \frac{\ud \chi}{\chi_\star} \,
        \left( \sqrt{\chi_\star / \chi} - \sqrt{\chi / \chi_\star} \right)^2
        \Delta_{\Psi}^2(\eta(\chi), L(\ell) / \chi)
    .
    \label{eqn:cl-kk-cmb-lensing}
\end{align}
The competition between the weight function (which diverges as $\chi / \chi_\star \to 0$) and the
shape of $\Delta_\Psi^2$ (which falls off as four powers of
$k / k_\mathrm{eq} = L(\ell) / \chi k_\mathrm{eq}$ at large $k$) places the bulk of the integral's
support roughly between redshifts $0$ and $5$~\cite{Lewis:2006fu}.
The CMB lensing spectrum thus retains mild sensitivity to the dark-energy era and thus can discern
to at least some degree its delayed beginning in heavy-neutrino cosmologies (which entail a larger
matter fraction $\Omega_m$ and equality scale factor $\amL$ when fixing $\theta_s$, per
\cref{sec:cmb-geometry}).
Since the growth factor's sensitivity to $f_\nu$ ranges between $-2$ and $-4$ over this redshift
interval (for all but the smallest masses), the net suppression due to neutrino free streaming at
large $\ell$ should asymptote to $1 - 6 f_\nu$ or so (ignoring any changes to $\Omega_m$).

\Cref{eqn:cl-kk-cmb-lensing} shows that scale-dependent features in the Weyl scalar appear in the
angular correlations of the CMB lensing potential at multipoles determined by the distance to last
scattering.
For instance, the horizon at matter-radiation equality $k_\mathrm{eq}$, at which the transfer
function turns over, corresponds to
\begin{align}
    \ell_\mathrm{eq}
    \equiv \chi_\star k_\mathrm{eq}
    &\approx 2 \sqrt{2} \left[ F_M(1; \amL) - F_M(a_\star; \amL)\right]
        \sqrt{\frac{1 - f_\nu}{a_\star x_\mathrm{eq}}}
    ,
    \label{eqn:ell-eq}
\end{align}
using \cref{eqn:comoving-distance-matter-Lambda} for $\chi_\star$ and approximating
$k_\mathrm{eq} = \omega_m H_{100} / \sqrt{\omega_r / 2}$ as for Universes with only massless
neutrinos.
The neutrino free-streaming scale \cref{eqn:neutrino-free-streaming-scale} imprints on sky
multipoles of order
\begin{align}
    \ell_\mathrm{fs}(a)
    \equiv \chi_\star k_\mathrm{fs}(a)
    &\sim
        81
        \frac{m_{\nu_i}}{0.02~\mathrm{eV}}
        \frac{2 \left[ F_M(1; \amL) - F_M(a_\star; \amL) \right]}{1.75 \sqrt{1 + z}}
        \sqrt{1 + ([1 + z] \amL)^{-3}}
    ,
    \label{eqn:ell-fs}
\end{align}
near $\ell_\mathrm{eq}\approx 144$ in \Planck{}'s best-fit, flat-\LCDM{} cosmology [although
$\ell_\mathrm{fs}(a)$ is smaller at earlier times where \cref{eqn:cl-kk-cmb-lensing} has most
support].
\Cref{eqn:ell-fs} is written relative to a fiducial $m_{\nu_i} = 0.02~\mathrm{eV}$, i.e., as for
degenerate neutrinos with the minimum mass sum of the normal hierarchy (for illustration).
Since both $k_\mathrm{fs}$ and $\chi_\star$ are largely determined by the late-time expansion
history, their product only depends on the Universe's late-time composition ($\Omega_m$ in flat
\LCDM{} cosmology) and the neutrinos' masses (relative to their average momentum, which we take as
fixed by the SM prediction for neutrino decoupling).
On the other hand, $k_\mathrm{eq}$ is a measure of the horizon at early times, and per
\cref{sec:mass-independent}, $\chi_\star k_\mathrm{eq}$ is fixed when $x_\mathrm{eq}$, $R_\star$,
and $\theta_s$ also are (i.e., the same quantity is independently constrained by the shape of the
primary CMB power spectra).
Both scales mark where power begins to drop and happen to take on similar values, and anyway their
corresponding features in wave number are blurred over multipole by the lensing kernel in the
line-of-sight integral \cref{eqn:cl-kk-cmb-lensing}; CMB lensing data by itself thus cannot easily
differentiate between the two angular scales for reasonable neutrino masses.

\begin{figure}[t!]
\begin{centering}
    \includegraphics[width=\textwidth]{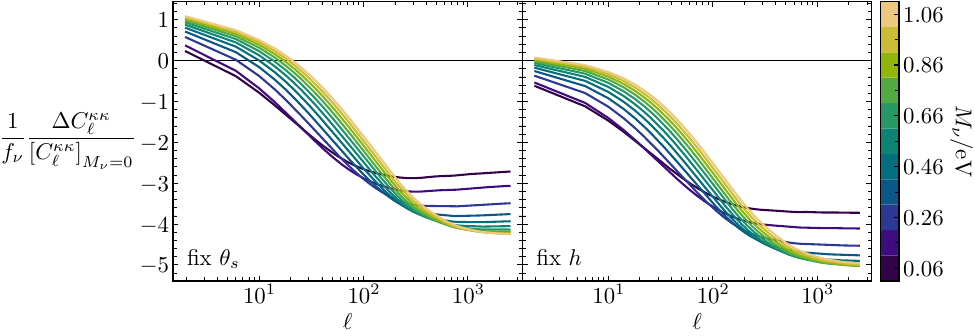}
    \caption{
        Residual of the CMB lensing spectrum for cosmologies varying the neutrino mass sum (by
        color) relative to a cosmology with massless neutrinos, measured in units of the
        neutrino fraction $f_\nu$ [\cref{eqn:neutrino-fraction}].
        Results neglect nonlinear structure growth for illustration.
        Results also take the normal mass hierarchy because the equal-mass approximation
        underestimates the suppression for the minimum mass sum; for all other curves, the
        differences between the normal, inverted, and equal-mass hierarchies are negligible.
        The left and right panels respectively fix the angular extent of the photon sound horizon
        $\theta_s$ and the Hubble constant $h$.
        In the former case, the matter fraction increases as $\Omega_m \propto (1 + f_\nu)^{5}$,
        delaying the dark energy epoch; above the neutrino free-streaming scale
        [\cref{eqn:neutrino-free-streaming-scale}] the lensing spectrum is thus enhanced by
        about one factor of $1 + f_\nu$, whereas on smaller scales the suppression is reduced to a
        similar degree compared to that if $\Omega_m$ were unchanged.
        In contrast, holding $h$ fixed (as commonly taken) yields no large-scale enhancement
        and a greater degree of suppression at high $\ell$.
    }
    \label{fig:cl-pp-residual-units-of-fnu-fix-ths-h}
\end{centering}
\end{figure}

\Cref{fig:cl-pp-residual-units-of-fnu-fix-ths-h} depicts the quantitative response of the CMB
lensing spectrum to massive neutrinos.
The fractional change of $C_\ell^{\kappa \kappa}$ divided by the neutrino fraction as depicted in
\cref{fig:cl-pp-residual-units-of-fnu-fix-ths-h} estimates the response exponent
$\partial \ln C_\ell^{\kappa \kappa} / \partial \ln (1 + f_\nu)$.
\Cref{fig:cl-pp-residual-units-of-fnu-fix-ths-h} compares the response for a variety of neutrino
mass sums when holding either $\theta_s$ or $h$ fixed; consistent with expectations, the increase in
the matter fraction $\Omega_m \propto (1 + f_\nu)^{5}$ incurred by the former indeed slightly
mitigates the degree to which structure is suppressed (and also enhances structure on large scales
where neutrinos do cluster~\cite{Pan:2015bgi}).\footnote{
    The sensitivity of large-scale structure is also often phrased as that for the present-day matter
    power spectrum, which overestimates that of CMB lensing because most of the lensing integral's
    support is at redshifts of order a few (rather than zero).
    The matter power spectrum is also dimensionful, adding complexity in its translation to the
    CMB lensing spectrum compared to \cref{eqn:C-ell-XY-limber,eqn:weyl-power-functional-dependence}.
}
The sensitivity of CMB lensing to neutrino masses near the minimum sum is yet weaker than commonly
quoted because of the logarithmic dependence of \cref{eqn:growth-neutrinos} on $z_{\nu_i}$: for the
normal hierarchy's minimum mass sum, the sensitivity is only about two-thirds of that in the
large-mass limit.

\subsubsection{Peak smearing and the integrated Sachs-Wolfe effect}\label{sec:peak-smearing-isw}

Structure at late times also distorts CMB temperature and polarization power spectra via weak
lensing~\cite{Seljak:1995ve, Zaldarriaga:1998ar, Stompor:1998zj}, the main effect of which is to
smooth or smear the acoustic peaks.
The degree of peak smearing inferred within a model thus probes massive neutrinos by effectively
measuring the amplitude of gravitational potential fluctuations at late
times~\cite{Kaplinghat:2003bh, Smith:2006nk}.
The impact of lensing on the primary anisotropies is partially degenerate with other physical
effects whose spectral shape is smooth (e.g., the modulation of the evolution of gravitational
potentials as the Universe transitions to matter domination) or that yield oscillatory residuals
(e.g., the amplitude of acoustic oscillations as controlled by the baryon density).
Inferring the degree of lensing-induced peak smearing therefore depends partly on early-Universe
physics.

The primary anisotropies independently constrain the six-parameter \LCDM{} model via early-time
effects well enough to tightly (but indirectly) constrain late-time cosmology and structure growth.
The degree to which the observed acoustic peaks appear smeared relative to the (unlensed) \LCDM{}
prediction provides a consistency test, one that \Planck{} data is known to not
pass~\cite{Planck:2018vyg}.
Introducing additional parameter freedom via a rescaling of the lensing potential
$A_L$~\cite{Calabrese:2008rt}, \Planck{} prefers excess peak smearing at the $2$ to $3 \sigma$
level~\cite{Planck:2018vyg}.
While the degree of this so-called lensing anomaly varies among recent \Planck{} maps and
likelihoods~\cite{Rosenberg:2022sdy, Tristram:2023haj} (e.g., the treatment of foregrounds), even
direct reconstructions of the CMB lensing spectrum from four-point statistics prefer an amplitude of
lensing larger than the \LCDM{} prediction.
This general lensing excess leaves little room for massive neutrinos to further suppress the
amplitude of structure.
Marginalizing over a parameter $A_\mathrm{smear}$, which rescales the amplitude of the lensing
spectrum insofar as it lenses the temperature and polarization anisotropies,\footnote{
    Typically, the parameter referred to as $A_L$ rescales the lensing spectrum consistently in its
    impact on temperature and polarization anisotropies as well as the predicted lensing spectrum
    itself (which is measured via the four-point function of CMB maps).
    Here by $A_\mathrm{smear}$ we denote only the former effect, which we use to derive constraints
    on structure from CMB lensing independent of the peak-smearing effect.
    Marginalizing over $A_\mathrm{smear}$ also yields CMB constraints on late-time cosmology that
    effectively only derive from geometry (i.e., $\theta_s$), as employed in
    \cref{sec:geometry-degeneracy-breaking}.
} effectively removes the information on late-time structure from the primary CMB.

Finally, massive neutrinos modify the time evolution of metric potentials as they become
nonrelativistic, both by altering the expansion rate (compared to that for a pure-matter Universe,
in which the potentials are constant) and the clustering of matter (compared to a Universe where all
matter clusters)~\cite{Dodelson:1995es, Lesgourgues:2012uu, Hou:2012xq}.
The incompatibility of measurements of the first acoustic peak (and lower multipoles) with these
effects requires that neutrinos become nonrelativistic well after recombination.
Since $T_\star \approx 0.3~\mathrm{eV}$, upper limits on $\summnu$ deriving only from this effect
are of order $1~\mathrm{eV}$.
In fact, by marginalizing over the amplitude of the lensing spectrum, the results in
\cref{fig:geometry-pr3-asmear-desi-des} using only \Planck{} temperature and polarization power
spectra are effectively driven only by the impact of an early-time nonrelativistic transition,
consistent with analyses of large-scale CMB anisotropies alone (e.g., from the Wilkinson Microwave
Anisotropy Probe~\cite{Ichikawa:2004zi, WMAP:2012nax}).
Because these effects have been excluded by multiple generations of CMB observations and are largely
irrelevant in the mass regimes allowed by current data, we do not discuss them further.
We note, however, that the $95\%$ upper limit $\summnu < 0.86~\mathrm{eV}$ (from
\cref{fig:geometry-pr3-asmear-desi-des} below) may be regarded as a cosmological bound largely
independent of the lensing excess and of late-time dynamics, including in particular modifications
to the dynamics of dark energy.

\section{Results}
\label{sec:results}

Using the physical understanding developed in \cref{sec:signatures} of neutrino cosmology and its
degeneracies with the geometry of the late Universe, we next investigate the neutrino mass
constraints derived from current data via geometry alone (\cref{sec:geometric-tensions}) and in
combination with probes of structure (\cref{sec:lensing-constraints}).
We study in detail the role of particular datasets, the sensitivity of results to potential
statistical or systematic effects, and the manner in which dataset combinations break parameter
degeneracies.
Before proceeding, we describe our analysis methods and informative summary statistics for
cosmological neutrino mass inference in light of the lower bounds imposed by neutrino oscillation
data.

The posterior distributions over $\summnu$ from (most) current cosmological data are not Gaussian,
and on physical grounds the masses of SM neutrinos are positive \textit{a priori}.
Measuring the degree to which posteriors are consistent with nonzero $\summnu$ is therefore
nontrivial.
For instance, the $z$ score of $\summnu = 0$ (i.e., quantifying the distance of the center from zero
in units of the standard deviation) is not particularly meaningful when posteriors are nonnormal,
let alone asymmetric.
These subtleties motivate a more nuanced quantification of a ``detection'' of nonzero neutrino
masses or the degree to which a particular cosmological model and set of data are compatible or not
with massive neutrinos.

Since the only physical model of neutrinos we consider is the standard one---described by SM physics
with mass terms for the neutrinos---we retain the physical prior that $\summnu \geq 0$, under which
``compatibility'' with negative values (as explored by recent work~\cite{Craig:2024tky,
Green:2024xbb, eBOSS:2020yzd, Allali:2024aiv, Naredo-Tuero:2024sgf, Elbers:2024sha}) cannot be
assessed.
But current laboratory probes place precise, nonzero lower bounds on $\summnu$ for the normal and
inverted hierarchies.
A more motivated metric for the compatibility of cosmological datasets with minimal, massive
neutrinos is the fraction of the posterior below these lower bounds, i.e., the cumulative
distribution function (CDF)
\begin{align}
    P(\summnu < {\summnu}_\mathrm{min} \vert \mathcal{D})
    &= \int_{0}^{{\summnu}_\mathrm{min}} \ud \summnu \,
        p(\summnu \vert \mathcal{D})
    \label{eqn:compatibility-metric}
\end{align}
where $p(\summnu \vert \mathcal{D})$ is the posterior density for dataset $\mathcal{D}$ marginalized
over all parameters but $\summnu$.
(All probabilities here are implicitly conditional on the assumed model.)
\Cref{eqn:compatibility-metric} thus measures what fraction of the posterior is incompatible with
a mass hierarchy whose minimum mass sum is ${\summnu}_\mathrm{min}$.
When $p(\summnu \vert \mathcal{D})$ is nearly normal, \cref{eqn:compatibility-metric} is of course
equivalent to a $z$ score (suitably translated to the CDF of a normal distribution).

In our analyses, we employ the 2018 \Planck{} (PR3) likelihoods via the \texttt{Plik\_lite} variants
that are marginalized over the parameters of the foreground models~\cite{Planck:2018vyg,
Planck:2019nip}; various combinations of BAO measurements from SDSS, including eBOSS
DR12 and DR16 data~\cite{eBOSS:2020yzd, eBOSS:2020lta, eBOSS:2020hur} and the Main Galaxy Sample
(MGS) from DR7~\cite{Ross:2014qpa}, and from DESI DR1~\cite{DESI:2024mwx, DESI:2024uvr,
DESI:2024lzq}; and supernova distance measurements from the Pantheon~\cite{Pan-STARRS1:2017jku},
Pantheon+~\cite{Brout:2022vxf, Scolnic:2021amr}, DES 5YR~\cite{DES:2024tys}, and
Union3~\cite{Rubin:2023ovl} results.
Except where noted, the SNe measurements are always marginalized over the fiducial magnitude rather
than calibrated, i.e., by SH0ES~\cite{Riess:2021jrx}.
We use CMB lensing data from the \Planck{} PR3 release~\cite{Planck:2018lbu} and the ACT DR6
measurement~\cite{ACT:2023dou, ACT:2023dou} combined with \Planck{} PR4 lensing
data~\cite{Carron:2022eyg}.\footnote{
    As noted by Ref.~\cite{Jiang:2024viw}, version 1.2 of these likelihoods~\cite{act-dr6-lenslike}
    has a moderate effect on results compared to prior versions (used, for instance, by
    Ref.~\cite{DESI:2024mwx}), depending on the likelihood configuration.
    We use the latest version (1.2).
}
Throughout we refer to the PR3 temperature and polarization likelihoods by ``PR3 CMB'' and
explicitly state whether and what CMB lensing likelihood is included.
We use \textsf{CLASS}~\cite{Blas:2011rf, Lesgourgues:2011re} and model nonlinear structure growth
with HMCODE-2016~\cite{Mead:2016zqy} (as used by \textsf{CLASS}, which has yet to update to the 2020
version~\cite{Mead:2020vgs}); we show in \cref{app:nonlinear} that neglecting nonlinear effects
biases neutrino mass constraints toward smaller values, even with current CMB lensing
datasets.\footnote{
    Ref.~\cite{Jiang:2024viw} found that the variation in upper limits on $\summnu$ varied
    negligibly between the 2016 and 2020 versions of HMCODE, but only for posteriors that
    include DESI DR1 data.
    As we discuss in \cref{app:nonlinear}, DESI data have a much stronger impact on neutrino mass
    constraints than do lensing datasets; investigating the impact of nonlinear modeling for
    dataset combinations that do not exclude the regime in which they matter is a worthwhile
    task we leave to future work.
}

In parameter inference, we sample over the standard set of \LCDM{} parameters with broad,
uniform priors: the present baryon and CDM densities,
$\omega_b \sim \mathcal{U}(0.005, 0.035)$ and $\omega_c \sim \mathcal{U}(0.01, 0.25)$; the Hubble
rate $h \sim \mathcal{U}(0.25, 1.1)$; the tilt and amplitude of the scalar power spectrum,
$n_s \sim \mathcal{U}(0.8, 1.2)$ and $A_s$ via $\ln(10^{10} A_s) \sim \mathcal{U}(1.61, 3.91)$; the
redshift of reionization $z_\mathrm{reion} \sim \mathcal{U}(4, 12)$; and the neutrino mass sum
$\summnu / \mathrm{eV} \equiv \sum_i m_{\nu_i} / \mathrm{eV} \sim \mathcal{U}(0, 1.5)$.
Here we use $\mathcal{U}(a, b)$ to denote a uniform prior between $a$ and $b$.
Some analyses also sample over a lensing amplitude $A_\mathrm{smear} \sim \mathcal{U}(0, 2)$ that
scales the impact of late-time structure on the primary temperature and polarization anisotropies
(see \cref{sec:peak-smearing-isw}).
We employ parameter sampling methods (using
\textsf{emcee}~\cite{Foreman-Mackey:2012any,Hogg:2017akh,Foreman-Mackey:2019}) and likelihood
implementations as described in Ref.~\cite{Baryakhtar:2024rky}.

\subsection{Geometric constraints and tensions}
\label{sec:geometric-tensions}

\Cref{sec:geometry} established the critical role played by probes of the large-scale geometry of
the Universe in neutrino mass inference.
Uncalibrated distance measurements from BAO and type Ia SNe are intrinsically sensitive to the
expansion history alone.
CMB anisotropies, on the other hand, provide essential geometric information by precisely measuring
the angular extent of the sound horizon, but they are simultaneously quite sensitive to the dynamics
of spatial perturbations (as discussed in \cref{sec:structure}).
In this section we explain how the CMB calibrates late-time geometry to constrain neutrino masses
(\cref{sec:calibrating-geometry}) and present results deriving from current low-redshift distance
datasets (\cref{sec:geometry-degeneracy-breaking}).
Motivated by discrepancies in these results, \cref{sec:desi} investigates the origin of DESI's
putative incompatibility with massive neutrinos, comparing to results from SDSS.

Uncalibrated distance measurements via the BAO scale in flat \LCDM{} are described by a
two-parameter model ($\Omega_m$ and $r_\mathrm{d} \sqrt{\omega_m}$); when deriving posteriors in
these parameters we simply evaluate likelihoods on a grid rather than sampling (in
\cref{sec:calibrating-geometry,sec:desi} only).
Similarly, when marginalizing over the overall amplitude of SNe distances $\log_{10} h - M_B / 5$,
we evaluate SNe likelihoods on a grid in $\Omega_m$.
To isolate the CMB's geometric information in the $r_\mathrm{d} \sqrt{\omega_m}$-$\Omega_m$ plane,
we derive from a complete analysis of \Planck{} data a marginal posterior over the angular extent of
the sound horizon to use in conjunction with low-redshift data; these results apply more generally
scenarios featuring modified early-time dynamics to the extent that the sound/drag horizon remains a
robust a predictor of the acoustic peak locations~\cite{Bernal:2020vbb}.
\Cref{app:bao-from-cmb} describes the derivation of a likelihood over the transverse BAO scale
from CMB data in more detail.

\subsubsection{Neutrino masses from calibrated geometry}\label{sec:calibrating-geometry}

Constraining the Universe's geometry---that is, dimensionless quantifiers of the background
metric---is insufficient to measure the neutrino masses.
Uncalibrated BAO and SNe distances directly constrain the late-time matter \emph{fraction}
$\Omega_m$ but are only sensitive to the absolute density $\omega_m$ relative to the drag
horizon;\footnote{
    The parameterization of the dimensionless ``amplitude'' of BAO angles [\cref{eqn:theta-bao}] as
    $r_\mathrm{d} \sqrt{\omega_m}$ is unitful---we drop the length unit $c / H_{100}$ to instead
    report the drag horizon relative to a fiducial \LCDM{} prediction.
}
either of these measurements may be calibrated to measure the dimensionful $\omega_m$---with a
prediction for $r_\mathrm{d}$ from early-Universe physics in the latter case or by calibrating the
brightness of Ia SNe with Cepheids~\cite{Riess:2016jrr, Riess:2019cxk, Riess:2021jrx,
Freedman:2024eph}, the tip of the red giant branch~\cite{Freedman:2019jwv, Freedman:2020dne,
Freedman:2021ahq, Freedman:2023jcz, Freedman:2024eph}, or other calibrators~\cite{Li:2024yoe,
Freedman:2024eph, DES:2024gpx} in the former.
Additional calibration is required because low-redshift distance measurements alone cannot determine
the partitioning of the late-time matter density: they only probe the Universe's geometry at times
long after neutrinos become nonrelativistic, when they are indistinguishable from other forms of
matter (at the background level).
A CMB-derived constraint on the density of the matter components that are present at early times
(baryons and CDM; see \cref{sec:mass-independent}) thus isolates the neutrinos' contribution to the
total matter density at late times.
While uncalibrated geometric information from CMB and BAO data provides constraints in the
$r_\mathrm{d} \sqrt{\omega_m}$-$\Omega_m$ plane without appealing to a model of early-Universe
physics [see \cref{eqn:theta-bao-flat-lcdm}], the CMB's calibration of the baryon and CDM densities
derives from the shape of the power spectra and therefore is sensitive to modified early-Universe
physics (e.g., in the manner captured by the parameters $R_\star$ and $x_\mathrm{eq}$ per
\cref{sec:mass-independent}).
The calibration of the drag horizon $r_\mathrm{d}$ also depends on early-Universe physics, of
course, as does is its relationship to the baryon and CDM densities.\footnote{
    Posteriors over $\theta_s$ or $\theta_{\mathrm{CMB}, \perp}(a_\mathrm{d})$ are moderately
    correlated with $R_\star$ and $x_\mathrm{eq}$, which only has a modest effect on inferring
    $\summnu$.
    When employing this reduced geometric information from \Planck{}, we only use its calibration
    of $\omega_b + \omega_c$ (within \LCDM{}) to illustrate how the
    $r_\mathrm{d} \sqrt{\omega_m}$-$\Omega_m$ parameter space translates to neutrino masses
    (i.e., on average) rather than to derive quantitative constraints (for which we use the full
    likelihoods).
}

In sum, the difference in the densities of early- and late-time matter components, once each is
calibrated, may be ascribed to massive neutrinos; $\omega_\nu$ may then be translated directly to
their summed masses via \cref{eqn:massive-neutrino-density} if their number density $n_\nu$ is
fixed, i.e., by the SM prediction for neutrino decoupling
[\cref{eqn:massive-neutrino-density}]:\footnote{
    \Planck{}'s preferred baryon and CDM densities are slightly sensitive to the late-time effects
    of lensing---for instance, $\omega_b + \omega_c$ shifts lower by about $1 \sigma$ when
    marginalized over $A_\mathrm{smear}$ (see \cref{sec:peak-smearing-isw} and
    Ref.~\cite{Planck:2018vyg}) without substantial degradation in precision.
    We take its central value of $\approx 0.142$ from an analysis of \LCDM{} (which holds with or
    without massive neutrinos) as a fiducial one in order to illustrate the impact of this
    calibration in standard analyses.
}
\begin{align}
    \summnu
    &= \frac{3 H_{100}^2 \Mpl^2}{n_\nu} \left( \omega_m - \omega_b - \omega_c \right)
    = 13.2~\mathrm{eV}
        \left( \frac{\omega_m}{\omega_b + \omega_c} - 1 \right)
        \frac{\omega_b + \omega_c}{0.142}
    .
    \label{eqn:summnu-from-matter-density-difference}
\end{align}
If a combination of observations infers a late-time matter abundance \textit{lower} than \Planck{}'s
early-time abundance, \cref{eqn:summnu-from-matter-density-difference} would suggest the neutrinos
have ``negative mass''.
Note, however, that the nonrelativistic energy density in neutrinos is truly
$\rho_{\nu_i} \approx \left\vert m_{\nu_i} \right\vert n_{\nu_i}$, since the corresponding limit of
each neutrino's energy is
$E(p) = \sqrt{p^2 + m_{\nu_i}^2} \approx \left\vert m_{\nu_i} \right\vert \left[ 1 + (p / m_{\nu_i})^2 / 2 \right]$.
Negative $\summnu$ should thus be interpreted as a would-be decrement in the late-time matter
abundance compared to the fiducial value of $\omega_b + \omega_c$ (which is itself measured with
$1\%$ precision), or as an incompatibility of standard neutrino physics with \LCDM{} (or vice
versa).

\begin{figure}[t!]
\begin{centering}
    \includegraphics[width=\textwidth]{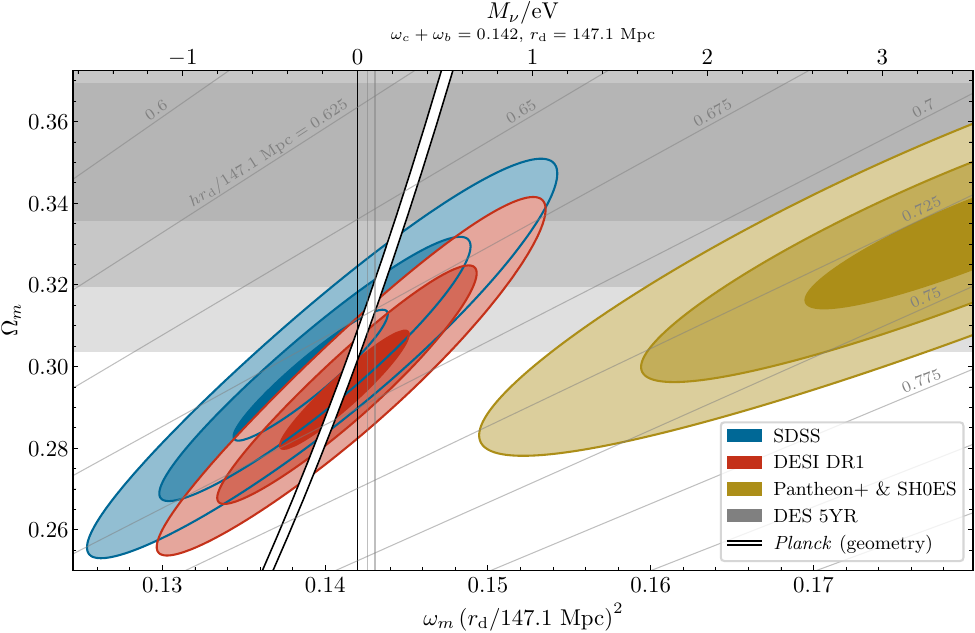}
    \caption{
        Geometric constraints on late-time \LCDM{} cosmology, parametrized in terms of the matter
        fraction $\Omega_m$ and the late-time matter abundance relative to the sound horizon at
        baryon decoupling, $r_\mathrm{d} \sqrt{\omega_m}$.
        The latter is reported relative to a fiducial drag horizon of $147.1~\mathrm{Mpc}$ as
        predicted in \LCDM{}.
        Colored regions depict $1$, $2$, and $3 \sigma$ contours of the two-dimensional joint
        posterior (i.e., the $39.3\%$, $86.5\%$, and $98.9\%$ mass levels thereof) from
        SDSS~\cite{eBOSS:2020yzd, eBOSS:2020lta, eBOSS:2020hur, Ross:2014qpa} (blue),
        DESI~\cite{DESI:2024mwx, DESI:2024uvr, DESI:2024lzq} (red), and
        Pantheon+~\cite{Brout:2022vxf, Scolnic:2021amr} using SH0ES calibrators~\cite{Riess:2021jrx}
        (gold).
        The white band outlined with black depicts the parameter space allowed by \Planck{}'s
        geometric information [that is, the $3 \sigma$ level of its measurement of
        $\theta_{\mathrm{CMB}, \perp}$, \cref{eqn:planck-theta-perp-measurement}].
        The grey regions indicate the $1$, $2$, and $3 \sigma$ ($68.3\%$, $95.4\%$, and $99.7\%$)
        intervals in $\Omega_m$ deriving from the calibration-marginalized DES 5YR supernova
        sample~\cite{DES:2024tys}.
        These geometric constraints make only minimal assumptions about physics before photon and
        baryon decoupling (see text for details); only the Pantheon+ \& SH0ES posterior moves if the
        drag horizon differs from the fiducial value.
        The top axis depicts the neutrino mass sum
        [\cref{eqn:summnu-from-matter-density-difference}] as attributed to the difference between
        $\omega_m$ and \Planck{}'s preferred abundance of matter components at early times,
        $\omega_b + \omega_c \approx 0.142$, taking the same fiducial value for $r_\mathrm{d}$.
        Both of these values are ``calibrated'' via the shape of the CMB power spectra within the
        \LCDM{} model.
        Vertical grey lines indicate the minimum mass sums for the normal and inverted hierarchies.
        Diagonal grey lines mark constant $h r_\mathrm{d} / 147.1~\mathrm{Mpc}$ as labeled.
    }
    \label{fig:geometry-wmrd2-Wm}
\end{centering}
\end{figure}

The CMB's geometric information, despite being one dimensional, is so precise that it effectively
selects a single line in the $r_\mathrm{d} \sqrt{\omega_m}$-$\Omega_m$ plane (within flat \LCDM{}
cosmologies)---one in which the shape of the expansion history ($\Omega_m$) varies rapidly with the
overall amplitude $r_\mathrm{d} \sqrt{\omega_m}$ [per \cref{eqn:cmb-geometric-degeneracy-general}].
Neutrino mass inference relies crucially on the interplay of this constraint with direct
measurements in the plane from uncalibrated BAO data or with measurements of $\Omega_m$ from SNe
distances, as depicted by \cref{fig:geometry-wmrd2-Wm}.
The discrepancy between SNe datasets and BAO datasets is immediately evident but is substantially
more severe when considering each's combination with \Planck{}---much of the parameter space in
which the BAO posteriors do overlap with DES's measurement of $\Omega_m$ is entirely incompatible
with \Planck{}.
Even ignoring the calibration information from SH0ES, the Pantheon+ SNe sample also measures larger
matter fractions than either BAO dataset (as does Union3, per
\cref{fig:matter-fraction-measurements} but not displayed in \cref{fig:geometry-wmrd2-Wm}); the
$\Omega_m$ preferences from all SNe datasets combine with \Planck{} to allow for relatively heavy
neutrinos ($\summnu \sim 0.2$ to $0.4~\mathrm{eV}$).

In contrast to any SNe result, BAO data from both DESI and SDSS favor \emph{lower} matter densities
than \Planck{} infers in baryons and CDM, leading each to prefer that neutrinos have ``negative''
mass on average, purely on geometric grounds.
This trend is compounded when combined with \Planck{}'s geometric information (as illustrated by the
marginal $\Omega_m$ posteriors in \cref{fig:matter-fraction-measurements}).
Despite measuring slightly larger matter densities than SDSS on its own, and despite its
measurements being of comparable precision to SDSS's~\cite{DESI:2024mwx}, DESI's first data release
exhibits an even stronger incompatibility with massive neutrinos when combined with \Planck{}.
Namely, because of its shift toward smaller matter \emph{fractions}, DESI's $1 \sigma$ mass level
overlaps with \Planck{}'s posterior fully where $\omega_m (r_\mathrm{d} / 147.1~\mathrm{Mpc})^2$ is
smaller than \Planck{}'s preferred abundance of baryons and CDM.
Indeed, because of the relative orientation of the CMB and BAO posteriors, artificially shifting the
latter to lower late-time matter densities $\omega_m r_\mathrm{d}^2$ would yield a preference for
heavier neutrinos!\footnote{
    In such a counterfactual, the joint posterior, if shifted sufficiently, would localize in
    parameter space further from that which maximizes the BAO likelihood---that is, the joint
    inference of heavier neutrinos could degrade the fit to BAO observations.
}
Interpreting the combination of this geometric data instead in terms of $h r_\mathrm{d}$ (which, per
\cref{fig:geometry-wmrd2-Wm}, DESI prefers to be larger than does SDSS) obfuscates the actual causal
relationship: per \cref{sec:cmb-geometry} \Planck{} cannot accommodate an increase in the matter
abundance from massive neutrinos without steeply increasing the matter fraction, and BAO posteriors
fall off rapidly in the parameter direction required to do so.

The SH0ES-derived calibration of the Pantheon+ distance ladder allows Hubble constants that, despite
corresponding to larger matter densities on average (via larger $h$), are entirely inconsistent with
both massive neutrinos and \Planck{} simultaneously.
A $\gtrsim 10\%$ miscalibration of $h$ by SH0ES would be required for consistency with massive
neutrinos (and the standard, \LCDM{}-derived drag horizon).
A comparable change in $r_\mathrm{d}$ is not in general equivalent, because scenarios that modify
the sound horizon may also require different baryon and CDM densities (and therefore alter what
remaining matter density may be attributed to neutrinos).
Likewise, joint constraints from \Planck{} and both BAO datasets cannot be reconciled with a
neutrino mass sum larger than $\sim 0.1~\mathrm{eV}$ without modifying the sound horizon (discussed
in \cref{sec:modified-cosmology}) or identifying an effect that would shift the relative positions
of their posteriors in \cref{fig:geometry-wmrd2-Wm} (which we consider in \cref{sec:desi}).

\subsubsection{Joint constraints from the CMB and low-redshift distances}\label{sec:geometry-degeneracy-breaking}

\begin{figure}[t!]
\begin{centering}
    \includegraphics[width=\textwidth]{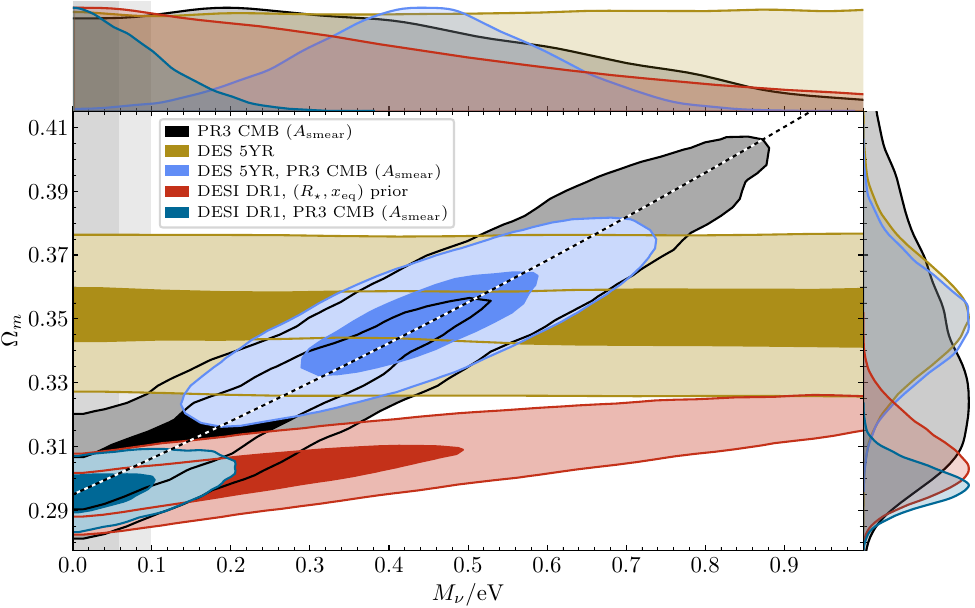}
    \caption{
        Geometric constraints on flat \LCDM{} cosmologies with massive neutrinos, depicting
        posterior distributions using \Planck{} PR3's primary CMB data~\cite{Planck:2018vyg,
        Planck:2019nip} (grey), DESI DR1 BAO data~\cite{DESI:2024mwx, DESI:2024uvr, DESI:2024lzq}
        (red), DES 5YR SNe data~\cite{DES:2024tys} (gold), and \Planck{} combined with each of DESI
        (dark blue) and DES (light blue).
        Results that employ \Planck{} data include the $A_\mathrm{smear}$ parameter in order to
        marginalize over the smearing of the primary CMB by gravitational lensing, while results
        that include only BAO data include a prior on the baryon and CDM densities derived from
        \Planck{} data (\cref{sec:mass-independent}).
        The lower left panel displays the 1 and $2 \sigma$ contours (i.e., the $39.3\%$ and $86.5\%$
        mass levels) of the two-dimensional marginal posterior density over the neutrino mass sum
        $\summnu$ and the matter fraction $\Omega_m$.
        Outer panels diagonal depict kernel density estimates of the one-dimensional marginal
        posteriors over each parameter normalized relative to the peak density.
        Vertical grey shading indicates neutrino mass sums incompatible with the normal and
        inverted hierarchies.
        The dashed black/white line indicates the expected degeneracy direction that fixes the
        distance to last scattering (\cref{sec:cmb-geometry}), $\Omega_m \propto (1 + f_\nu)^{5}$
        where the neutrino fraction $f_\nu$ is defined in \cref{eqn:neutrino-fraction}.
        This figure quantitatively demonstrates how ``calibrating'' geometry with CMB data informs
        neutrino mass constraints, i.e., by marginalizing over $r_\mathrm{d}$ and $\omega_b +
        \omega_c$ within \LCDM{} cosmologies (whereas \cref{fig:geometry-wmrd2-Wm} fixed values
        thereof for illustration).
        The two-dimensional posteriors demonstrate that the discordant geometric preferences of
        DESI BAO and DES SNe data combine with \Planck{}'s measurement of the distance to last
        scattering to yield completely distinct inference of the neutrino mass sum (dark and light
        blue posteriors).
    }
    \label{fig:geometry-pr3-asmear-desi-des}
\end{centering}
\end{figure}

To illustrate the quantitative impact of combining different low-redshift datasets with the CMB's
geometric information and its \LCDM{}-based calibration of the baryon and CDM densities,
\cref{fig:geometry-pr3-asmear-desi-des} depicts the posteriors from each of DESI BAO, DES SNe, and
\Planck{} PR3 individually and combined.
We now employ the full \Planck{} temperature and polarization likelihoods (rather than a reduced
geometric one as in \cref{sec:calibrating-geometry}).
By marginalizing over the degree to which the primary CMB anisotropies are lensed
(via the parameter $A_\mathrm{smear}$; see \cref{sec:peak-smearing-isw}) and excluding four-point
reconstructions of the lensing potential itself, the \Planck{} results in
\cref{fig:geometry-pr3-asmear-desi-des} effectively remove neutrinos' impact on structure at late
times.
(Reference~\cite{BOSS:2014hhw} performed similar analyses using early BOSS data.)

\Cref{fig:geometry-pr3-asmear-desi-des} shows that mass sums as large as an $\mathrm{eV}$ or so are
allowed by \Planck{} when marginalized over $A_\mathrm{smear}$, beyond which the early-time impact
of neutrinos' transition to the nonrelativistic regime (via, for instance, the ISW effect; see
\cref{sec:peak-smearing-isw}) remains precluded.
The \Planck{}-only posterior in \cref{fig:geometry-pr3-asmear-desi-des} thus follows the geometric
degeneracy of \cref{eqn:matter-fraction-fnu-degeneracy-cmb} up to matter fractions as large as
$0.41$ in its $2 \sigma$ mass level.
[The Hubble constant $h$ for cosmologies with neutrino masses and matter fractions so large is as
low as $0.6$ or so, consistent with \cref{eqn:hubble-fnu-degeneracy-cmb}.]
The degeneracies with $\summnu$ for BAO data are qualitatively similar but shallower because many
DESI tracers reside at low redshift.
SNe distances, by contrast, simply measure the matter fraction independently of $h$ when
marginalized over their calibration.
As anticipated from \cref{fig:geometry-wmrd2-Wm}, joint posteriors from \Planck{} combined with
either DESI or DES localize in completely distinct parameter space, with dramatic consequences for
neutrino mass inference.

\begin{figure}[t!]
\begin{centering}
    \includegraphics[width=\textwidth]{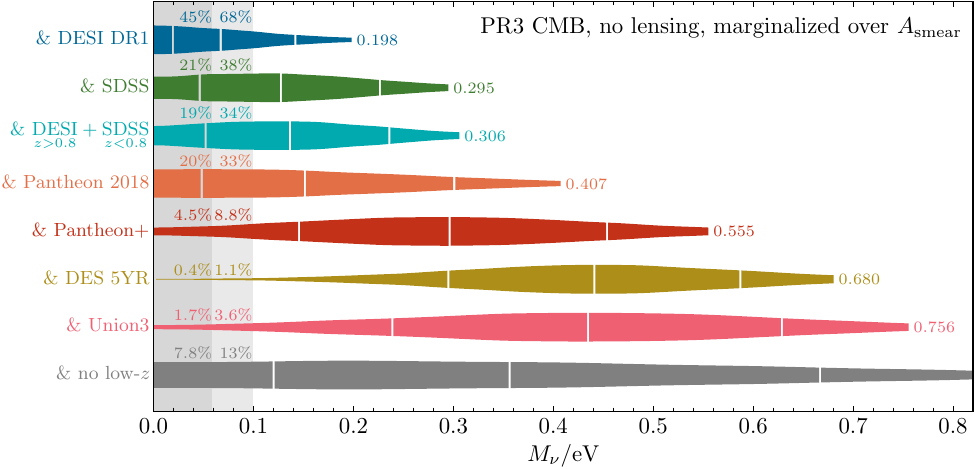}
    \caption{
        Marginal posterior distributions over the neutrino mass sum $\summnu$ deriving largely from
        geometric information in \Planck{} PR3 temperature and polarization power
        spectra~\cite{Planck:2018vyg, Planck:2019nip} combined individually with various BAO and
        (uncalibrated) SNe datasets as labeled.
        Results marginalize over the peak-smearing amplitude $A_\mathrm{smear}$ so that the CMB
        constraints are effectively independent of neutrinos' effects on late-time structure (see
        \cref{sec:peak-smearing-isw}).
        The shaded bands depict the range of masses that are incompatible with the normal and
        inverted hierarchies, and the fraction of each posterior within these ranges
        [\cref{eqn:compatibility-metric}] is annotated on the figure.
        Posteriors are truncated at the $95$th percentile (whose value is also labeled); vertical
        white lines indicate the median and $\pm 1 \sigma$ quantiles.
        Note that the combination of DESI and SDSS data differs from that considered in
        Ref.~\cite{DESI:2024mwx}; see \cref{sec:desi} for elaboration.
    }
    \label{fig:summnu-violin-geometry-asmear-vary-low-z}
\end{centering}
\end{figure}

\Cref{fig:summnu-violin-geometry-asmear-vary-low-z} summarizes the neutrino mass constraints from
combinations of \Planck{} (still marginalized over $A_\mathrm{smear}$) with all recent BAO and SNe
datasets, reporting the hierarchy compatibility metric [\cref{eqn:compatibility-metric}] and the
$95$th percentile of each posterior.
As expected from \cref{fig:geometry-wmrd2-Wm}, joint constraints with DESI DR1 localize close to
zero (with $95\%$ of the marginal posterior at $\summnu \lesssim 0.2~\mathrm{eV}$).
In sharp contrast, DES SNe's preference for substantially larger $\Omega_m$ combines with \Planck{}
to \emph{detect} nonzero $\summnu / \mathrm{eV} \approx 0.44 \pm 0.15$ (a Gaussian $z$ score of
nearly $3$).
While only $32\%$ and $55\%$ of the posterior for \Planck{} combined with DESI are compatible with
the inverted and normal hierarchies, respectively, only $1\%$ and $0.4\%$ of the DES \& \Planck{}
posterior are \emph{in}compatible with each.
The recent Pantheon+ dataset (when uncalibrated) also yields a preference for nonzero masses,
$\summnu / \mathrm{eV} \approx 0.30_{-0.15}^{+0.16}$; its spread is comparable to DES's but centered
at lower masses, which simply reflects the $\sim 1 \sigma$ offset in their measurements of
$\Omega_m$ (see \cref{fig:matter-fraction-measurements}).
Both SNe posteriors are inconsistent with those from any BAO dataset in
\cref{fig:summnu-violin-geometry-asmear-vary-low-z}, but especially so for the full combination of
DESI DR1 measurements.
However, \cref{fig:summnu-violin-geometry-asmear-vary-low-z} also shows that replacing DESI's
low-redshift measurements with SDSS's (which are of comparable precision) substantially weakens its
limits on $\summnu$.
In fact, SDSS data by itself yields an extremely similar posterior that of this combination.
We investigate what BAO measurements from individual tracers drive DESI's strong preference for
negligible neutrino masses next.

\subsubsection{Impact of DESI luminous red galaxy measurements}
\label{sec:desi}

Neutrino mass constraints are highly sensitive to any bias in low-redshift distance measurements
because of the crucial role they play in breaking degeneracies in late-time cosmology, as well as
the close proximity of current constraints to the prior boundary $M_\nu = 0$.
\Cref{fig:geometry-wmrd2-Wm} shows that the tightened upper limit on $\summnu$ from DESI compared to
that from SDSS (each combined with \Planck{}) derives only from a shift in their preferred matter
fractions; both surveys at present constrain similar volumes in parameter space (i.e., at a given
posterior mass level).
While the DESI DR1 measurements pass numerous robustness tests~\cite{DESI:2024uvr, DESI:2024lzq,
DESI:2024ude}, the difference between DESI and SDSS measurements for similar populations of tracers
motivates scrutinizing the discrepancies carefully.

\begin{figure}[t!]
\begin{centering}
    \includegraphics[width=\textwidth]{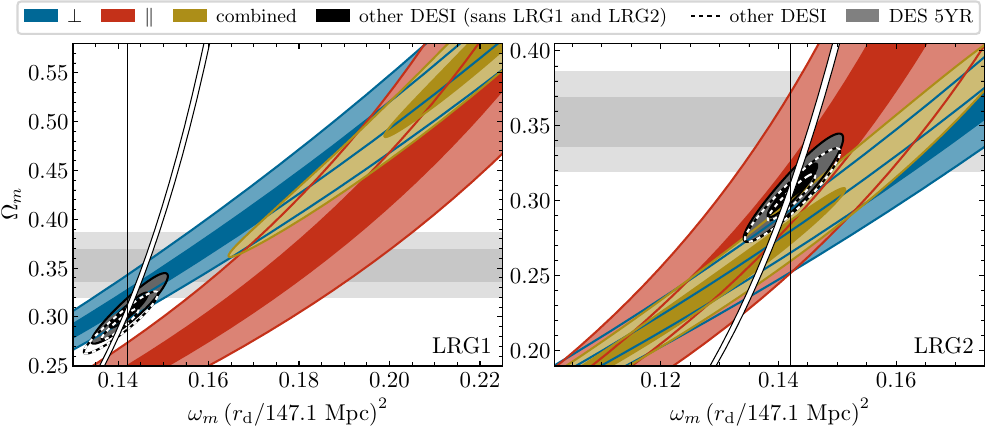}
    \caption{
        Comparison of \LCDM{} parameter space compatible with DESI DR1's BAO
        measurements~\cite{DESI:2024mwx, DESI:2024uvr, DESI:2024lzq} from luminous red galaxy
        samples at effective redshifts $0.51$ (LRG1, left) and $0.706$ (LRG2, right), displaying
        posteriors deriving from transverse (blue) and line-of-sight (red) measurements and their
        combination (gold).
        Appearing in addition are posteriors from all other DESI tracers (dashed black/white lines),
        from all other DESI tracers excluding both LRG bins (black), from the DES 5YR SNe
        sample~\cite{DES:2024tys} (grey), and from \Planck{} (white band outlined in black with
        extent covering its $5 \sigma$ interval; see \cref{app:bao-from-cmb}).
        Results are presented as in \cref{fig:geometry-wmrd2-Wm} but instead displaying only $1$ and
        $2 \sigma$ mass levels.
        While the $\text{LRG1}_\perp$ and $\text{LRG2}_\parallel$ measurements coincide quite well
        with the joint measurement from other tracers, the $\text{LRG1}_\parallel$ and
        $\text{LRG2}_\perp$ ones are each offset from this parameter space by several standard
        deviations.
        Comparing the filled black and unfilled dashed contours indicates the individual impact of
        the LRG2 (left) and LRG1 (right) measurements on joint constraints; both measurements
        truncate the joint posteriors at lower $\Omega_m$, reducing the allowed contribution to the
        matter density from neutrinos (and increases the degree of tension with the range of
        $\Omega_m$ preferred by SNe datasets).
    }
    \label{fig:desi-para-perp-compare-lrg}
\end{centering}
\end{figure}

\Cref{fig:desi-para-perp-compare-lrg} shows that, when interpreted in the flat \LCDM{} model, DESI's
DR1 measurements from both LRG bins are nonnegligibly discrepant with all other tracers.
In fact, just the line-of-sight measurement in the $z = 0.51$ bin ($\text{LRG1}_\parallel$) and the
transverse one in the $z = 0.706$ bin ($\text{LRG2}_\perp$) are in tension; the posteriors from the
other measurement in each bin coincide extremely well with the joint posterior from all other
tracers (see \cref{fig:desi-sdss-para-perp-compare-other-tracers}).\footnote{
    We obtain separate constraints from transverse and line-of-sight measurements simply by
    marginalizing over the other at the level of the likelihood; their correlation is accounted for
    in the joint likelihood as usual, but neglecting it only slightly distorts the shape of the
    resulting posterior.
}
Moreover, these other tracers are far more compatible with the $\Omega_m$ measurements deriving from
SNe datasets like DES 5YR and Pantheon+.
Observe that, though the joint LRG1 posterior itself localizes at substantially larger
$\omega_m r_\mathrm{d}^2$ and $\Omega_m$ than all other tracers, it only overlaps with the bulk of
the posterior from all other tracers rather far in its tail, driving combined constraints toward
lower matter fractions.
This contrast in concordance is even more dramatic when considering their intersections with
\Planck{}'s geometric constraint (even without calibrating a value for the drag horizon).
DESI's other tracers, its $\text{LRG1}_\perp$ and $\text{LRG2}_\parallel$ measurements, and
\Planck{} all intersect in precisely the same region of parameter space, while
$\text{LRG1}_\parallel$ and $\text{LRG2}_\perp$'s $2 \sigma$ mass levels both intersect with the
\Planck{} constraint at much lower $\Omega_m$ and $r_\mathrm{d} \sqrt{\omega_m}$.\footnote{
    See \cref{fig:geometry-wmrd2-Wm-desi-lrg-subsets} in \cref{app:lowz-consistency} for further
    illustration of this point and \cref{fig:desi-sdss-para-perp-compare-other-tracers} for
    consistency tests analogous to \cref{fig:desi-para-perp-compare-lrg} for all other tracers from
    DESI and eBOSS.
    \Cref{fig:desi-sdss-para-perp-compare-other-tracers} in particular shows that eBOSS's
    Ly-$\alpha$ measurements are slightly anomalous and also shift joint constraints toward
    parameter space that is less compatible with massive neutrinos.
}

\begin{figure}[t!]
\begin{centering}
    \includegraphics[width=\textwidth]{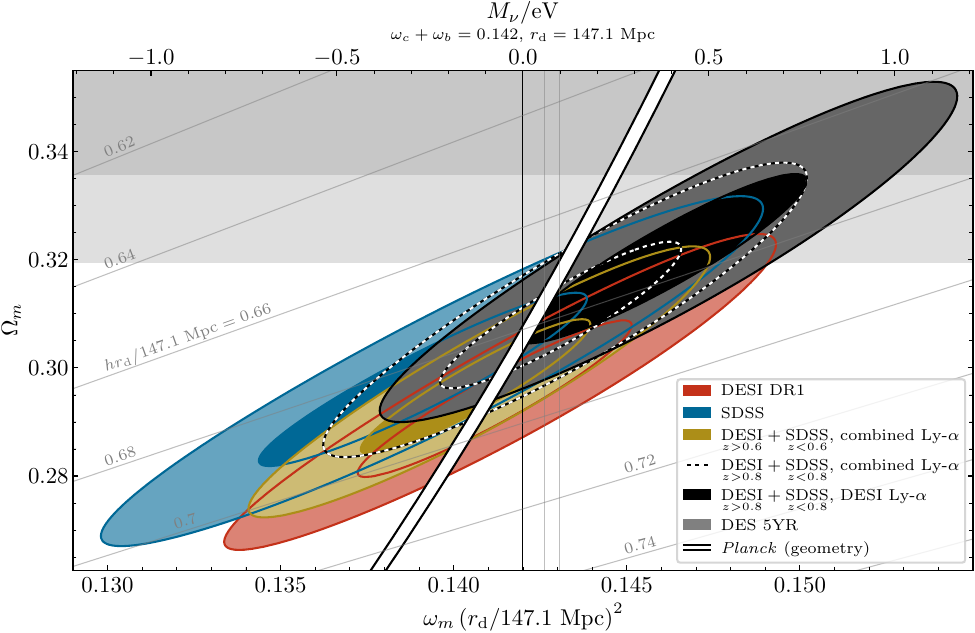}
    \caption{
        Geometric constraints on late-time \LCDM{} cosmology from various combinations of
        SDSS~\cite{eBOSS:2020yzd, eBOSS:2020lta, eBOSS:2020hur, Ross:2014qpa} and
        DESI~\cite{DESI:2024mwx, DESI:2024uvr, DESI:2024lzq} data, presented as in
        \cref{fig:geometry-wmrd2-Wm} but instead displaying only $1$ and $2 \sigma$ mass levels (and
        the $2 \sigma$ interval for \Planck{}; see \cref{app:bao-from-cmb}).
        Results include all DESI tracers (red); all SDSS tracers, including both the DR7 MGS and
        eBOSS (blue); the ``DESI+SDSS'' combination from Ref.~\cite{DESI:2024mwx} which includes
        SDSS data at $z < 0.6$, DESI data at higher redshift, and the joint DESI+SDSS Ly-$\alpha$
        dataset (gold); the same combination but instead split at $z = 0.8$ (dashed black/white);
        and the $z = 0.8$ split that only uses DESI's Ly-$\alpha$ measurement rather than the joint
        DESI+SDSS analysis thereof.
        Note that the DESI/SDSS split at $z = 0.6$ (gold), as employed in Ref.~\cite{DESI:2024mwx},
        includes DESI's LRG2 bin, which is replaced by eBOSS DR16's measurement at a similar
        redshift in the other combinations split at $z = 0.8$; the latter combinations also exclude
        DESI's LRG3+ELG bin (covering $0.8 < z < 1.1$) to be conservative about double counting
        information.
        Comparing these results shows that all BAO measurements from SDSS and DESI that could be
        interpreted as outliers within \LCDM{} (LRG1 and LRG2 for DESI and Ly-$\alpha$ for SDSS)
        shift posteriors toward lower matter fraction $\Omega_m$ and amplitude $r_\mathrm{d}
        \sqrt{\omega_m}$, directions that each disfavor massive neutrinos.
    }
    \label{fig:geometry-wmrd2-Wm-desi-sdss-combos}
\end{centering}
\end{figure}

\Cref{fig:geometry-wmrd2-Wm-desi-sdss-combos} illustrates the impact of these potential outliers on
parameter inference by comparing various combinations of BAO measurements from DESI and SDSS.
Reference~\cite{DESI:2024mwx} combined SDSS data below redshift $0.6$ with DESI data above it, using
a joint DESI+SDSS Ly-$\alpha$ measurement~\cite{DESI:2024lzq}.
When combined with \Planck{} data, the resulting neutrino mass constraints are slightly weaker than
those from the DESI-only dataset in spite of the combination having slightly greater constraining
power in the $r_\mathrm{d} \sqrt{\omega_m}$-$\Omega_m$ plane than the latter.
Reference~\cite{DESI:2024mwx} identified the offset in $h r_\mathrm{d}$ preferred by each
combination as the origin of this finding, but \cref{fig:geometry-wmrd2-Wm} and surrounding
discussion show that this is only a happenstance consequence of the offset in $\Omega_m$ which, when
projected along \Planck{}'s constraint, produces a tighter constraint on $\summnu$.
This particular combination, however, includes both DESI's LRG2 measurement and eBOSS's Ly-$\alpha$
data.
\Cref{fig:geometry-wmrd2-Wm-desi-sdss-combos} shows that splitting data from the surveys at
$z = 0.8$ (i.e., using none of DESI's LRG data) instead shifts posteriors toward higher $\Omega_m$
and $\omega_m r_\mathrm{d}^2$.
Swapping the combined Ly-$\alpha$ measurement for DESI's alone further shifts the posteriors along
this direction toward parameter space much more compatible with SNe data; however, these
combinations prefer the same $\Omega_m$ at fixed amplitude $\sqrt{\omega_m} r_\mathrm{d}$.
Any dataset combination that uses SDSS's LRG data rather than any of DESI's---that is, both
alternative combinations and the SDSS-only dataset---intersects with \Planck{} at roughly the same
point in parameter space (at $2 \sigma$), explaining the close agreement of posteriors over
$\summnu$ in \cref{fig:summnu-violin-geometry-asmear-vary-low-z}.

In future data releases DESI's apparent outliers could well shift back to concordant parameter space
by pure statistical chance.
To the extent that future data releases can be considered independent measurements with much higher
precision (a decent approximation, given that the effective volume for LRGs will increase by about a
factor of three~\cite{DESI:2024uvr}) and no systematic differences, their central values may be
considered a random draw from the DR1 likelihoods.
The probability that the LRG1 line-of-sight measurement drops by more than $1 \sigma$ while the
transverse one shifts by no more than $1 \sigma$ in either direction\footnote{
    The reported probabilities increase by a factor of $1.5$ to $2$ if allowing the transverse
    measurement to instead shift by as much as $\pm 2 \sigma$.
    The transverse and line-of-sight measurements would have to be much more correlated
    (than their $\sim -0.4$ coefficient) for this marginalization to have a more substantive effect.
    Measurements in other bins are independent; the additional penalty in probability incurred by
    requiring they not become apparent outliers is likewise independent of any shift in the
    particular bins we call out.
} is about $10\%$; a shift more than $- 2 \sigma$ has $1\%$ probability.
The LRG1 and LRG2 bins (which are independent of each other) each have nearly the same correlation
coefficient, so the same quantities apply to possible shifts in the LRG2 transverse measurement.

These likelihood-level metrics, however, ignore additional prior information from SDSS LRG
measurements in a similar redshift range, which do prefer parameter space consistent with all other
DESI tracers.\footnote{
    As discussed in Ref.~\cite{DESI:2024mwx}, the two surveys overlap, making a quantitative
    comparison nontrivial; our discussion is meant only to be illustrative.
}
Reference~\cite{DESI:2024mwx} identified $\text{LRG2}_\perp$ as most discrepant from the most
directly comparable SDSS measurement, though their line-of-sight measurements at $z \approx 0.51$
also do not agree well.
And in contrast to Ref.~\cite{Naredo-Tuero:2024sgf}, we find that one measurement from each redshift
bin matters, not both in only the LRG2 bin: both outliers ($\text{LRG1}_\parallel$ and
$\text{LRG2}_\perp$) have the same effect on joint posteriors (i.e., posteriors derived by excluding
each are nearly identical).
The measurements at lower redshift are especially important because the orientation of their
posteriors in the $r_\mathrm{d} \sqrt{\omega_m}$-$\Omega_m$ plane differs the most from those at
high redshift (e.g., from the CMB), making them crucial for breaking degeneracies and determining
the parameter values at which joint constraints localize.

Reference~\cite{DESI:2024uvr} found negligible shifts in results when applying DESI's new analysis
pipeline to the BOSS/eBOSS catalogs, suggesting that no presently known systematic can adjudicate
these measurements.
Reference~\cite{DESI:2024ude} studies the sensitivity of DESI's measurements to the fiducial
cosmology employed in the analysis with mock data, finding in particular that a $14\%$ reduction in
$\Omega_m$ can lead to some of the largest scatter (up to $1-2\%$) among the alternative cosmologies
tested.
Given that the matter fraction preferred by DES SNe is $12\%$ larger than the fiducial value used in
DESI's analysis, it would be interesting to further investigate the sensitivity thereof and test how
the individual shifts in BAO measurements within each alternative cosmology jointly propagate to
cosmological parameters.
In addition, DESI's likelihoods are specified by Gaussian approximations in
$1/\theta_{\mathrm{BAO}, \perp}$ and $1/\theta_{\mathrm{BAO}, \parallel}$; it might also be worth
checking this approximation, since the joint parameter space preferred by all DESI data lie rather
far in the tail of, e.g., the $\text{LRG1}_\parallel$ measurement [see
\cref{fig:desi-para-perp-compare-lrg}].

\subsection{Impact of CMB lensing}\label{sec:lensing-constraints}

We now explore the impact of CMB lensing data on neutrino mass inference.
In \cref{sec:no-geometry} we study constraints that effectively remove the CMB's geometric
information, seeking to illustrate how low-redshift distance datasets break the degeneracy between
$\summnu$ and the matter fraction $\Omega_m$ that derives purely from the growth of structure.
\Cref{sec:joint-constraints} then presents results from the full combination of CMB temperature and
polarization spectra, CMB lensing, and low-redshift distance datasets, sans any nonstandard
modifications.

\subsubsection{CMB lensing without CMB geometry}\label{sec:no-geometry}

Large-scale structure observations alone are sensitive to the combined effect of the expansion
history and clustering of matter components, but direct probes of late-time geometry can break this
degeneracy.
As an approximate means to remove geometric information from CMB likelihoods, we construct a
multivariate normal approximation to posteriors derived using uniform priors over \LCDM{} parameters
and \Planck{} temperature and polarization data (by computing the mean and covariance of a sample
thereof) and marginalize over $\theta_s$ (i.e., by removing that dimension from the approximate
posterior).
While this procedure is rather \textit{ad hoc}, it simply serves as a means to take \Planck{}'s
information on the shape of the CMB anisotropy spectra as a prior for CMB lensing likelihoods.
To showcase the constraining power of CMB lensing maps in probing structure, we also marginalize the
\Planck{} posteriors over $A_\mathrm{smear}$.

\begin{figure}[t!]
\begin{centering}
    \includegraphics[width=\textwidth]{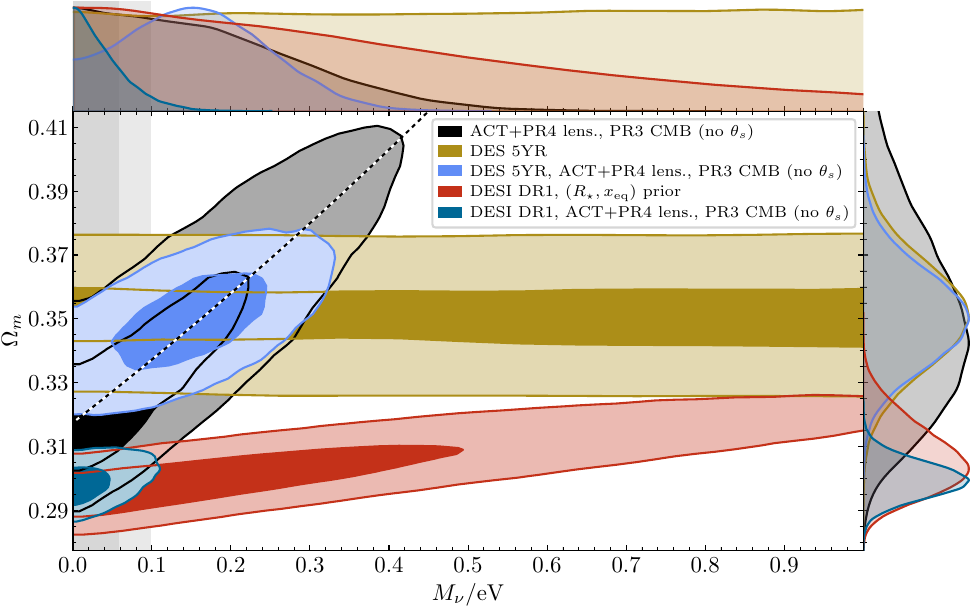}
    \caption{
        Posterior distribution over the neutrino mass sum $\summnu$ and the matter fraction
        $\Omega_m$, ACT + \Planck{} PR4 lensing data~\cite{ACT:2023dou, ACT:2023dou, Carron:2022eyg}
        (grey), DESI DR1 BAO data~\cite{DESI:2024mwx, DESI:2024uvr, DESI:2024lzq} (red), DES 5YR SNe
        data~\cite{DES:2024tys} (gold), and the lensing dataset combined with each of DESI (dark
        blue) and DES (light blue).
        Results that include lensing data impose a prior on all \LCDM{} parameters except
        $\theta_s$, deriving from the posterior using \Planck{} PR3 CMB data~\cite{Planck:2018vyg,
        Planck:2019nip} (also marginalized over $A_\mathrm{smear}$), while results that exclude
        \Planck{} data instead include a prior on the baryon and CDM densities derived from
        \Planck{} data (\cref{sec:mass-independent}).
        The dashed black/white line indicates the parameter direction
        $\Omega_m \propto (1 + f_\nu)^{8}$ (chosen to match by eye the orientation of the posterior
        using only CMB and lensing data).
        Results are otherwise presented as in \cref{fig:geometry-pr3-asmear-desi-des}.
    }
    \label{fig:structure-pr3-no-theta-s-act-pr4-lens-desi-des}
\end{centering}
\end{figure}

We take this posterior approximation as a ``geometry-free'' \Planck{} prior over $\omega_b$,
$\omega_c$, $\tau_\mathrm{reio}$, $n_s$, and $A_s$ in combination with measurements of the CMB
lensing power spectrum from ACT and \Planck{} PR4 in
\cref{fig:structure-pr3-no-theta-s-act-pr4-lens-desi-des}.
Without \Planck{}'s geometric information to pin the dark energy density and fix the distance to
last scattering, the matter fraction is only constrained with $\sim 5\%$ precision (at any
particular value of $\summnu$, increasing to $\sim 9\%$ when marginalized).
The suppression of structure by massive neutrinos can thus be partly compensated for by delaying the
dark energy era, leading to a degeneracy direction $\Omega_m \propto (1 + f_\nu)^8$ or so.
The posterior for lensing data alone falls off at a 95th percentile of $\summnu \approx
0.4~\mathrm{eV}$, at which point neutrinos suppress the amplitude of structure to an extent
intolerable to the ACT+PR4 lensing dataset.

The steep correlation between the neutrino mass sum and the matter fraction again poises
low-redshift distances to contribute substantially to neutrino mass constraints.
\Cref{fig:structure-pr3-no-theta-s-act-pr4-lens-desi-des} also displays the same posteriors using
DESI DR1 BAO and DES 5YR SNe from \cref{fig:geometry-pr3-asmear-desi-des} and the combination of
these data with ACT+PR4 lensing and the geometry-free \Planck{} prior.
Once again, the inconsistent geometric preferences of these two distance datasets yield strikingly
different inference of neutrino masses---a tight upper limit from DESI compared to a marginal
preference for $\summnu \approx 0.15~\mathrm{eV}$ from DES SNe.
The latter dataset requires so late an onset of dark-energy domination that massive neutrinos are in
fact required to fit the amplitude of structure measured directly by the CMB lensing power spectrum.
In contrast, DESI infers a matter fraction substantially lower than the CMB lensing dataset even
without massive neutrinos; this apparent tension drives their combination to hardly allow for
massive neutrinos at all.

\subsubsection{Joint constraints}\label{sec:joint-constraints}

\begin{figure}[t!]
\begin{centering}
    \includegraphics[width=\textwidth]{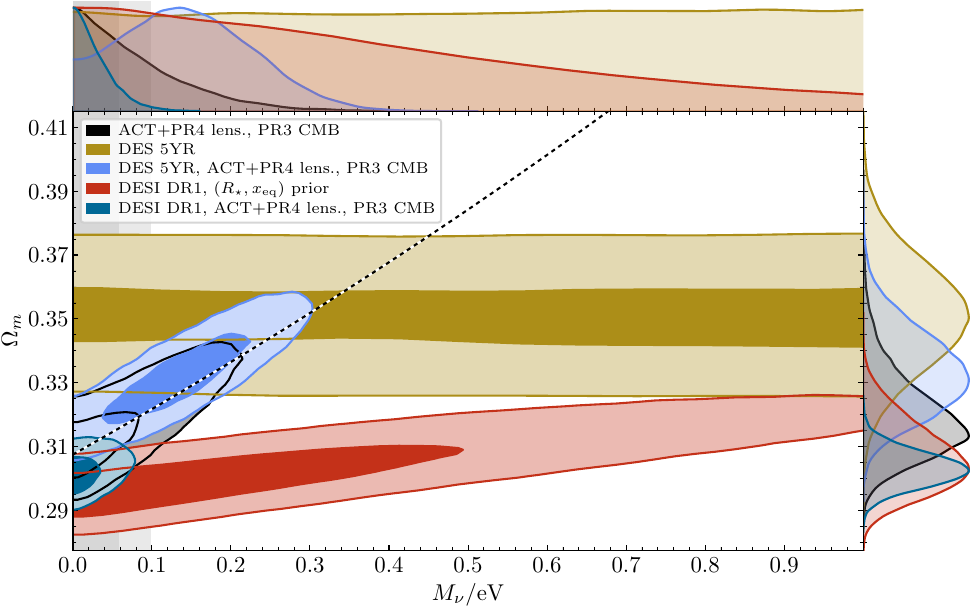}
    \caption{
        Constraints on flat \LCDM{} cosmology with massive neutrinos using information from both
        geometry and structure, displaying posterior distributions over the neutrino mass sum
        $\summnu$ and the matter fraction $\Omega_m$ using \Planck{} PR3 CMB~\cite{Planck:2018vyg,
        Planck:2019nip} and ACT + \Planck{} PR4 lensing data~\cite{ACT:2023dou, ACT:2023dou,
        Carron:2022eyg} (grey), DESI DR1 BAO data~\cite{DESI:2024mwx, DESI:2024uvr, DESI:2024lzq}
        (red), DES 5YR SNe data~\cite{DES:2024tys} (gold), and the CMB and lensing datasets combined
        with each of DESI (dark blue) and DES (light blue).
        The dashed black/white line indicates the parameter direction
        $\Omega_m \propto (1 + f_\nu)^{6}$ (chosen to match by eye the orientation of the posteriors
        using only CMB and lensing data).
        Results are otherwise presented as in
        \cref{fig:geometry-pr3-asmear-desi-des,fig:structure-pr3-no-theta-s-act-pr4-lens-desi-des}.
    }
    \label{fig:full-pr3-act-pr4-lens-desi-des}
\end{centering}
\end{figure}

Having established that low-redshift distances also impact neutrino mass inference via the amplitude
of structure, we now study constraints deriving jointly from low-redshift distances and CMB lensing
data (as well as primary CMB data).
\Cref{fig:full-pr3-act-pr4-lens-desi-des} displays the analog of
\cref{fig:structure-pr3-no-theta-s-act-pr4-lens-desi-des} including the full \Planck{} likelihoods
(rather than artificially removing \Planck{}'s geometric information).
Clearly, DESI BAO and DES SNe's distinct preferences for $\Omega_m$ still lead to discrepant
neutrino mass inference, even when consistently accounting for all of their effects.
However, though both geometry and structure yield a positive correlation between the neutrino mass
sum and the matter fraction, the degree to which these combinations disagree is less severe than
when constraining geometry (\cref{fig:geometry-pr3-asmear-desi-des}) or structure
(\cref{fig:structure-pr3-no-theta-s-act-pr4-lens-desi-des}) separately.
Neutrino mass bounds are also tighter for \Planck{} CMB data combined with ACT+PR4 lensing without
any low-redshift data, signaling that \Planck{}'s geometric preferences alone prefer smaller matter
fractions than CMB lensing data does by itself.
Nevertheless, whereas the combination with DESI data yields posteriors that are largely incompatible
with either mass hierarchy, the posterior using DES SNe data peaks not just at nonzero $\summnu$ but
at $\summnu \approx 0.14~\mathrm{eV}$, greater than the minimum mass for either hierarchy.

The apparent tension between structure and geometry evident in the comparison of the CMB-only
posteriors in
\cref{fig:geometry-pr3-asmear-desi-des,fig:structure-pr3-no-theta-s-act-pr4-lens-desi-des} could be
interpreted as a manifestation of the lensing excess.
Note however that \cref{fig:summnu-violin-vary-lensing-asmear-desi-des-none} shows that
marginalizing over the amplitude of lensing that smears the acoustic peaks ($A_\mathrm{smear}$) only
marginally weakens neutrino mass bounds.
In particular, the direct reconstruction of the CMB lensing spectrum from ACT and \Planck{} PR4
yields nearly identical posteriors, indicating that its information on late-time structure
dominates that inferred from the smearing of the acoustic peaks.
These results should therefore be robust to any systematic explanation for the lensing anomaly
from \Planck{}'s PR3 power spectra, but they still suggest that four-point reconstructions of
CMB lensing prefer more structure growth than \Planck{} data would otherwise predict.
A preference for lensing amplitudes larger than predicted in \LCDM{} also persists in the South Pole
Telescope's newest reconstructed lensing spectrum~\cite{SPT-3G:2024atg}.

\begin{figure}[t!]
\begin{centering}
    \includegraphics[width=\textwidth]{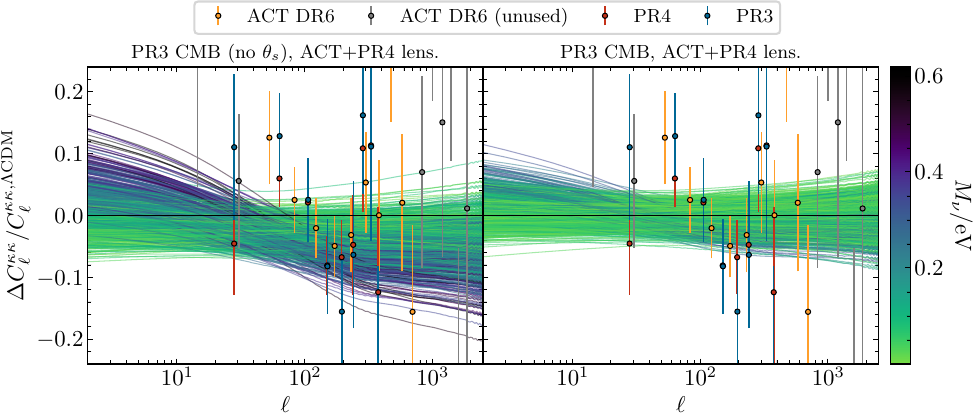}
    \caption{
        Residuals of the CMB lensing convergence relative to the \LCDM{} best fit (using \Planck{}
        PR3 CMB and ACT+PR4 lensing data) for samples from posteriors that use the full
        \Planck{} likelihoods (right) and that artificially remove its geometric information (left),
        as described in \cref{sec:no-geometry}.
        Note that the reference result includes no massive neutrinos.
        Curves are colored by their value of $\summnu$ and data from various observations are also
        depicted as labeled in the legend; note that the data included for the posteriors depicted
        here are labeled ACT DR6 (yellow) and PR4 (red).
    }
    \label{fig:mnu3-lens-spaghetti-geometry-vs-structure}
\end{centering}
\end{figure}

To illustrate the impact of \Planck{}'s geometric information on predictions for the CMB lensing
spectra, \cref{fig:mnu3-lens-spaghetti-geometry-vs-structure} depicts residuals of posterior samples
relative to the \LCDM{} best-fit cosmology (with massless neutrinos only) for results using all CMB
and CMB lensing data (as depicted in \cref{fig:full-pr3-act-pr4-lens-desi-des}) and that with
\Planck{}'s geometric information artificially removed
(\cref{fig:structure-pr3-no-theta-s-act-pr4-lens-desi-des}).
\Cref{fig:mnu3-lens-spaghetti-geometry-vs-structure} also superimposes CMB lensing measurements from
various \Planck{} and ACT datasets, also relative to the best-fit cosmology with massless neutrinos.
Intriguingly, the data skew above this best-fit prediction at $\ell \lesssim 100$ and below
at $100 \lesssim \ell \lesssim 300$---a feature \cref{fig:cl-pp-residual-units-of-fnu-fix-ths-h}
suggests could possibly be accounted for by massive neutrinos, given the correlated increase in
$\Omega_m$ (at fixed $\theta_s$) that enhances structure on large scales.
This scale-dependent trend indeed appears to partially drive the geometry-free result in
\cref{fig:mnu3-lens-spaghetti-geometry-vs-structure}: the correlated increase of $\summnu$ and
$\Omega_m$ in \cref{fig:structure-pr3-no-theta-s-act-pr4-lens-desi-des} roughly fixes power at
moderate scales (near $\ell \sim 100$) while enhancing it above the neutrino free-streaming scale
and suppressing it below.
When employing the full primary CMB likelihood, however, heavier neutrinos can only partially
explain the shape of the data's residuals because \Planck{}'s geometric information
(\cref{fig:geometry-pr3-asmear-desi-des}) requires smaller matter fractions at a given $\summnu$.
Note that \LCDM{} without massive neutrinos has no freedom to explain such scale-dependent
features---samples of the model's posterior, subjected to the same datasets, only yield residuals
that are effectively flat across $\ell$.

Given the close coincidence of $\ell_\mathrm{fs}$ [\cref{eqn:ell-fs}, below which neutrino
free-streaming imprints] and $\ell_\mathrm{eq}$ [\cref{eqn:ell-eq}, the multipole at which the
turnover in the transfer function imprints, per \cref{sec:growth-with-neutrinos}], it is unclear
what this trend may be attributed to.
Since primary CMB data is also strongly sensitive to $\ell_\mathrm{eq}$, it could arise due to an
internal tension in the scale of matter-radiation equality preferred by each observable.
On the other hand, neutrino free streaming provides a plausible explanation for such a mismatch.
Regardless, \cref{fig:mnu3-lens-spaghetti-geometry-vs-structure} shows that standard cosmology with
standard massive neutrinos cannot explain this feature in lensing data sufficiently well to actually
detect nonzero neutrino masses.
We comment on what hypothetical new physics could improve massive neutrino's capacity to better
improve upon \LCDM{}'s fit to these data in \cref{sec:modifications-to-neutrinos}.
Finally, we note that, because lensing data no longer systematically skew below the best-fit \LCDM{}
prediction at $\ell \gtrsim 300$, modeling of nonlinear structure growth (or neglecting it) has a
nonnegligible impact on neutrino mass inference, a point elaborated on in \cref{app:nonlinear}.

\begin{figure}[t!]
\begin{centering}
    \includegraphics[width=\textwidth]{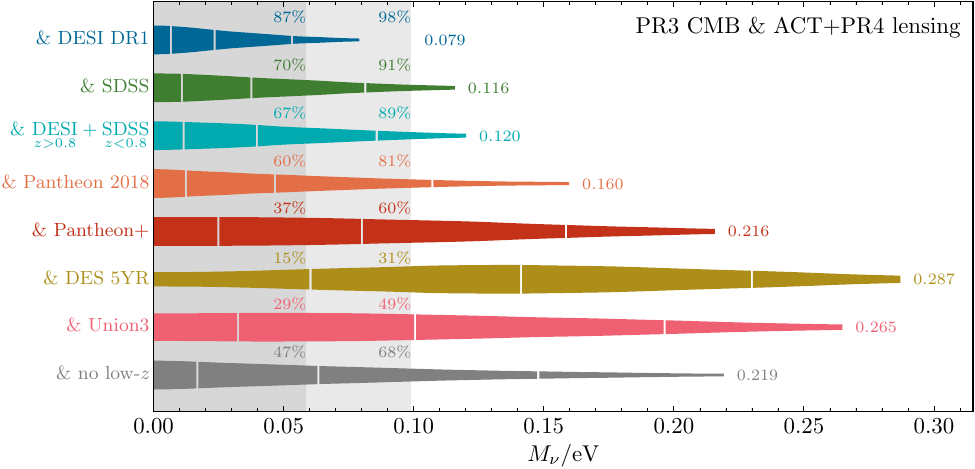}
    \caption{
        Marginal posterior distribution over the neutrino mass sum $\summnu$ for \Planck{} PR3
        CMB~\cite{Planck:2018vyg, Planck:2019nip} and ACT+PR4 CMB lensing data~\cite{ACT:2023dou,
        ACT:2023dou, Carron:2022eyg} combined with various BAO and uncalibrated SNe datasets as
        labeled.
        Results thus consistently derive from massive neutrinos' effects on both geometry and
        structure (in contrast to \cref{fig:summnu-violin-geometry-asmear-vary-low-z}, which
        effectively derives only from geometry).
        The shaded bands depict the range of masses that are incompatible with the normal and
        inverted hierarchies, and the fraction of each posteriors within these ranges
        [\cref{eqn:compatibility-metric}] is annotated on the figure.
        Posteriors are truncated at the $95$th percentile (whose value is also labeled);
        vertical white lines indicate the median and $\pm 1 \sigma$ quantiles.
    }
    \label{fig:summnu-violin-act-pr4-vary-lowz}
\end{centering}
\end{figure}

To close, in \cref{fig:summnu-violin-act-pr4-vary-lowz} we present final constraints on the neutrino
mass sum deriving from the full \Planck{} CMB and ACT+PR4 CMB lensing datasets when combined with
the same set of low-redshift datasets displayed in
\cref{fig:summnu-violin-geometry-asmear-vary-low-z}.
Beyond the near-incompatibility of DESI with massive neutrinos and DES SNe's measurement of
$\summnu / \mathrm{eV} = 0.141_{-0.081}^{+0.089}$, \cref{fig:summnu-violin-act-pr4-vary-lowz}
demonstrates again how varied neutrino mass bounds are between BAO and SNe datasets.
The differences between these posteriors in $\summnu$ correlate strongly to the dispersion in
$\Omega_m$ measurements displayed in \cref{fig:matter-fraction-measurements}, corroborating the
unique importance of late-time geometry to neutrino mass inference explained in
\cref{sec:signatures}.
Likewise, the influence of outliers in DESI's DR1 data as identified in \cref{sec:desi} carries over
to analyses including structure as well as geometric information: the SDSS/DESI combination that
excludes these outliers again yields a posterior nearly identical to that from SDSS alone.
Finally, we note that the neutrino mass bounds originally derived from ACT's DR6 lensing data
included BAO data~\cite{ACT:2023kun}; given the disparate inference among various late-time probes
in \cref{fig:summnu-violin-act-pr4-vary-lowz}, the results presented in this section that combine
with \Planck{}'s primary CMB data alone (and no late-time dataset) perhaps represent a more
conservative bound on neutrino masses from CMB lensing as a probe of structure.
\Cref{fig:summnu-violin-vary-lensing-asmear-desi-des-none} in \cref{app:lensing-asmear} includes
results derived from other CMB lensing datasets combined with \Planck{}.

\section{Discussion}
\label{sec:discussion}

In this section we discuss the broader implications of the discordance in neutrino mass inference
established by \cref{sec:results}.
We first comment on the discrepant inference of the low-redshift expansion history from distance
datasets and other longer-standing dataset tensions in \cref{sec:tensions}.
These tensions may well signal a need for extending standard cosmology, which we consider in
\cref{sec:modified-cosmology}.
Neutrino mass constraints in cosmologies with modified late-time dynamics are well studied, but new
physics at early times yields qualitatively distinct possibilities that have received less
attention~\cite{Baryakhtar:2024rky}.
Finally, the apparent incompatibility of DESI's BAO data (combined with CMB data) with massive
neutrinos has prompted consideration of new neutrino physics.
In \cref{sec:modifications-to-neutrinos} we comment on the ramifications of our findings for
interpreting evidence for nonminimal neutrino dynamics in current data.

\subsection{Dataset tensions}\label{sec:tensions}

The qualitatively distinct preferences for massive neutrinos between current BAO and SNe distance
datasets underscores the growing tension in their inference of the Universe's composition, even
independent of its overall scale (i.e., of the Hubble tension).
Within the flat \LCDM{} model with massive neutrinos, both observables are individually consistent
with the CMB's geometric information, but in highly separated regions of parameter space (as evident
in \cref{fig:geometry-wmrd2-Wm}).
The most recent BAO dataset from DESI, combined with CMB and CMB lensing data, is nominally almost
completely incompatible with the inverted mass hierarchy and perhaps even the minimum mass sum
required by either hierarchy.
\Cref{fig:desi-para-perp-compare-lrg,fig:geometry-wmrd2-Wm-desi-sdss-combos}, however, show that
inference within \LCDM{} using BAO data from DESI and SDSS is strongly influenced by data points
that appear to be outliers compared to most other data within each survey---measurements that
moreover are inconsistent with comparable ones from the other survey.
\Cref{fig:desi-para-perp-compare-lrg} (and extended discussion in \cref{app:lowz-consistency})
indicates that all BAO measurements that could be interpreted as outliers turn out to drive
constraints away from parameter space that is both consistent with SNe data and with massive
neutrinos.
Notably, the SDSS/DESI split at $z = 0.8$ described in \cref{sec:desi}, which excludes these
outliers (and differs from the combined dataset considered in Ref.~\cite{DESI:2024mwx}) measures
$\Omega_m = 0.320^{+0.016}_{-0.015}$ (see \cref{fig:matter-fraction-measurements}); the Gaussian
tension of this combination with DES 5YR drops from $2.6$ to $1.4 \sigma$ compared to DESI DR1 alone
(and from $1.6$ to $0.5 \sigma$ with Pantheon+).

The tension between SNe and BAO distances, regardless of their calibration, cannot be reconciled by
massive neutrinos, nor by any modification of physics at early times.
Understanding why direct probes of distances over a similar range of redshifts infer different
matter fractions is thus of paramount importance to arbitrate between their respective implications
for neutrino masses.
(See \cref{fig:sne-vs-bao-distances} in \cref{app:lowz-consistency} for a comparison of transverse
distance measurements from SNe and BAO surveys across redshift.)
While recent combined analyses of these SNe, BAO, and CMB datasets suggest a preference for
nonstandard dark energy~\cite{DES:2024tys, DESI:2024mwx}, it is unclear whether it is driven by
genuine physical features evinced by individual datasets or simply by the model's expanded freedom
to superficially improve the consistency of BAO and SNe distances.
Indeed, the preference for evolving dark energy is stronger the more the combined datasets are in
tension (within \LCDM{}).

Beyond dimensionless measures of geometry, the longer-standing tension between CMB- and
distance-ladder--based inference of the Hubble constant is only worsened by massive neutrinos: the
CMB's geometric degeneracy requires $h \propto (1 + f_\nu)^{-2}$
[\cref{eqn:hubble-fnu-degeneracy-cmb}], all else equal.
Among attempts to resolve the tension with new early-time physics, modified recombination exhibits a
unique interplay with massive neutrinos~\cite{Baryakhtar:2024rky}, which we discuss in
\cref{sec:modified-cosmology}.
The (milder) tension in the amplitude of structure, however, could more plausibly relate to massive
neutrinos.
References~\cite{Amon:2022azi, Preston:2023uup} suggest that this tension may arise from mismodeling
of nonlinear structure (to which weak galaxy lensing surveys are sensitive), i.e., if dynamical
effects suppress structure to a greater degree than assumed.
Sufficiently heavy neutrinos could provide a natural realization of this hypothesis.

The amplitude of structure inferred by the CMB, whether through lensing reconstruction or its impact
on the power spectra, is inextricably tied to the optical depth to reionization; indeed the optical
depth has long been recognized as a main source of uncertainty in neutrino mass constraints based
purely on the growth of structure~\cite{Kaplinghat:2003bh, CMB-S4:2016ple}.
\Cref{fig:mnu3-lens-spaghetti-geometry-vs-structure} suggests that CMB lensing data prefer a greater
amplitude of structure than predicted by \LCDM{} on length scales larger than the neutrino
free-streaming scale and smaller than the nonlinear scale,\footnote{
    As mentioned previously, the lensing amplitude as inferred from the two-point statistics of the
    temperature and polarization spectra~\cite{Planck:2018lbu, Planck:2018vyg} is even larger than
    that inferred from lensing spectra reconstructed via four-point statistics.
    The analysis of \cref{fig:summnu-violin-vary-lensing-asmear-desi-des-none} in
    \cref{app:lensing-asmear}, however, shows that current CMB lensing datasets yield independent
    neutrino mass constraints that are consistent with (if not more constraining than) that inferred
    from the smearing of the acoustic peaks alone, confirming their robustness to a possible lensing
    anomaly inferred from peak smearing~\cite{Green:2024xbb}.
}
which could be explained if \Planck{}'s measurement of $\tau_\mathrm{reio}$~\cite{Planck:2018vyg}
were biased low.
Namely, the small-scale anisotropies precisely measure the amplitude
$A_s e^{- 2 \tau_\mathrm{reio}}$, whereas late-time structure depends only on $A_s$.
Inference of the lensing amplitude parameter $A_L$ different from unity could be explained if
\Planck{}'s measurement were biased by $\Delta \tau_\mathrm{reio} = - (A_L - 1) / 2$.

\begin{figure}[t!]
\begin{centering}
    \includegraphics[width=\textwidth]{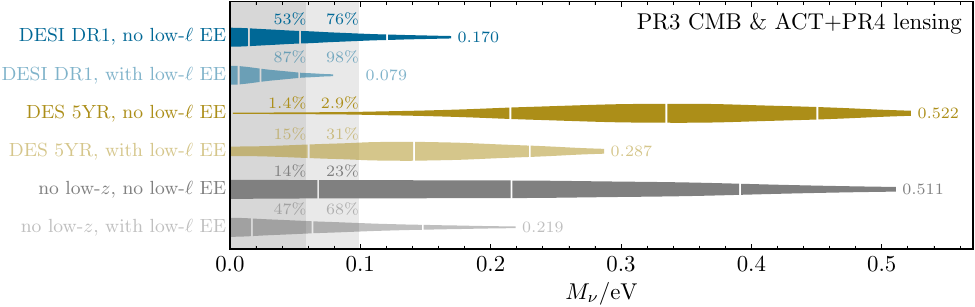}
    \caption{
        Impact of direct measurements from large-scale CMB polarization of the optical depth to
        reionization $\tau_\mathrm{reio}$ on neutrino mass inference, displaying marginal posteriors
        over the neutrino mass sum $\summnu$ for \Planck{} PR3 CMB data~\cite{Planck:2018vyg,
        Planck:2019nip} combined with various BAO and SNe datasets as labeled either including
        (transparent) or excluding (opaque) \Planck{}'s low-$\ell$ polarization data.
        The shaded bands depict the range of masses that are incompatible with the normal and
        inverted hierarchies, and the fraction of each posteriors within these ranges
        [\cref{eqn:compatibility-metric}] is annotated on the figure.
        Posteriors are truncated at the $95$th percentile (whose value is also labeled);
        vertical white lines indicate the median and $\pm 1 \sigma$ quantiles.
    }
    \label{fig:summnu-violin-act-vary-lowz-low-E}
\end{centering}
\end{figure}

\Cref{fig:summnu-violin-act-vary-lowz-low-E} displays the impact of large-scale polarization data
from \Planck{}, which directly measure $\tau_\mathrm{reio}$, on neutrino mass inference.
Excluding these data has a similar effect to marginalizing over the lensing amplitude $A_L$ and
illustrates the extent to which the lensing excess is alleviated in the absence of direct
constraints on $\tau_\mathrm{reio}$ via the resulting impact on neutrino mass inference.
Large enough $\tau_\mathrm{reio}$ does have directly observable impacts on other CMB data,
which limits the extent to which the lensing excess could be explained by a biased
inference of $\tau_\mathrm{reio}$ from large-scale polarization data; without this data,
\Planck{} (combined with ACT+PR4 lensing data) measures
$\tau_\mathrm{reio} = 0.094 \pm 0.023$, nearly twice as large as the measurement
$\tau_\mathrm{reio} = 0.0542^{+ 0.0075}_{- 0.0072}$ inferred in combination with low-$\ell$
polarization.\footnote{
    The lensing anomaly has a nonnegligible effect on the latter measurement---the full set of
    \Planck{} CMB data excluding lensing and marginalized over $A_\mathrm{smear}$ yields
    $\tau_\mathrm{reio} = 0.0486^{+0.0077}_{-0.0080}$.
}
This difference of $0.04$ is quite consistent with the preferences for $A_L \sim 1.08$ in most
recent CMB dataset~\cite{SPT-3G:2024atg}, but would amount to more than a $5 \sigma$ offset from the
measurement from \Planck{}'s low-$\ell$ polarization data.

\Cref{fig:summnu-violin-act-vary-lowz-low-E} shows that neutrino mass bounds become substantially
weaker without large-scale polarization data (by about a factor of two), no matter what low-redshift
dataset is used.
In particular, while the combination with DESI data still exhibits no preference for nonzero
$\summnu$ because of its geometric incompatibility with massive neutrinos, the combination with DES
SNe instead demonstrates a detection thereof comparable to its pure-geometry measurement
[\cref{fig:summnu-violin-geometry-asmear-vary-low-z}].
\Cref{fig:summnu-violin-act-vary-lowz-low-E} thus illustrates the significant sensitivity of
neutrino mass bounds deriving from CMB lensing to possible biases in the optical depth.

Finally, identifying the implications of individual datasets for the physical signatures of massive
neutrinos (or lack thereof) requires a judicious combination of datasets and comparative analysis
thereof, not simply combining as many as possible in the name of breaking degeneracies.
The previous era of cosmological data may have only evinced a consistent cosmological model because
of the granularity afforded by the data's precision; the choice of data is as much of an assumption
in Bayesian analyses as the choice of a model and therefore necessitates scrutiny, even if only to
recognize that parameter constraints may be artificially strong because of unexplained
incompatibilities in the data.
Determining whether joint parameter constraints derive from genuine physical predictions of a model
is only possible through comparisons in the space of the data---that is, by comparing the posterior
predictive distribution (not just a single best-fit prediction) to the data itself.\footnote{
    A possible exception is models of sufficiently low dimensionality whose predictions are fully
    encoded by an easy-to-visualize parameter space---e.g., the two-dimensional plane $(r_\mathrm{d}
    \sqrt{\omega_m}, \Omega_m)$ that fully specifies the flat-\LCDM{} predictions for the BAO scale
    (and the CMB's geometric information), as displayed in
    \cref{fig:geometry-wmrd2-Wm,fig:desi-para-perp-compare-lrg,fig:geometry-wmrd2-Wm-desi-sdss-combos,fig:desi-sdss-para-perp-compare-other-tracers}.
}
For instance, the analysis of \cref{fig:mnu3-lens-spaghetti-geometry-vs-structure} reveals that
current CMB lensing data \textit{do} exhibit trends that evoke the scale-dependent suppression
characteristic to massive neutrinos; that the posterior predictive distribution does not reproduce
these features may suggest some degree of model tension (whether or not it is a genuine signal of
massive neutrinos), as discussed in \cref{sec:joint-constraints}.

\subsection{Implications for cosmology}\label{sec:modified-cosmology}

Because the observable signatures of massive neutrinos are (individually) largely degenerate with
other cosmological parameters, inference of their masses is sensitive to modifications to the
baseline cosmological model.
Nonminimal physics relevant at late times---e.g., nonzero spatial curvature or dark energy beyond a
cosmological constant---introduces additional freedom that in principle relaxes the close
relationship between massive neutrinos and the matter fraction, both geometrically and within the
growth of structure.
Such scenarios have been studied extensively~\cite{Hannestad:2005gj, Allison:2015qca,
Poulin:2018zxs, Vagnozzi:2018jhn, RoyChoudhury:2019hls, Jiang:2024viw, Reboucas:2024smm,
Shao:2024mag}.
We leave a comparison of geometry- and growth-of-structure--based constraints on neutrinos (and the
individual role of current datasets) in modified late-time cosmologies to future work.

The implications of modified early-time physics, on the other hand, are qualitatively different
because such scenarios can alter the CMB's calibration of both the sound horizon and the density of
early-time matter (that in baryons and CDM)~\cite{Baryakhtar:2024rky}.
To study the cosmology dependence of neutrino mass inference more generally,
rewrite \cref{eqn:summnu-from-matter-density-difference} in terms of
$\omega_b + \omega_c = \omega_r / a_\star x_\mathrm{eq}$ [see \cref{eqn:x-eq}] and
$\omega_m r_{s, \star}^2$,\footnote{
    We write \cref{eqn:neutrino-mass-inference-general} in terms of the sound horizon at photon
    rather than baryon decoupling for simplicity, which makes no material difference because
    $a_\star / a_\mathrm{d}$ and $r_{s, \star} / r_\mathrm{d}$ are precisely measured and insensitive
    to modified physics~\cite[see also \cref{app:bao-from-cmb}]{Lin:2021sfs}.
}
and express
$r_{s, \star} \equiv a_\star / (\sqrt{\omega_r} H_{100} / c) \cdot F_{r_s}(R_\star, x_\mathrm{eq})$
[see \cref{eqn:rs-in-matter-radiation-ito-wr-R_star}]:
\begin{align}
    \summnu
    &= \frac{3 H_{100}^2 \Mpl^2}{n_\nu}
        \frac{\omega_r}{a_\star x_\mathrm{eq}}
        \left(
            \frac{\omega_m r_{s, \star}^2}{a_\star}
            \frac{(H_{100} / c)^2}{F_{r_s}(R_\star, x_\mathrm{eq})^2 / x_\mathrm{eq}}
            - 1
        \right)
    .
    \label{eqn:neutrino-mass-inference-general}
\end{align}
The neutrino mass sum is thus inferred via three factors encoding the number density of neutrinos
$n_\nu$ contributing to their net density $\omega_\nu$, the overall density of matter components
present during recombination, and the relative increase in the matter abundance afterward.
The neutrino number density is largely fixed within the SM and is also probed via their contribution
to the relativistic energy density at early times; we comment on physics that modify this prediction
in \cref{sec:modifications-to-neutrinos}.
The density in radiation (or equivalently $N_\mathrm{eff}$) at early times could also be increased
by modifying the physics of neutrino decoupling, but such effects would presumably also impact their
number density; a simpler possibility is that new, light degrees of freedom increase the radiation
density~\cite{Dvorkin:2022jyg}.
These scenarios are strongly constrained by CMB anisotropies on small scales via their effect on
diffusion damping~\cite{Planck:2018vyg}.
Early recombination---mechanized for instance by supposing the electron mass varies in
time~\cite{Hart:2019dxi, Sekiguchi:2020teg}---also increases this factor, since fixing
$x_\mathrm{eq}$ requires increasing density of baryons and CDM (so that matter-radiation equality
occurs at the same moment relative to recombination).

The final factor in \cref{eqn:neutrino-mass-inference-general} shows that, at fixed
$\omega_m r_{s, \star}^2$, early recombination incurs an additional increase in $\summnu$ not
present when varying $N_\mathrm{eff}$.
What distinguishes early-recombination and extra-radiation scenarios in this term is that, at fixed
$\theta_s$, varying the radiation density itself does not traverse the CMB's geometric degeneracy,
\cref{eqn:cmb-geometric-degeneracy-general,eqn:matter-fraction-fnu-degeneracy-cmb}.\footnote{
    The partitioning of the energy density in this case is unchanged because fixing $x_\mathrm{eq}$
    requires $\omega_b + \omega_c \propto \omega_r$, i.e., leaves
    $\left( \omega_b + \omega_c \right) r_\mathrm{d}^2$ invariant; the same matter fraction is then
    required to fix $\theta_s$
    [\cref{eqn:theta-s-degeneracy}].
    The geometric degeneracy between $\Omega_m$ and $1 + f_\nu$, however, holds regardless of the
    overall density of the Universe.
}
As shown in Ref.~\cite{Baryakhtar:2024rky} (and reviewed in \cref{sec:geometry}), varying the
electron mass and the neutrino mass sum (or, more generally, the ratio of the late- and early-time
matter abundances) and other \LCDM{} parameters yields a novel geometric degeneracy.\footnote{
    Varying the present-day CMB temperature $T_0$~\cite{Ivanov:2020mfr} (i.e., ignoring precise
    measurements thereof~\cite{Fixsen:1996nj, Fixsen:2009ug}) also mechanizes early recombination
    (or rather, ``late present''), but with different consequences for the present-day matter
    density and neutrino number density.
    Both $n_\nu$ and $\omega_b + \omega_c$ drop with $T_0^3$ simply due to redshifting;
    larger $\summnu$ are thus enabled only via the third factor in
    \cref{eqn:neutrino-mass-inference-general}, which varies even while compensatory changes to
    $a_\star$ and $1 + f_\nu$ keep the matter fraction fixed.
}
Namely, proportional increases of the Hubble constant $h$, the neutrino fraction $1 + f_\nu$,
$1/a_\star$, $\omega_b$, and $\omega_c$ leave BAO distances and the primary CMB anisotropies
invariant---that is, the matter fraction, the product $h r_\mathrm{d}$ or
$r_\mathrm{d} \sqrt{\omega_m}$, $\theta_s$, $R_\star$, and $x_\mathrm{eq}$ all remain unchanged.
This degeneracy is notable in contrast to \LCDM{}, in which neutrino masses and $h$ are
anticorrelated at fixed $\theta_s$.

The potential of early recombination with massive neutrinos to solve the Hubble tension, however, is
ultimately limited by the impact of neutrino free-streaming on the growth of structure allowed by
\Planck{} (i.e., via secondary effects at late times)~\cite{Baryakhtar:2024rky}.
Following the discussion in \cref{sec:cmb-lensing}, the angular scale of the horizon at
matter-radiation equality [\cref{eqn:ell-eq}] is unchanged along the degeneracy.
The neutrino free-streaming scale [\cref{eqn:neutrino-free-streaming-scale}], on the other hand,
scales with $\summnu \sqrt{\omega_m} \propto \summnu / r_\mathrm{d}$, meaning
$\ell_\mathrm{fs} \propto \summnu \chi_\star / r_\mathrm{d}
\propto \summnu / \theta_{\mathrm{CMB}, \perp}(a_\mathrm{d})$ scales simply with $\summnu$.
The neutrino mass sum increases superlinearly with $1/a_\star$
[\cref{eqn:neutrino-mass-inference-general}], however, and the redshift at which they become
nonrelativistic is unchanged, so their suppression of structure increases relatively rapidly
along the degeneracy.

\begin{figure}[t!]
\begin{centering}
    \includegraphics[width=\textwidth]{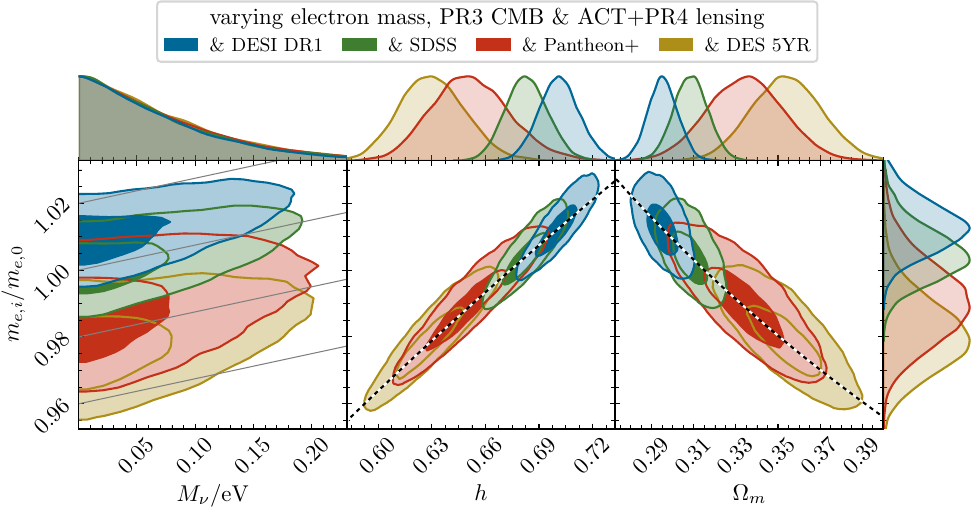}
    \caption{
        Constraints on early-recombination cosmologies via a varying electron mass including massive
        neutrinos, displaying joint posteriors over the early-time electron mass $m_{e, i}$ with the
        neutrino mass sum $\summnu$, the Hubble constant $h$, and the matter fraction $\Omega_m$
        (presented as in \cref{fig:geometry-pr3-asmear-desi-des}).
        All results use \Planck{} PR3 CMB~\cite{Planck:2018vyg, Planck:2019nip} and ACT + \Planck{}
        PR4 lensing data~\cite{ACT:2023dou, ACT:2023dou, Carron:2022eyg}, combined with either BAO
        data from DESI DR1 (blue) or SDSS~\cite{eBOSS:2020yzd, eBOSS:2020lta, eBOSS:2020hur,
        Ross:2014qpa} (green) or uncalibrated SNe data from Pantheon+~\cite{Brout:2022vxf,
        Scolnic:2021amr} (red) or DES~\cite{DES:2024tys} (gold).
        Dashed black/white lines indicate the parameter directions $h \propto m_{e, i}^{3.16}$ and
        $\Omega_m \propto m_{e, i}^{-5.32}$ expected from the CMB's geometric degeneracy at fixed
        neutrino fraction
        [\cref{eqn:hubble-fnu-degeneracy-cmb,eqn:matter-fraction-fnu-degeneracy-cmb}].
        The grey lines in the leftmost panel lie along $m_{e, i} \propto 1 + f_\nu$, the degeneracy
        between early recombination and late dark matter presented by Ref.~\cite{Baryakhtar:2024rky}
        that would fix $\Omega_m$ [see \cref{eqn:matter-fraction-fnu-degeneracy-cmb}].
        For these results (and any of the other low-redshift datasets presented in
        \cref{fig:summnu-violin-act-pr4-vary-lowz}), the marginal posteriors over $\summnu$ all have
        95th percentiles at $\approx 0.2~\mathrm{eV}$, and about $50\%$ and $75\%$ of the posteriors
        are incompatible with the normal and inverted hierarchies, respectively.
    }
    \label{fig:vary-me-vs-summnu-h-Wm-vary-lowz}
\end{centering}
\end{figure}

Reference~\cite{Baryakhtar:2024rky} studied this scenario in detail in application to a hyperlight
scalar field (in this case postulated to mechanize the time variation of fundamental constants) and
showed that the results are qualitatively similar with massive neutrinos instead.
The resulting posteriors, displayed in \cref{fig:vary-me-vs-summnu-h-Wm-vary-lowz} for the latter
case, therefore predominantly exhibit the degeneracy $h \propto a_\star^{-3.16}$ evident in
\cref{eqn:hubble-fnu-degeneracy-cmb} at fixed neutrino fraction.
Moreover, the matter fraction is as sensitive to the scale factor of recombination as to neutrino
masses [\cref{eqn:matter-fraction-fnu-degeneracy-cmb}], making parameter inference thereof just as
sensitive to the geometric tensions in low-redshift distance datasets~\cite{Baryakhtar:2024rky}.
Despite the discordance in preferred low-redshift expansion histories, bounds on the neutrino mass
sum are robust across dataset combinations.
The geometric degeneracy between $a_\star$ and the neutrino fraction is still at play, as evident in
\cref{fig:vary-me-vs-summnu-h-Wm-vary-lowz}---just not to an extent that qualitatively alters the
marginal posteriors over other parameters.
The scenario effectively marginalizes out geometric sensitivity to neutrino masses by adjusting
$a_\star$ as needed to match a given dataset's preferred $\Omega_m$; in analogy to the
$A_\mathrm{smear}$-marginalized results of
\cref{fig:geometry-pr3-asmear-desi-des,fig:summnu-violin-geometry-asmear-vary-low-z}, the posteriors
in \cref{fig:vary-me-vs-summnu-h-Wm-vary-lowz} over $\summnu$ effectively derive only from the
impact of neutrinos on the growth of structure (as constrained by \Planck{} CMB data and the ACT+PR4
lensing dataset).\footnote{
    Marginalizing over lensing amplitude parameters ($A_L$ or $A_\mathrm{smear}$) as well therefore
    completely relaxes constraints on neutrino masses, up to the point that they become
    nonrelativistic during recombination (\cref{sec:peak-smearing-isw}).
}

While the treatment of Ref.~\cite{Baryakhtar:2024rky} (and \cref{sec:geometry}) applies to a number
of extended cosmologies (namely, those whose additional free parameters are nominally fixed by SM
predictions, like the fundamental constants and $N_\mathrm{eff}$), it does not precisely capture
cases that introduce new degrees of freedom featuring qualitatively distinct dynamics. One such
example is early dark energy~\cite{Poulin:2018cxd, Hill:2020osr}.
The parameter relations found in models with extra radiation, however, may extend to this case to
the extent that its effects on the background cosmology reduce to an increase in the density at
recombination.
Indeed, Ref.~\cite{Reeves:2022aoi} found that varying the neutrino masses had little impact on the
preferred parameter space for axionlike early dark energy (and vice versa).
Early recombination thus appears unique in its consequences for massive neutrinos (a finding
observed but not explained in Ref.~\cite{Khalife:2023qbu}).

\subsection{Implications for neutrino physics}
\label{sec:modifications-to-neutrinos}

Establishing whether cosmological data truly call for new physics ultimately requires testing
concrete models---not just to ensure analyses capture the full set of predicted physical effects,
but also to place results in the context of theoretical priors and external constraints.
In this paper we treat neutrinos consistently within the predictions of the Standard Model, in which
they both contribute to the matter density and also resist gravitational clustering.
A number of recent articles, on the other hand, explored the putative preference for ``negative''
neutrino mass deriving from DESI data, taking alternative approaches that seek to emulate different
new-physics scenarios.
For example, Refs.~\cite{Craig:2024tky, Green:2024xbb} did so solely via the $A_L$ parameter (i.e.,
by attributing a rescaling of the lensing spectrum to neutrino masses, positive or negative); this
analysis, however, neglects the contribution of neutrinos to the matter density in the cosmological
background, which our results show is critical to both geometry and structure.
Namely, neglecting the neutrino's background effect overestimates their net impact on structure
because they also require substantially larger matter fractions (at fixed $\theta_s$).
The identified preference for $A_L > 1$ bears no impact on DESI's geometric incompatibility with
massive neutrinos (as illustrated by \cref{fig:geometry-wmrd2-Wm,fig:geometry-pr3-asmear-desi-des}),
though it may mimic the outcome of a scenario in which new physics solely affects the growth of
structure.
Of course, the preference for excess lensing in CMB data has long been
appreciated~\cite{Planck:2018vyg, Planck:2018lbu, Planck:2013pxb}; the only substantive difference
made by recent data is DESI's nominally stronger preference for even lower matter fractions than
\Planck{} would itself prefer.

Reference~\cite{Craig:2024tky} also surveys a number of new-physics scenarios that might modify the
dynamics of neutrinos and their inferred masses as a result.
In one category, the neutrino abundance is depleted via decays or annihilations into radiation after
recombination but before neutrinos would become nonrelativistic; while the possibility would remove
the prolonged period over which standard neutrinos suppress structure growth, it would also
eliminate their impact on the cosmological background.
The late-time matter density would then be just that from baryons and CDM; to the extent that the
expansion history reproduces that in \LCDM{} with massless neutrinos, the matter fraction inferred
from the CMB's geometric information would then agree with that measured directly via BAO or SNe
distances.
Models with time-varying neutrino masses could yield similar phenomenology, although the scalar
field responsible for the variations would have its own gravitational
effects~\cite{Baryakhtar:2024rky}.

Another category attempts to modify the scale on which neutrinos freely stream, e.g., by reducing
their phase-space--averaged momentum after recombination [to reduce $c_{\nu_i}$ in
\cref{eqn:neutrino-free-streaming-scale}] without affecting their number density.
In this case, geometry would play a larger role in neutrino mass inference as discussed in
\cref{sec:geometry,sec:geometric-tensions}; \cref{fig:summnu-violin-geometry-asmear-vary-low-z}
suggests that such neutrinos could still have masses compatible with either hierarchy or that mass
sums $\summnu \sim 0.3$ to $0.5~\mathrm{eV}$ would be preferred, depending on the low-redshift
distance dataset.
Mechanizing a smaller free-streaming scale, moreover could also enable massive neutrinos to better
fit features in current CMB lensing data that \LCDM{} cannot explain (see
\cref{fig:mnu3-lens-spaghetti-geometry-vs-structure} and surrounding discussion).
Finally, Ref.~\cite{Craig:2024tky} considered modified physics outside of the neutrino sector that
could explain the lensing excess, which would also leave neutrino masses to be probed via geometry.
One example---new long-range forces between dark matter particles---would again entail additional
dynamics from the mediator; Refs.~\cite{Archidiacono:2022iuu, Bottaro:2023wkd, Bottaro:2024pcb},
however, show that this scenario does not ultimately affect the amplitude of structure due to an
effective cancellation between the clustering rate and the modified redshifting of dark matter.

Other authors extrapolated to negative masses by simply fitting a normal distribution to posteriors
inferred using the physical prior that $\summnu > 0$~\cite{eBOSS:2020yzd, Allali:2024aiv,
Naredo-Tuero:2024sgf} or extrapolated at the level of the theoretical model by subtracting the
residual of a positive-$\summnu$ cosmology relative to that with
$\summnu = 0$~\cite{Elbers:2024sha}.
We emphasize that a posterior peaked at the boundary of a parameter's prior is entirely expected if
a given dataset is unable to resolve the impact of its deviation from that boundary (i.e., if the
true value of the parameter is close to the boundary compared to the spread of the posterior).
For example, posterior distributions for the tensor-to-scalar ratio $r$ from current data take the
same form for the same reason: both $\summnu$ and $r$ are nonnegative \textit{a priori} on grounds
of theoretical consistency (the latter because an autopower spectrum cannot be negative).
The more curious feature of neutrino mass constraints that include BAO data is not that their
marginal posteriors appear as if they would peak at unphysical, negative values, but rather that
they peak below the minimum mass sum for either hierarchy and with substantially higher density
than at ${\summnu}_\mathrm{min}$.
(\Cref{fig:summnu-violin-act-pr4-vary-lowz} demonstrates that this feature is not universal across
dataset combinations, however, since it derives from offsets in $\Omega_m$ measurements.)
Assessing whether such a distribution indicates a systematic inconsistency or tension within the
data or calls for a different physical model requires both testing viable candidate models and
carefully examining what in the data drives such preferences.

\section{Conclusions}\label{sec:conclusions}

Cosmological observations definitively confirm the role of relativistic neutrinos predicted by the
Standard Model.
The current interpretation of cosmological evidence for or against their masses, however, is
nebulous, if not entirely inconsistent.
The signatures of massive neutrinos in both the Universe's geometry and its large-scale structure
are partially degenerate with the composition of the Universe at late times, positioning direct
probes of the late-time expansion history as crucial in pinpointing the neutrino mass scale.
By carefully considering what relevant quantities are actually constrained by observations---in
particular, the angular extent of the photon-baryon sound horizon $\theta_s$, and therefore the
comoving distance to last scattering in standard cosmology---we showed that heavier neutrinos imply
an increase in the present-day \textit{fraction} of energy in all matter ($\Omega_m$) that is five
times the neutrino's relative contribution to the matter \textit{density}.
In other words, the well-known geometric degeneracy between the neutrino mass sum and the Hubble
constant entails a surprisingly large modification to the composition of the Universe.

Dimensionless quantifiers such as $\Omega_m$ are not just those that directly control the late-time
growth of structure (independent of neutrinos' resistance to clustering) but are also what
low-redshift distance datasets directly measure, even without calibration.
Geometric probes, moreover, are immune to the physical effects that challenge the detection of
neutrinos' suppression of large-scale structure growth (e.g., the optical depth
$\tau_\mathrm{reio}$, which is difficult to measure independently with sufficient
precision)~\cite{CMB-S4:2016ple}.
\Cref{sec:signatures} shows that indeed, contrary to common thought, measurements of the matter
fraction via cosmological distances are more sensitive to neutrino masses than equally precise
measurements of the amplitude of structure below the free-streaming scale via CMB lensing.

The same opportunity afforded by geometric probes to inform neutrino mass measurements also presents
a vulnerability to unknown systematics and inconsistencies between datasets.
A growing number of cosmological observations yield inconsistent parameter inference within the
standard cosmological model; we showed that the strikingly discrepant preferences in late-time
geometries (i.e., matter fractions $\Omega_m$) between the most recent BAO and (uncalibrated) SNe
datasets beget a starkly divergent compatibility with massive neutrinos, both on purely geometric
grounds (\cref{fig:summnu-violin-geometry-asmear-vary-low-z}) and in combination with observations
of structure (\cref{fig:summnu-violin-act-pr4-vary-lowz}).
Resolving this geometric tension, even independently of the tension in calibration (i.e., $H_0$), is
thus paramount to place robust constraints on all of massive neutrinos' cosmological effects.
Ultimately, a genuine measurement of neutrino masses from CMB data and low-redshift distances
derives from concordant inference of their independent effects on geometry and structure in the
$\Omega_m$-$\summnu$ plane---only their combination can break their respective degeneracies
$\Omega_m \propto (1 + f_\nu)^5$ (\cref{fig:geometry-pr3-asmear-desi-des}, deriving from $\theta_s$)
and $\Omega_m \propto (1 + f_\nu)^8$ (\cref{fig:structure-pr3-no-theta-s-act-pr4-lens-desi-des},
independent of $\theta_s$).

Because posteriors over the neutrino mass sum from numerous contemporary datasets peak near the
lower boundary of the prior ($\summnu = 0$), upper limits on $\summnu$---measures of the tail of its
one-sided distribution---can be exponentially sensitive even to small shifts in central values from
data, whatever their origin.
In \cref{sec:desi}, we identified two individual data points among DESI's measurements that strongly
influence its apparent preference for ``negative neutrino mass'' and its discrepancy with SNe
datasets.
The thorough analyses of systematics and robustness in Refs.~\cite{DESI:2024uvr, DESI:2024lzq,
DESI:2024ude} make it challenging to interpret the impact of these measurements and their
inconsistency with those made previously by SDSS.
Nevertheless, when a small subset of the data drive strong conclusions about models, a Bayesian must
weigh the possibility of statistical effects (as discussed in \cref{sec:desi}) and unknown
systematics (which are more challenging to na\"ively quantify) against their theoretical prior.
Neutrinos are indisputably massive today, though their mass mechanism is yet unknown; the
theoretical plausibility and cosmological viability of nonminimal neutrino physics can only be
fairly assessed in the context of concrete microphysical models with a consistent implementation of
their predicted phenomenology.
Of course, too strong a prior defeats the purpose of taking data in the first place, but open
mindedness to hints of new physics must be tempered by careful scrutiny of the data that might drive
them.
In \cref{sec:modified-cosmology} we studied the impact of modified early-Universe physics on
neutrino mass inference, of which early recombination provides a particularly intriguing example
(see also Ref.~\cite{Baryakhtar:2024rky}).
We leave dedicated analyses of possible nonminimal neutrino physics, like those commented upon in
\cref{sec:modifications-to-neutrinos}, to future work.

Our findings likely generalize to other scenarios featuring light relics~\cite{Dvorkin:2022jyg},
whether hyperlight scalars~\cite{Hlozek:2014lca, Hlozek:2016lzm, Hlozek:2017zzf, Lague:2021frh,
Rogers:2023ezo, Baryakhtar:2024rky} or thermal relics like hot axions~\cite{Giare:2020vzo,
Ferreira:2020bpb, DEramo:2022nvb, Notari:2022ffe, Bianchini:2023ubu}, since such scenarios can also
increase the matter abundance after recombination and entail qualitatively similar effects on
structure growth.
Trends in recent data also motivate revisiting the growth versus geometry analyses of
Refs.~\cite{Wang:2007fsa, Ruiz:2014hma, Bernal:2015zom, Lin:2017ikq, DES:2020iqt, Andrade:2021njl,
Zhong:2023how}.
The implications of modified late-time cosmologies (such as evolving dark energy) for neutrino mass
inference may well also be sensitive to the choice of low-redshift dataset (if studied individually
rather than combined).

Despite the contemporary ambiguity in their implications, the advantage of geometric probes is that
they more or less directly measure the late-time expansion history.
Future data with increased sample size and redshift coverage, like BAO data from
Euclid~\cite{Euclid:2024yrr, Euclid:2024imf}, Roman~\cite{Eifler:2020vvg}, and subsequent DESI
releases~\cite{DESI:2016fyo} and SNe data from the Zwicky Transient Facility~\cite{Rigault:2024kzb}
and the Vera Rubin Observatory~\cite{LSSTDarkEnergyScience:2018jkl}, can definitively test the
significance of the trends in current data.
These datasets themselves provide the means to constrain late-time expansion histories beyond
\LCDM{} that can confound neutrino mass inference from both geometry and structure.
Moreover, it is fortuitous to have multiple independent methods of measuring late-time geometry in
the first place.
An abundance of future surveys will also probe the suppression of structure characteristic to
neutrinos, including not just DESI, Euclid, Roman, and Rubin but also CMB-S4~\cite{CMB-S4:2016ple} and
Simons Observatory~\cite{SimonsObservatory:2018koc}.
Synergy---and concordance---between these distinct observables will ultimately be imperative to
conclusively measure the neutrino mass scale with cosmological data.

\begin{acknowledgments}
We thank Masha Baryakhtar, Stephen Chen, Marco Costa, Jessie Muir, Caio Nascimento, Murali
Saravanan, and Olivier Simon for helpful discussions.
M.L.\ and Z.J.W.\ are supported by the Department of Physics and College of Arts and Science at the
University of Washington and the Dr. Ann Nelson Endowed Professorship.
M.L. is also supported by the Department of Energy grants DE-SC0023183 and DE-SC0011637.
M.L. is grateful for the hospitality of Perimeter Institute where part of this work was carried out.
Research at Perimeter Institute is supported in part by the Government of Canada through the
Department of Innovation, Science and Economic Development and by the Province of Ontario through
the Ministry of Colleges and Universities.
This research was also supported in part by the Simons Foundation through the Simons Foundation Emmy
Noether Fellows Program at Perimeter Institute.
This work made use of the software packages
\textsf{emcee}~\cite{Foreman-Mackey:2012any,Hogg:2017akh,Foreman-Mackey:2019},
\textsf{corner.py}~\cite{corner}, \textsf{NumPy}~\cite{Harris:2020xlr},
\textsf{SciPy}~\cite{Virtanen:2019joe}, \textsf{matplotlib}~\cite{Hunter:2007ouj},
\textsf{xarray}~\cite{hoyer2017xarray}, \textsf{ArviZ}~\cite{arviz_2019},
\textsf{SymPy}~\cite{Meurer:2017yhf}, and \textsf{CMasher}~\cite{cmasher}.
\end{acknowledgments}

\appendix

\section{Transverse BAO scale from CMB analyses}\label{app:bao-from-cmb}

In this appendix we describe the construction of likelihoods for the CMB-derived transverse BAO
scale, $\theta_{\mathrm{CMB}, \perp}(a_\mathrm{d})$, used in \cref{sec:geometric-tensions}.
The angular scale $\theta_s$ measured by the CMB \textit{is} the transverse BAO scale, except
that it measures the sound horizon when photons decoupled rather than when baryons decoupled (i.e.,
because it quantifies the angular correlation of the photon distribution).
The two epochs are separated by a rather narrow interval of expansion,
$a_\star / a_\mathrm{d} \approx 1.0268 \pm 6 \times 10^{-4}$, that is determined by the
baryon-to-photon ratio $R_\star$ (i.e., the ratio of their number densities), which itself is well
constrained by the shape of the acoustic peaks (\cref{sec:mass-independent}).
As such, $\theta_s$ and $\theta_{\mathrm{CMB}, \perp}(a_\mathrm{d})$ are closely
related~\cite{Lin:2021sfs}---namely, $\theta_{\mathrm{CMB}, \perp}(a_\mathrm{d})$ should be a
derived quantity as model independent as $\theta_s$ itself.
Phrasing the CMB's geometric information in terms of $\theta_{\mathrm{CMB}, \perp}(a_\mathrm{d})$
enables a direct comparison to BAO measurements without assumptions on the magnitude of the
sound/drag horizon, i.e., in the common parameter space $(r_\mathrm{d} \sqrt{\omega_m}, \Omega_m)$.

Reference~\cite{Lin:2021sfs} inferred geometric constraints on $h r_\mathrm{d}$ and $\Omega_m$ from CMB
experiments by marginalizing over the uncertainty in $a_\star / a_\mathrm{d}$ and
$r_{s, \mathrm{d}} - r_{s, \star}$ to translate a measurement of $\theta_s$ to one of
$\theta_{\mathrm{CMB}, \perp}(a_\mathrm{d})$.
We take an approach that is more direct and precise by simply calculating
$\theta_{\mathrm{CMB}, \perp}(a_\mathrm{d})$ itself from posterior samples.
The posterior over $\theta_\mathrm{d}$ deriving from \Planck{} PR3 data (excluding lensing and
marginalizing over $\summnu$ and $A_\mathrm{smear}$) is well approximated by
\begin{align}
    100 \theta_{\mathrm{CMB}, \perp}(a_\mathrm{d})
    &\sim \mathcal{N}(1.06061, 4.2 \times 10^{-4}),
    \label{eqn:planck-theta-perp-measurement}
\end{align}
where $\mathcal{N}(\mu, \sigma)$ denotes a normal distribution with mean $\mu$ and standard
deviation $\sigma$.
For comparison, $100 \theta_s \sim \mathcal{N}(1.04198, 3.1 \times 10^{-4})$ from the same
posterior, a measurement only marginally more precise.

To demonstrate the robustness of \cref{eqn:planck-theta-perp-measurement} to early-Universe physics,
\cref{tab:theta-bao-from-cmb} tabulates results for a number of additional models and dataset
combinations.
\begin{table*}[t!]
    \setlength{\tabcolsep}{12pt}
    \centering
    \begin{tabular}[t]{rrll}
        \toprule
        $\mu$ & $10^3 \sigma$
        & additional parameters & data
        \\
        \midrule
        $1.06081$ & $0.39$ &  & \Planck{} PR3
        \\
        $1.06077$ & $0.38$ &  & \Planck{} PR3 (with lensing)
        \\
        $1.06051$ & $0.42$ & $A_\mathrm{smear}$ & \Planck{} PR3
        \\
        $1.06083$ & $0.39$ & $\summnu$ & \Planck{} PR3
        \\
        $1.06080$ & $0.40$ & $\summnu$ & \Planck{} PR3 (with lensing)
        \\
        $1.06061$ & $0.42$ & $\summnu$, $A_\mathrm{smear}$ & \Planck{} PR3
        \\
        $1.06090$ & $0.39$ & $m_{e, i}$ & \Planck{} PR3 (with lensing)
        \\
        $1.06167$ & $1.2$ & $N_\mathrm{eff}$, $Y_\mathrm{He}$ & \Planck{} PR3 (with lensing)
        \\
        \bottomrule
    \end{tabular}
    \label{tab:theta-bao-from-cmb}
    \caption{
        Mean and standard deviation of $100 \theta_{\mathrm{CMB}, \perp}(a_\mathrm{d})$ inferred
        from \LCDM{} and a number of extensions (whose additional free parameters are enumerated),
        including \Planck{} PR3 CMB data and PR3 lensing data where indicated.
    }
\end{table*}
Notably, the measurement is identical in \LCDM{} and cosmologies that vary the electron mass; since
early recombination by varying the electron mass (or the present-day CMB
temperature~\cite{Ivanov:2020mfr}) is the only (known) model that completely opens a geometric
degeneracy in the distance to last scattering without modifying the flat \LCDM{} model at
late times, this finding corroborates the robustness of the angular drag scale even if the absolute
length scale is substantially different from the \LCDM{} prediction.
The result is also robust to varying the neutrino mass sum, $A_\mathrm{smear}$, or both.
As another test, the precision of \cref{eqn:planck-theta-perp-measurement} degrades to the same degree
as the measurement of $\theta_s$ itself does (by a factor of 3) when varying both $N_\mathrm{eff}$
and the helium yield $Y_\mathrm{He}$.
In this case, the free-streaming fraction varies, which induces a phase shift in the acoustic
peaks~\cite{Bashinsky:2003tk, Baumann:2015rya}, while the helium yield is free to compensate for the
effect of additional radiation on the damping tail.
In spite of this, the mean $\theta_{\mathrm{CMB}, \perp}(a_\mathrm{d})$ hardly moves, and permille
precision is still superior by an order of magnitude to any single BAO measurement.

The precise redshift of baryon drag (or of peak photon visibility, for that matter) is constrained
with high (but not infinite) precision by \Planck{} data (in cosmologies that do \emph{not} vary the
fundamental constants).
For the same posterior from which \cref{eqn:planck-theta-perp-measurement} is derived,
$1 + z_\mathrm{d} \approx 1061.13 \pm 0.33$ (similar precision to that for $z_\star$).
Because decoupling occurs so early, the relative uncertainty propagated to the comoving distance
$\chi(a_\mathrm{d})$ is $\lesssim 10^{-4}$---small compared to the precision of
\cref{eqn:planck-theta-perp-measurement}.
Decoupling occurs so early, however, that the impact of radiation on expansion leads to
$\sim 0.5\%$ corrections to $F_M(a_\star; \amL)$ evaluated with
\cref{eqn:comoving-distance-matter-Lambda}.
The comoving distance in a Universe with radiation as well as matter and cosmological constant does,
however, have an analytic form in terms of elliptic integrals~\cite{Karchev:2022mcv} which are often
available as special functions in numerical software.
These functions provide an efficient means to calculate
$\theta_{\mathrm{CMB}, \perp}(a_\mathrm{d}) / r_\mathrm{d} \sqrt{\omega_m}$, and simply evaluating
the comoving distance at fixed $a_\mathrm{d} = 1/1061.1$ provides sufficient precision for
measurements as precise as \cref{eqn:planck-theta-perp-measurement}.

\section{Supplementary results}\label{app:supplemental-results}

\subsection{Constraints on structure from lensing reconstruction and peak smearing}\label{app:lensing-asmear}

\begin{figure}[t!]
\begin{centering}
    \includegraphics[width=\textwidth]{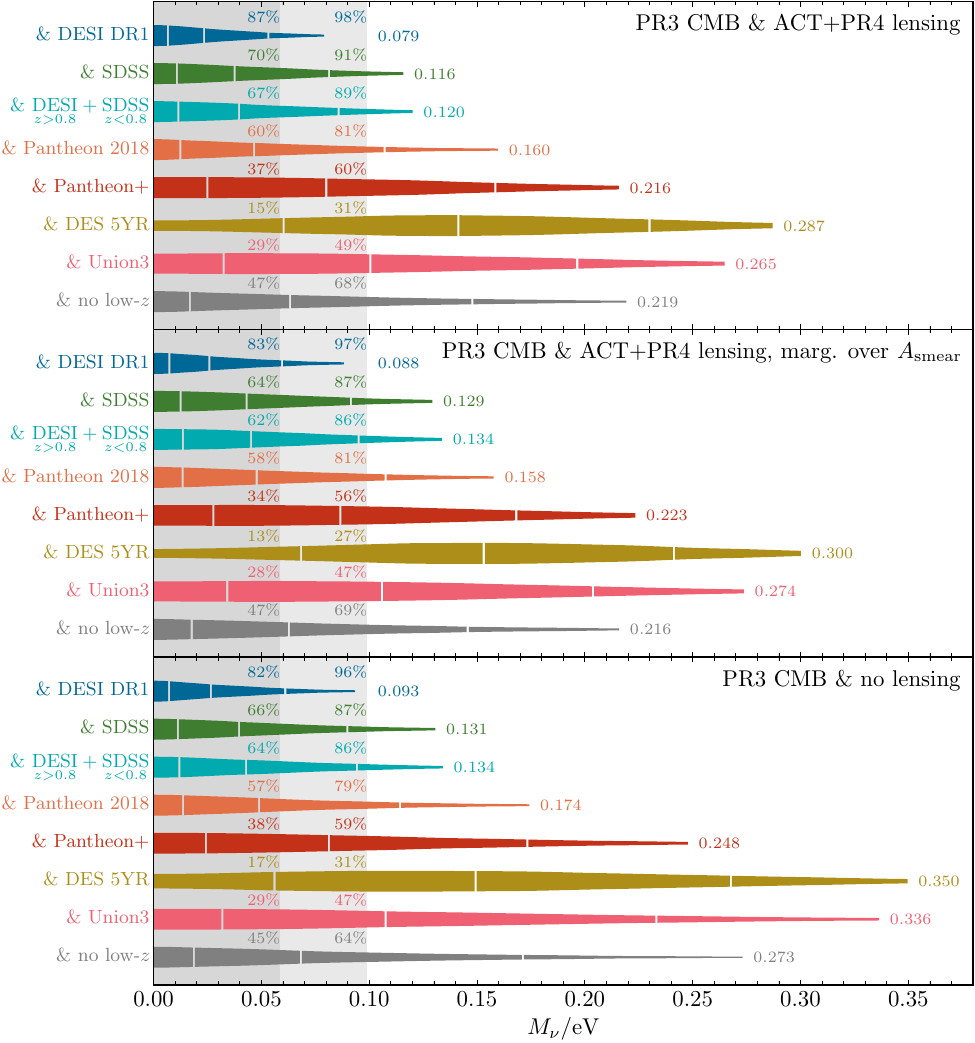}
    \caption{
        Marginal posterior distribution over the neutrino mass sum $\summnu$ for \Planck{} PR3 CMB
        combined with various BAO and uncalibrated SNe datasets (by color) and no lensing data
        (bottom panel) and ACT+PR4 lensing data excluding (top) and including (middle) the parameter
        $A_\mathrm{smear}$.
        The shaded bands depict the range of masses that are incompatible with the normal and
        inverted hierarchies, and the fraction of each posterior within these ranges
        [\cref{eqn:compatibility-metric}] is annotated on the figure.
        Posteriors are truncated at the $95$th percentile (whose value is also labeled);
        vertical white lines indicate the median and $\pm 1 \sigma$ quantiles.
    }
    \label{fig:summnu-violin-vary-lens-asmear-low-z}
\end{centering}
\end{figure}

\Cref{fig:summnu-violin-vary-lens-asmear-low-z} displays posterior distributions over the neutrino
mass sum $\summnu$ analogous to
\cref{fig:summnu-violin-geometry-asmear-vary-low-z,fig:summnu-violin-act-pr4-vary-lowz} for
additional analysis variants.
In particular, \cref{fig:summnu-violin-vary-lens-asmear-low-z} includes results combining various
low-redshift distance datasets with \Planck{} CMB data alone, combined with ACT+PR4 lensing data
(the same results as \cref{fig:summnu-violin-act-pr4-vary-lowz}), and combined with the same lensing
data but marginalized over $A_\mathrm{smear}$.
These results thus compare the amount of information added by the ACT+PR4 reconstructed lensing map
both with and without the influence of \Planck{}'s own inference of lensing via its temperature and
polarization power spectra.

\begin{figure}[t!]
\begin{centering}
    \includegraphics[width=\textwidth]{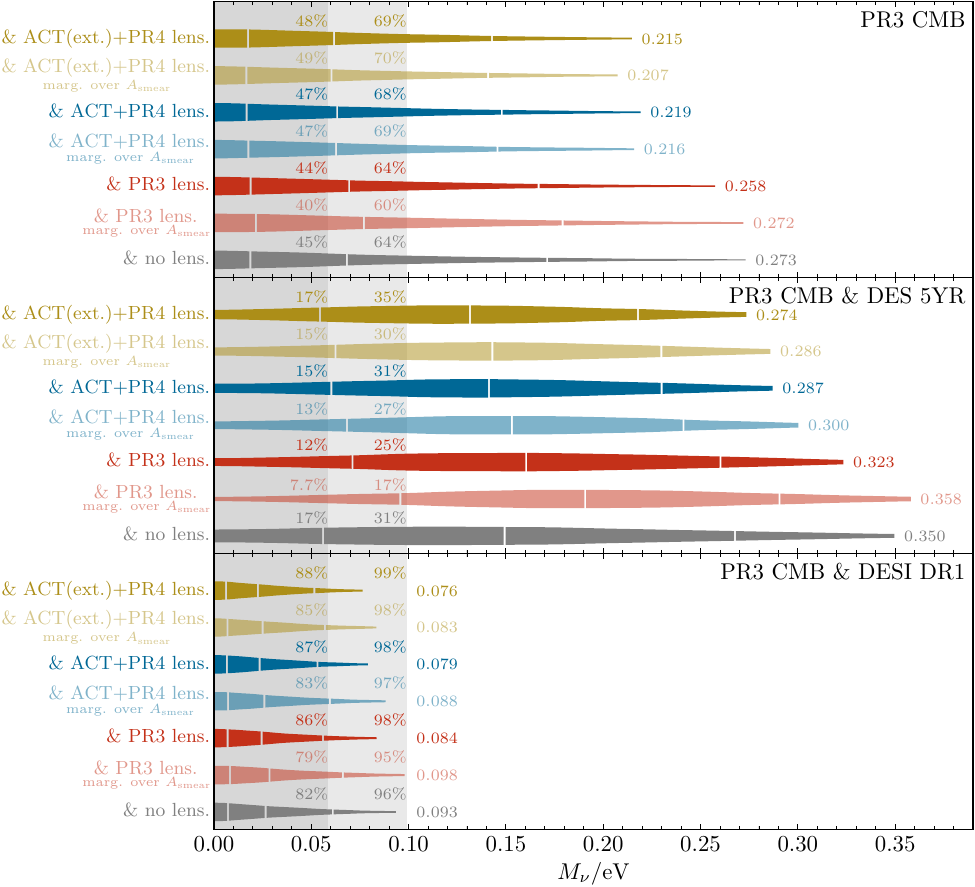}
    \caption{
        Marginal posterior distribution over the neutrino mass sum $\summnu$ for \Planck{} PR3 CMB
        data combined with various lensing datasets (by color), each marginalized over
        $A_\mathrm{smear}$ (transparent, lighter colors) or not (opaque, darker colors).
        Results are presented as in \cref{fig:summnu-violin-vary-lens-asmear-low-z}.
        Here ACT(ext.)+PR4 denotes the ACT+\Planck{} lensing combination that includes ACT data
        up to a larger maximum multipole (1250).
        The middle and bottom panels respectively include DES 5YR SNe data and DESI DR1 BAO data.
    }
    \label{fig:summnu-violin-vary-lensing-asmear-desi-des-none}
\end{centering}
\end{figure}

\Cref{fig:summnu-violin-vary-lensing-asmear-desi-des-none} again displays posteriors over $\summnu$
deriving from \Planck{} CMB data, various lensing datasets, and either no low-redshift distances,
DESI's BAO data, or DES's SNe data.
Each panel compares results marginalized or not over $A_\mathrm{smear}$ as a test of the sensitivity
of $\summnu$ bounds to the \Planck{} lensing anomaly.
\Cref{fig:summnu-violin-vary-lensing-asmear-desi-des-none} includes results using not just the
ACT+PR4 lensing dataset but also an extended variant thereof (including multipoles up to 1250) as
well as the \Planck{} lensing data from 2018 (i.e., PR3).
Comparing pairs that do and do not marginalize over $A_\mathrm{smear}$ shows that the information
content in PR3 temperature and polarization power spectra only marginally tightens neutrino mass
bounds from PR3 lensing data and has an even more negligible effect on inferences using ACT DR6 +
\Planck{} PR4 lensing; neutrino mass constraints from LSS are thus largely robust to any potential
systematic lensing anomaly in \Planck{}'s primary CMB data.
In addition, comparing to results with no lensing dataset (grey) shows that the peak-smearing effect
by itself yields similar constraints to the PR3 lensing reconstruction.
\Cref{fig:summnu-violin-vary-lensing-asmear-desi-des-none} also shows that this conclusion holds
regardless of what (if any) low-redshift data are included.

\subsection{Consistency of low-redshift distance measurements}\label{app:lowz-consistency}

\begin{figure}[t!]
\begin{centering}
    \includegraphics[width=\textwidth]{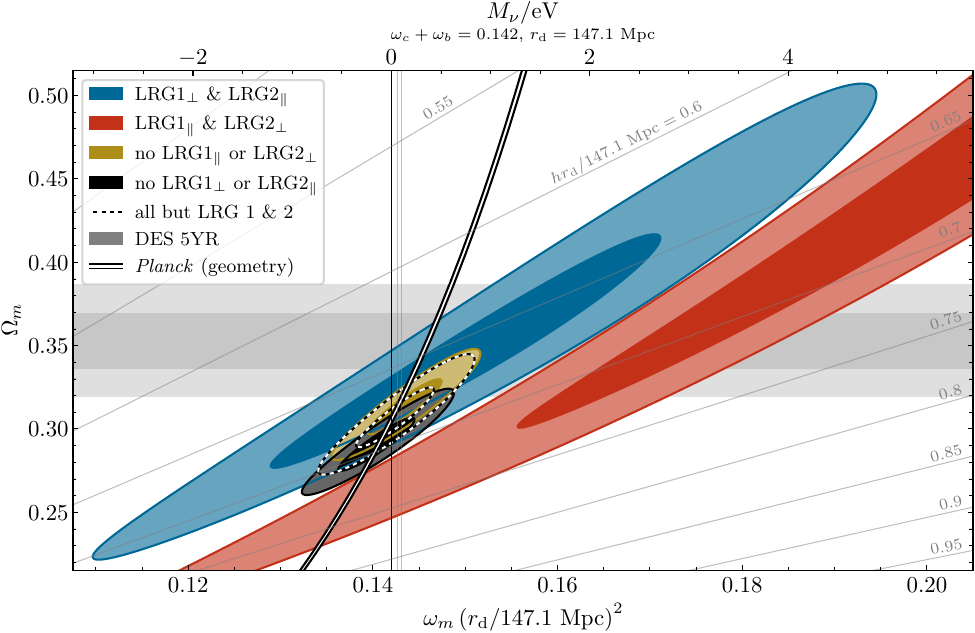}
    \caption{
        Geometric constraints on late-time \LCDM{} cosmology from various subsets of DESI data,
        presented as in \cref{fig:geometry-wmrd2-Wm} but instead displaying only $1$ and $2 \sigma$
        mass levels (and just \Planck{}'s $2 \sigma$ interval).
        Posteriors from the line-of-sight measurement in the $z = 0.51$ LRG bin
        ($\text{LRG1}_\parallel$) combined with transverse one in the $z = 0.706$ LRG bin
        ($\text{LRG2}_\perp$) appears in red and the combination of all DESI data except these two
        points in gold.
        Constraints from the other measurements in the LRG bins ($\text{LRG1}_\perp$ and
        $\text{LRG2}_\parallel$) appear in blue and that from all measurements except these two in
        black.
        The posteriors from DESI excluding the LRG1 and LRG2 bins entirely appear as unfilled,
        dashed black/white lines.
        While the $\text{LRG1}_\perp$ and $\text{LRG2}_\parallel$ combination (in blue) is
        quite consistent with all other tracers and only marginally shrinks the posterior when
        combined with them (comparing the black/white dashed lines to the gold contours), the
        apparent outliers (red) are in clear tension with all other measurements and shift the
        combined posterior (black) substantially in the direction that disfavors massive neutrinos.
    }
    \label{fig:geometry-wmrd2-Wm-desi-lrg-subsets}
\end{centering}
\end{figure}

\Cref{fig:geometry-wmrd2-Wm-desi-lrg-subsets} displays joint posteriors from DESI BAO data that
include and exclude the outliers identified in \cref{fig:desi-para-perp-compare-lrg}.
The $2 \sigma$ mass level of their combination does not overlap with that from other tracers
or even that from the concordant $\text{LRG1}_\perp$ and $\text{LRG2}_\parallel$ measurements.
Taking \Planck{}'s shape-based calibration of $\omega_b + \omega_c \approx 0.142$ in standard
cosmology, \cref{fig:desi-para-perp-compare-lrg,fig:geometry-wmrd2-Wm-desi-lrg-subsets} shows that
these outliers clearly bear most of the responsibility for favoring \LCDM{} parameter space that is
incompatible with the effect of massive neutrinos on geometry and structure.
Excluding these outliers yields a constraint that agrees better with recent SNe datasets (that all
prefer larger matter fractions) and allows for substantially larger neutrino masses.

\begin{figure}[t!]
\begin{centering}
    \includegraphics[width=\textwidth]{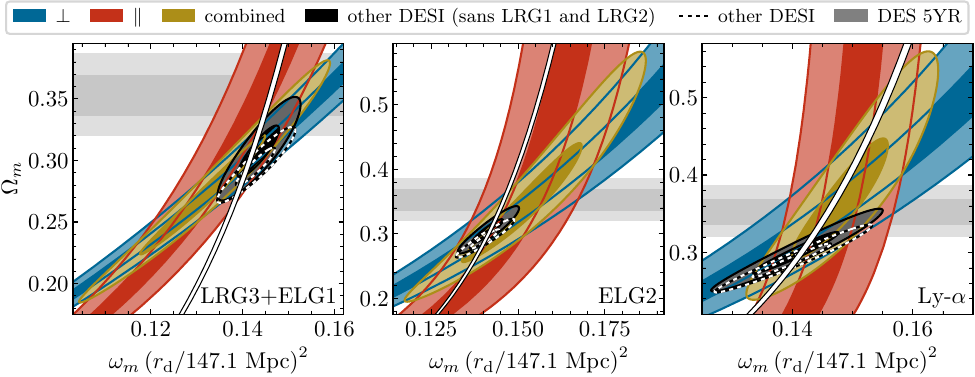}
    \includegraphics[width=\textwidth]{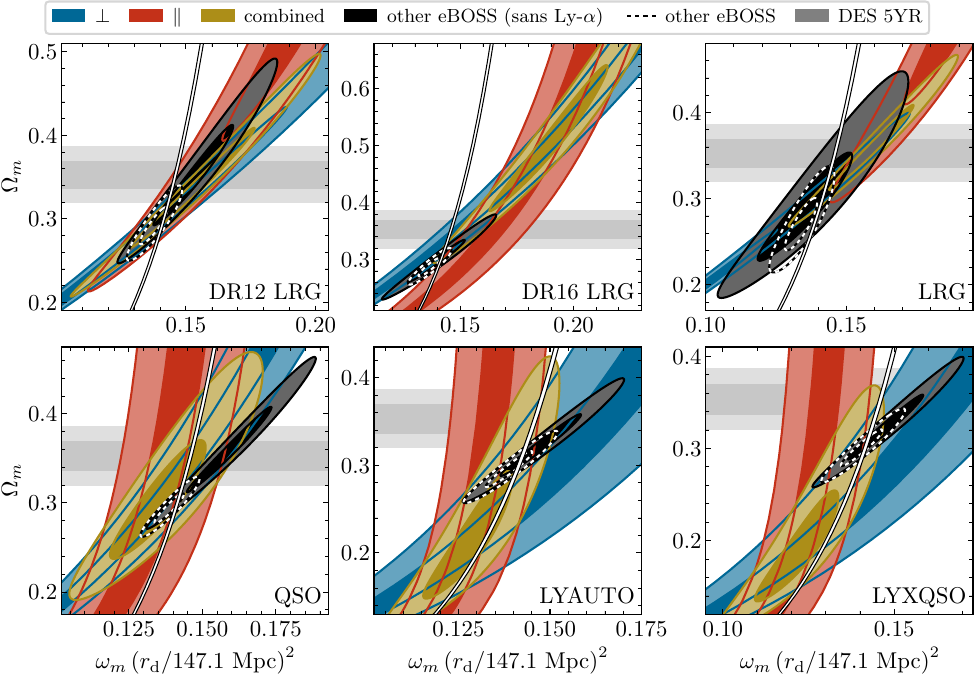}
    \caption{
        Comparison of \LCDM{} parameter space compatible with BAO measurements from individual
        tracers (by panel), displaying posteriors deriving from transverse (blue) and line-of-sight
        (red) measurements and their combination (gold), presented as in
        \cref{fig:desi-para-perp-compare-lrg}.
        The top figure display results from DESI DR1, including posteriors from all other DESI
        tracers (dashed black/white lines), from all other DESI tracers excluding both LRG bins
        (black), from the DES 5YR SNe sample (grey), and from \Planck{} (white band with extent
        covering its $5 \sigma$ interval for visibility).
        The bottom figure displays analogous results for the final eBOSS BAO dataset, including results from all other eBOSS tracers (dashed black/white lines) and
        all other eBOSS tracers excluding both Ly-$\alpha$ bins as well (black).
    }
    \label{fig:desi-sdss-para-perp-compare-other-tracers}
\end{centering}
\end{figure}

\Cref{fig:desi-sdss-para-perp-compare-other-tracers} displays the same internal consistency test
(within flat \LCDM{} cosmologies) from \cref{fig:desi-para-perp-compare-lrg} for all other DESI and
eBOSS tracers that include separate transverse and line-of-sight measurements.
These results demonstrate the internal consistency of all other tracers from DESI DR1 (i.e., except
for the LRG1 and LRG2 measurements displayed in \cref{fig:desi-para-perp-compare-lrg}), as well as
most eBOSS tracers.
The eBOSS results that include Ly-$\alpha$ measurements (LYAUTO and LYXQSO), however, are moderately
discrepant with the joint result from other tracers, and, like DESI's LRG measurements, prefer
parameter space that is largely incompatible with massive neutrinos.
(The LRG measurements from SDSS DR16 [at $z = 0.698$] are also slightly offset from all other
tracers, but in the opposite direction in parameter space.)
\Cref{fig:desi-sdss-para-perp-compare-other-tracers} thus demonstrates the sensitivity of neutrino
mass constraints to potential outliers (again, when interpreted in flat \LCDM{} cosmologies) from
individual tracers and explains our combination of SDSS data below and DESI data above redshift
$0.8$ (using DESI-only Ly-$\alpha$) as an illustrative example.

\begin{figure}[t!]
\begin{centering}
    \includegraphics[width=\textwidth]{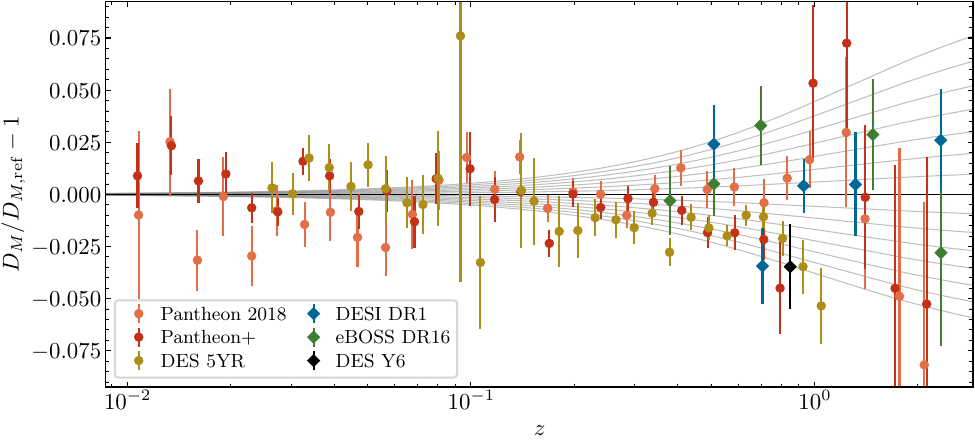}
    \caption{
        Comparison of transverse distance measurements $D_M$ versus redshift from various SNe
        (circles) and BAO (diamonds) surveys, normalized to a fiducial cosmology with
        $\Omega_m = 0.31$ and $h = 0.685$.
        Results take \Planck{}'s best-fit $r_\mathrm{d} = 147.1~\mathrm{Mpc}$ and fix
        the fiducial magnitude $M_B$ to the best fit for each individual SNe survey (to remove any
        calibration difference---e.g., datasets differ in their conventions for normalizing
        magnitudes).
        The SNe data are binned as described in the text.
        Grey lines indicate the residuals for cosmologies with different $\Omega_m$, spaced by
        increments of $0.01$.
        Note that, on a bin-by-bin basis, these comparisons do depend on the fiducial calibration
        taken, but different calibrations only have a uniform impact across redshifts.
    }
    \label{fig:sne-vs-bao-distances}
\end{centering}
\end{figure}

Finally, \cref{fig:sne-vs-bao-distances} compares transverse distance measurements from SNe (i.e.,
translated from their luminosity distances) to those from BAO as a function of redshift, displayed
relative to a fiducial cosmology with $\Omega_m = 0.31$.
To facilitate visualization, the SNe data are binned into 30 logarithmically spaced intervals
between the minimum and maximum redshift of each survey in a manner similar to
Ref.~\cite{Poulin:2024ken}, computing inverse-variance--weighted redshifts and magnitudes using the
sub-block of the covariance matrix within each bin.
That is, the measurements $i$ within a bin with subcovariance $C$ have weights
$w_i = [C^{-1}]_{ii}$ and combine into a magnitude $\sum_i w_i \hat{m}_i / \sum_i w_i$ at redshift
$\sum_i w_i z_i / \sum_i w_i$ with variance $1 / \sum_i w_i$.
These binned magnitudes are then translated to transverse distances and their errors propagated
accordingly.
Cross-bin covariance is neglected, but it cannot be depicted in \cref{fig:sne-vs-bao-distances}
anyway.
At $z \gtrsim z_{m-\Lambda} \approx 0.3$, measurements from the most recent SNe datasets
(Pantheon+ and DES 5YR) in \cref{fig:sne-vs-bao-distances} systematically trend low, indicative of
their preference for higher matter fractions $\Omega_m$, i.e., lower $z_{m-\Lambda}$; most of those
from eBOSS and DESI, on the other hand, are consistent or skew slightly higher than the fiducial
cosmology (since each prefers $\Omega_m$ near but slightly below $0.31$).

\section{Impact of nonlinear evolution on neutrino mass inference}\label{app:nonlinear}

\begin{figure}[t!]
\begin{centering}
    \includegraphics[width=\textwidth]{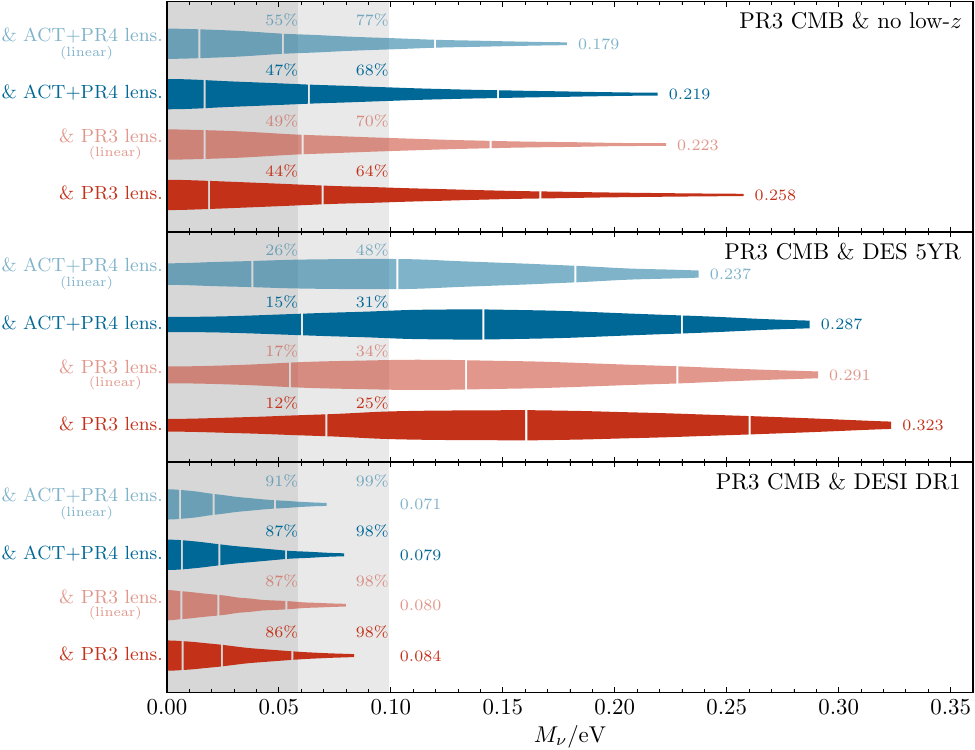}
    \caption{
        Marginal posterior distribution over the neutrino mass sum $\summnu$ when including (opaque)
        and excluding (transparent) nonlinear contributions to the growth of structure.
        Each posterior uses \Planck{} PR3 CMB data combined with various BAO and SNe datasets
        (labeled on each panel) and PR3 lensing (red) or ACT+PR4 lensing (blue).
        Results are presented as in \cref{fig:summnu-violin-vary-lens-asmear-low-z}.
    }
    \label{fig:summnu-violin-linear-vs-nonlinear}
\end{centering}
\end{figure}

In this appendix we demonstrate the impact of nonlinear structure growth on neutrino mass inference
with current datasets.
\Cref{fig:summnu-violin-linear-vs-nonlinear} compares the one-dimensional posteriors over $\summnu$
for various dataset combinations when including and excluding nonlinear effects (modeled with
HMCODE-2016~\cite{Mead:2016zqy}).
Neglecting nonlinear evolution clearly biases neutrino mass constraints, artificially shifting
posteriors to lower values to a modest degree.
The bias varies between $0.03~\mathrm{eV}$ (for PR3 lensing data) and $0.05~\mathrm{eV}$ (for
ACT+PR4) except when combining with DESI data.
In that case, DESI's constraints on the expansion history (combined with \Planck{}'s primary CMB
data) disfavor the larger matter fractions that are required to accommodate massive neutrino's
geometric effects and that would give more room for neutrinos to suppress structure growth,
independent of whether nonlinear effects are included.
Though the posteriors do shift marginally, accounting for nonlinear evolution does not alter any
qualitative conclusions for dataset combinations that include DESI DR1 (like, for instance, their
incompatibility with the inverted hierarchy).

\begin{figure}[t!]
\begin{centering}
    \includegraphics[width=\textwidth]{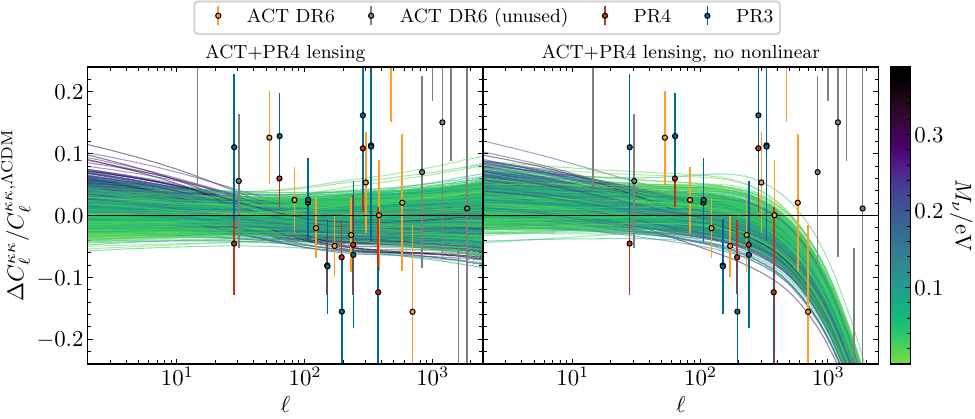}
    \caption{
        Residuals of the CMB lensing convergence relative to the \LCDM{} best fit (using \Planck{}
        PR3 CMB and ACT+PR4 lensing data) for samples from various posteriors that include (left)
        and exclude (right) nonlinear corrections to the growth of structure.
        Note that the reference result includes no massive neutrinos but does account for nonlinear
        evolution.
        Curves are colored by their value of $\summnu$ and data from various observations are also
        depicted as labeled in the legend; note that the data included for the posteriors depicted
        here are labeled ACT DR6 (yellow) and PR4 (red).
    }
    \label{fig:mnu3-lens-spaghetti-linear-vs-nonlinear}
\end{centering}
\end{figure}

To illustrate what in the lensing data allows for heavier neutrinos when nonlinear effects are
accounted for, \cref{fig:mnu3-lens-spaghetti-linear-vs-nonlinear} displays the residuals of
posterior samples relative to the \LCDM{} best-fit cosmology (with massless neutrinos only) with
current lensing data superimposed.
Though nonlinear effects play no role at the multipoles $30 \lesssim L \lesssim 200$ where the data
exhibit a clear downward trend relative to the \LCDM{} best-fit, they have a significant impact at
higher multipoles where the data skew back toward positive residuals.
Though the measurements at $L \gtrsim 300$ are noisy and do not incontrovertibly demonstrate a
systematic upward trend, the majority of the data points do indeed skew upward.
In this range, the posterior samples for fully linear predictions drop precipitously and are unable
to accommodate these data points, especially for heavier neutrino masses.
Because nonlinear evolution enhances small-scale structure, one can only expect this trend to become
more important as future observations improve the accuracy of measurements at high multipoles,
underscoring the importance of nonlinear modeling to unbiased neutrino mass constraints using such
data.

\bibliography{references,manual}

\end{document}